\documentclass[aps,superscriptaddress,nofootinbib,floatfix,A4]{revtex4}

\usepackage{amsfonts}
\usepackage{amsmath}
\usepackage{amssymb}
\usepackage{amsthm}
\usepackage{graphicx}
\usepackage{booktabs}
\usepackage{theorem}
\usepackage{array}
\usepackage{color}

\usepackage{graphicx}
\usepackage{amssymb}
\usepackage{amsmath}
\usepackage{axodraw}

\allowdisplaybreaks[1]

\numberwithin{equation}{section}

\newcommand{\be}{\begin{equation}}
\newcommand{\ee}{\end{equation}}
\newcommand{\beq}{\begin{equation}}
\newcommand{\eeq}{\end{equation}}
\newcommand{\bea}{\begin{eqnarray}}
\newcommand{\eea}{\end{qnarray}}

\begin{document}


\begin{flushright}
SHEP-10-36\\
\end{flushright}
\vspace*{0.7cm}

\title{$Z'$ physics with early LHC data}

\author{Elena Accomando
\footnote{E-mail: \texttt{e.accomando@soton.ac.uk}}}
\affiliation{School of Physics \& Astronomy, University of Southampton,\\ 
        Highfield, Southampton SO17 1BJ, UK}
\affiliation{Particle Physics Department, Rutherford Appleton Laboratory,  
       \\Chilton, Didcot, Oxon OX11 0QX, UK} 
	
\author{Alexander Belyaev
\footnote{E-mail: \texttt{a.belyaev@soton.ac.uk}}}
\affiliation{School of Physics \& Astronomy, University of Southampton,\\ 
        Highfield, Southampton SO17 1BJ, UK}
\affiliation{Particle Physics Department, Rutherford Appleton Laboratory, 
       \\Chilton, Didcot, Oxon OX11 0QX, UK}
       
\author{Luca Fedeli
\footnote{E-mail: \texttt{fedeli@fi.infn.it}}}
\affiliation{INFN, 50019 Sesto F., Firenze, Italy and \\
Department of Physics and Astrophysics, University of Florence, \\
50019 Sesto F., Firenze, Italy}
       
\author{Stephen~F.~King
\footnote{E-mail: \texttt{king@soton.ac.uk}}}
\affiliation{School of Physics \& Astronomy, University of Southampton,\\ 
        Highfield, Southampton SO17 1BJ, UK}
	
\author{Claire Shepherd-Themistocleous
\footnote{E-mail: \texttt{C.H.Shepherd-Themistocleous@rl.ac.uk}}}
\affiliation{Particle Physics Department, Rutherford Appleton Laboratory, 
       \\Chilton, Didcot, Oxon OX11 0QX, UK}

\begin{abstract}
\noindent

We discuss the prospects for setting limits on or discovering spin-1 $Z'$ bosons
using early LHC data at 7 TeV.  Our results are based on the narrow width approximation
in which the leptonic Drell-Yan $Z'$ boson production cross-section
only depends on the $Z'$ boson mass together with two parameters $c_u$ and $c_d$.
We carefully discuss the experimental cuts that should be applied
and tabulate the theoretical next-to-next-to-leading order corrections which must be included.
Using these results the approach then provides a safe, convenient and unbiased way of comparing experiment to
theoretical models which avoids any built-in model dependent assumptions.
We apply the method to three classes of perturbative $Z'$ boson benchmark models: 
$E_6$ models, left-right symmetric models and sequential standard models.
We generalise each class of model in terms of  {mixing angles which continuously
parametrize} linear combinations of pairs of generators and lead to distinctive
orbits in the $c_u-c_d$ plane. We also apply this method to the strongly coupled
four-site benchmark model in which two $Z'$ bosons are predicted. 
By comparing the experimental limits or 
discovery bands to the theoretical predictions on the $c_u$-$c_d$ plane,
we show that the LHC at 7 TeV with integrated luminosity 
of 500 pb$^{-1}$ will greatly improve on current Tevatron mass limits for the benchmark models.
If a $Z'$ is discovered our results show that measurement of the mass and cross-section will provide a powerful
discriminator between the benchmark models using this approach.  
\end{abstract}


\maketitle

\setcounter{footnote}{0}


\section{Introduction}

The end of the first decade of the millenium is an exciting time in particle
physics, with the CERN LHC enjoying an extended run at 7 TeV, and the Fermilab
Tevatron collecting unprecedented levels of integrated luminosity, eventually up to
perhaps 10 fb$^{-1}$, in the race to discover the first signs of new physics Beyond
the Standard Model (BSM).  Since spin-1 $Z'$ bosons are predicted by dozens of such
models, and  are very easy to discover in the leptonic Drell-Yan mode, this makes
them good candidates for an early discovery at the LHC. For a review see Refs.
\cite{Langacker:2008yv,Erler:2009jh,Nath:2010zj} and references therein. 
Furthermore high mass $Z'$
bosons are more likely to be discovered at the LHC than the Tevatron
\cite{Carena:2004xs}, since energy is more important than luminosity for the
discovery of high mass states. This makes the study of $Z'$
bosons both timely and promising and has led to widespread recent interest in this
subject (see for example 
\cite{Chen:2008zz,Coriano:2008wf,King:2005jy,King:2005my,Howl:2007zi,%
Li:2009xh,Diener:2010sy,Erler:2010uy,Petriello:2008zr,%
Accomando:2010ir,Basso:2010pe,Athron:2009ue,%
Athron:2009bs,Appelquist:2002mw,Hays:2009dh,Salvioni:2009mt,Salvioni:2009jp,Coriano:2008wf}). 


Since one of the purposes of this paper is to facilitate the connection between experiment and theory,
it is worth being clear at the outset precisely what we shall mean by a $Z'$ boson.
To an experimentalist a $Z'$ is a resonance ``bump'' more massive than the $Z$ of the
Standard Model (SM) which can be observed in Drell-Yan production followed by its decay
into lepton-antilepton pairs. To a phenomenologist a $Z'$ boson is a new massive electrically
neutral, colourless boson (equal to its own antiparticle) which couples to SM 
matter. To a theorist it is useful to classify the
$Z'$ according to its spin, even though actually measuring its spin will require high
statistics. For example a spin-0 particle could correspond to a sneutrino in R-parity
violating supersymmetric (SUSY) models. A spin-2 resonance could be identified as a
Kaluza-Klein (KK) excited graviton in Randall-Sundrum models.  However a spin-1 $Z'$ is
by far the most common possibility usually considered, and this is what we shall mean by 
a $Z'$ boson in this paper.

In this paper, then, we shall discuss electrically neutral colourless spin-1 $Z'$ bosons, which are produced
by the Drell-Yan mechanism and decay into lepton-antilepton pairs, yielding a resonance bump more massive
than the $Z$. We shall be particularly interested in the prospects for discovering or setting limits on such
$Z'$ bosons using early LHC data. By early LHC data we mean the present 2010/11 run at the LHC at 7 TeV,
which is anticipated to yield an integrated luminosity approximately of 1 fb$^{-1}$. Since the present LHC
schedule involves a shut-down during 2012, followed by a restart in 2013, the early LHC data will provide the
best information possible about $Z'$ bosons over the next three years, so in this paper we shall focus
exclusively on what can be achieved using these data, comparing the results with current Tevatron limits.  In
order to enable contact to be made between early LHC experimental data and theoretical models, we advocate
the narrow width approximation, in which the leptonic Drell-Yan $Z'$ boson production cross-section only
depends on the $Z'$ boson mass together with two parameters $c_u$ and $c_d$ \cite{Carena:2004xs}. Properly
defined experimental information on the $Z'$ boson cross-section may then be recast as limit or discovery
contours in the $c_u-c_d$ plane, with a unique contour for each value of $Z'$ boson mass. In order to
illustrate how this formalism enables contact to be made with theoretical models we study three classes of
$Z'$ boson benchmark models:  $E_6$ models, left-right (LR) symmetric models and sequential standard models
(SSM). We also apply this method to the strongly coupled four-site benchmark model in which two $Z'$ bosons
are predicted
\cite{Accomando:2010ir}.
Each benchmark model may be expressed in the $c_u-c_d$ plane which enables contact to be made with the
experimental limit or discovery contours. Working to next-to-next-leading order (NNLO) we show that the LHC
at 7 TeV with as little data as 500 pb$^{-1}$ can either greatly improve on current Tevatron mass limits, or
discover a $Z'$, with a measurement of the mass and cross-section providing powerful resolving power between
the benchmark models using this approach. We also briefly discuss the impact of the $Z'$ boson width on
search strategies. Although the width $\Gamma_{Z'}$ into standard model particles may in principle be
predicted as a function of $M_{Z'}$, in practice there may be considerable uncertainty concerning
$\Gamma_{Z'}$ due for example to the possible decay into other non-standard particles, including
supersymmetric (SUSY) partners and exotic states, for example. 

At the outset we would like to highlight some of the strengths and 
limitations of our approach to the benchmark models. One of the strengths of 
our approach is that the considered benchmark models encompass two quite 
different types of $Z'$ models: perturbative gauge models and strongly 
coupled models, where the perturbative models generally involve relatively 
narrow widths (which however can get larger if SUSY and exotics are included 
in the decays in addition to SM particles), while the strongly coupled models 
involve multiple $Z'$ bosons with rather broad widths. The perturbative 
benchmark gauge models are defined in terms 
of {continuous mixing angles, 
in analogy to the $E_6$ class of models 
expressed through the linear combinations of $\chi$ and $\psi$ generators}. This 
approach is generalised to the case of LR models involving linear 
combinations of the generators $R$ and  $B-L$
\footnote{We remark that in 
\cite{Salvioni:2009mt} the authors focussed on models obtained by taking a linear
combination of $Y$ and $B-L$ which is related  to the generalized LR models that we also consider,
however the choice of generator basis and parametrisation is different
and they did not use {the mixing angle parametrisation} that we propose here.}, and 
similarly we define a new class of SM-like $Z^\prime$ models involving linear 
combinations of $L$ and $Q$ generators, with the precise linear combination 
in each case parameterised by a separate angle. 
The strength of this approach is that it enables a finite number of classes, 
each containing an infinite number of benchmark models, to be defined, rather 
than just a finite number of models. For each class of models, the
respective angle serves to parametrize the specific orbit which represents 
that class in the $c_u$-$c_d$ plane. The different orbits turn out to be 
non-overlapping for the abovementioned three classes of models which in the 
following we label as: $E_6(\theta )$, $GLR(\phi )$ and $GSM(\alpha )$ 
respectively. A limitation of the approach is that it ignores the effect of 
the SUSY and exotics on the width $\Gamma_{Z'}$. In
addition it also ignores the effects of $Z-Z'$ mixing since this is model dependent. However any such mixing
must be small due to the constraints from electroweak precision measurements, 
and we refer to such constraints on the mixing angle where possible. As regards the strongly
coupled models, in principle $Z'$ bosons could emerge from techni-rho bound states in Walking
Technicolour models, or a series of strongly interacting resonances such KK excitations of the $Z$ boson. A limitation of our approach here is that we only 
consider the four-site Higgless model which contains just two excitations of 
the $Z$ (and $W$) bosons, as a simple
approximation to both the Walking technicolour models and  the KK excitations of the $Z$. 
Neverthless the four-site model is representative of the physics of a typical strongly
interacting $Z'$ model and by representing it for the first time in the $c_u$-$c_d$ plane, it
is clearly seen that the associated  $Z'$ bosons may easily be distinguished from those of the
perturbative gauge models.

The layout of the rest of the paper is as follows. In section \ref{II} we
describe our model independent approach based on the narrow width
approximation and the variables $M_{Z'}$, $c_u$ and $c_d$. Higher order
corrections to the cross-section are tabulated in the narrow width
approximation and new K factors are defined which enable the $c_u$ and $c_d$
approach to be reliably extended to NNLO. In this section we consider finite
width effects and discuss the choice of invariant mass window around
$M_{Z'}$ which matches the narrow width approximation, showing that in the
case of the LHC this choice is only weakly constrained. We also comment on
the effects of interference and show that they may become important for
invariant mass cuts close to 100 GeV but are negligible for the a suitable
invariant mass cut around $M_{Z'}$, which we therefore advocate.  In section
\ref{III} we define our benchmark models based on  $E_6$, LR and SSM,
generalised using  {variables which continuously parametrize mixing of
the respective $U(1)$ generators}. 
 
For these perturbative models we specify the gauge coupling and  calculate the vector 
and axial couplings in 
terms of the {mixing angle} variable. 
We tabulate the results for some special choices of the {mixing angle} variable which reproduce the
models commonly quoted and analysed in the
literature. 
For the four-site model we discuss the parameter space describing the two 
$Z'$ bosons allowed by electroweak precision measurements and unitarity. In 
section \ref{IV} we apply the model independent approach to the benchmark 
models discussing first the current Tevatron limits using the latest $D0$ 
results with collected luminosity $L=5.4 fb^{-1}$ , then the expected LHC 
potential based on the projected CMS limits on the $Z^\prime$ boson cross section 
normalized to the SM $Z$-boson one for an integrated luminosity 
$L=500 pb^{-1}$. We use the $D0$ results rather than $CDF$ since they are 
more closely related to the narrow width approximation that we advocate, and 
we use CMS rather than ATLAS since the projected limits are publicly 
available. In both cases we express the experimental cross-section limits in 
the $c_u$-$c_d$ plane and compare these limits to the benchmark models also 
displayed in the $c_u$-$c_d$ plane. In the case of CMS we also show the 
discovery limits. We tabulate some of the results for some special choices of 
the {mixing angle} variable, including the width, the leptonic branching ratio, 
$c_u$ and $c_d$ values, our derived current direct limit on the $Z^\prime$ mass 
based on $D0$ results, as well as other indirect and mixing limits where 
available.
Finally in section \ref{V} we discuss the impact of $\Gamma_{Z'}$ on search 
strategies.
Section \ref{VI} summarises and concludes the paper.


\section{Model Independent Approach \label{II}}

\subsection{Couplings}

At collider energies, the gauge group of a typical model predicting a single 
extra $Z^\prime$ boson is:
\beq
SU(3)_C\times SU(2)_L\times U(1)_Y \times U^\prime (1)
\label{low}
\eeq
where the Standard Model is augmented by an additional $U^\prime (1)$ gauge 
group. The $U^\prime (1)$ gauge group is broken near the TeV scale 
giving rise to a massive $Z^\prime$ gauge boson with couplings to a SM 
fermion $f$ given by:
$$
{\mathcal L}_{NC}=\frac{g'}{2}Z^\prime_{\mu}\bar{f}\gamma^\mu (g_V^f  - g_A^f\gamma^5) f .
$$
The values of $g_V^f,g_A^f$ depend on the particular choice of $U^\prime (1)$
and on the particular fermion $f$. We assume universality amongst the three 
families,
$g_V^u=g_V^c=g_V^t$, and $g_V^d=g_V^s=g_V^b$, as well as
$g_V^e=g_V^{\mu}=g_V^{\tau}$, and similarly for the $g_A^f$ couplings,
which means that there are eight model dependent couplings to SM fermions
$g_{V,A}^f$, with $f=u,d,e,\nu$. These eight couplings are not all independent since
they are related to the left (L) and right (R) couplings as follows:
$$
{\mathcal L}_{NC}=\frac{g'}{2}Z^\prime_\mu\bar{f}\gamma^{\mu}(g_V^f-g_A^f\gamma^5) f
 = {g'} Z^\prime_{\mu}\bar{f}\gamma^{\mu}(\epsilon_L^fP_L + \epsilon_R^fP_R ) f .
$$
where $P_{L,R}=(1\mp \gamma_5)/2$, and 
$g_{V,A}^f=\epsilon_L^f\pm \epsilon_R^f$.
The couplings are not all independent since the left-handed fermions
are in doublets with the same charges $\epsilon_L^u=\epsilon_L^d$
and $\epsilon_L^e=\epsilon_L^{\nu}$. Excluding the right-handed neutrinos
(which we assume to be heavier than the $Z'$) there are really five independent
couplings $\epsilon_L^{e=\nu},\epsilon_L^{u=d}, \epsilon_R^u, \epsilon_R^d, 
\epsilon_R^e$. However we prefer to work with the eight vector and axial 
couplings $g_{V,A}^f$. In addition, the strength of the gauge coupling $g'$ 
is model dependent. Throughout, we follow the conventions of 
\cite{Langacker:2008yv}.

A slightly more complicated setup is needed to describe the four-site model
which, in this paper, has been chosen to represent Higgsless multiple 
$Z^\prime$-boson theories. The corresponding framework will be given in
Sec.\ref{4site-Model}.

Throughout the paper we shall ignore the couplings of the $Z'$ to beyond SM 
particles such as right-handed neutrinos, SUSY partners and any exotic 
fermions in the theory, which all together may increase the width of the $Z'$ 
by up to about a factor of five \cite{Kang:2004bz} and hence lower the 
branching ratio into leptons by the same factor.

\subsection{$Z'$ production and decay in the narrow width approximation}
\label{sec:nwa}

The $Z'$ contribution to the Drell-Yan production cross-section of 
fermion-antifermion pairs in a symmetric mass window around the $Z^\prime$ mass 
($|M-M_{Z^\prime}|\le \Delta$) may be written as:
\beq
\sigma_{f\overline{f}} = \int_{(M_{Z'}-\Delta)^2}^{(M_{Z'}+\Delta)^2}
\frac{d\sigma }{dM^2}(pp\rightarrow Z'\rightarrow f\overline{f}X)dM^2.
\eeq
In the narrow width approximation (NWA), it becomes
\beq
\sigma_{f\overline{f}} \approx \left( \frac{1}{3}\sum_{q=u,d}
\left( \frac{dL_{q\overline{q}}}{dM_{Z'}^2}\right) 
\hat{\sigma}(q \overline{q}\rightarrow Z')\right)
\times Br(Z' \rightarrow f\overline{f})
\label{ff}
\eeq
where the parton luminosities are written as $\left(
\frac{dL_{q\overline{q}}}{dM_{Z'}^2}\right)$ and $ \hat{\sigma}(q
\overline{q}\rightarrow Z')$ is the peak cross-section given by:
\beq
\hat{\sigma}(q \overline{q}\rightarrow Z')=
\frac{\pi}{12}{g'}^2[(g_V^q)^2 + (g_A^q)^2 ].
\eeq
The branching ratio of the $Z'$ boson into fermion-antifermion pairs is
\beq
Br(f\bar f) \equiv Br(Z' \rightarrow f\bar f)= 
\frac{\Gamma(Z' \rightarrow f\bar f)}{\Gamma_{Z'}}
\eeq
where $\Gamma_{Z'}$ is the total $Z'$ width and the partial widths
into a particular fermion-antifermion pair of $N_c$ colours is given by
\beq
\Gamma(Z' \rightarrow f\overline{f})=
N_c\frac{{g'}^2}{48\pi}M_{Z'}[(g_V^f)^2 + (g_A^f)^2 ].
\label{width_ff}
\eeq
Assuming only SM fermions in the final state and neglecting in first 
approximation their mass, one finds the total width:
\beq
\Gamma_{Z'}=
\frac{{g'}^2}{48\pi}M_{Z'}\left[
9({g_V^u}^2 + {g_A^u}^2)+9({g_V^d}^2 + {g_A^d}^2)+3({g_V^e}^2 + {g_A^e}^2)+
3({g_V^{\nu}}^2 + {g_A^{\nu}}^2)
\right].
\label{total_width}
\eeq

Specializing to the charged lepton pair production cross-section relevant for
the first runs at the LHC, Eq.\ref{ff} may be written at the leading order (LO)
as \cite{Carena:2004xs}:
\beq
\sigma_{\ell^+\ell^-}^{LO} = \frac{\pi}{48 s}
\left[ c_uw_u(s,M_{Z'}^2)+ c_dw_d(s,M_{Z'}^2)\right]
\label{eq:ll}
\eeq
where the coefficients $c_u$ and $c_d$ are given by:
\beq
c_{u}=\frac{{g'}^2}{2}({g_V^u}^2+{g_A^u}^2)Br(\ell^+\ell^-), \ \ \ \
c_{d}=\frac{{g'}^2}{2}({g_V^d}^2+{g_A^d}^2)Br(\ell^+\ell^-),
\label{eq:cucd}
\eeq
and $w_u(s,M_{Z'}^2)$ and $w_d(s,M_{Z'}^2)$ are related to the parton
luminosities $\left( \frac{dL_{u\overline{u}}}{dM_{Z'}^2}\right)$ and $\left(
\frac{dL_{d\overline{d}}}{dM_{Z'}^2}\right)$ and therefore only depend on the
collider energy and the $Z'$ mass. All the model dependence of the 
cross-section is therefore contained in the two coefficients, $c_u$ and $c_d$. 
These parameters can be calculated from $g_V^f, g_A^f$ and $g'$, assuming   
only SM decays of the $Z'$ boson. The corresponding values for all models 
which predict a single $Z^\prime$ boson purely decaying into SM fermions are 
given in Table \ref{models2}. 

A slight complication arises in Higgsless theories, which in the present 
paper are represented by the four-site model. Here in fact the two neutral 
extra gauge bosons, $Z_{1,2}$, decay preferebly into di-boson intermediate 
states. Their total width gets therefore two contributions:
\beq
\Gamma_{Z_i}=\Gamma_{Z_i}^{f\bar f}+\Gamma_{Z_i}^{VV}\ \ \ \ \ \ (i=1,2)
\eeq
where the two terms on the right-hand side represent the fermionic and bosonic
decay, respectively. More in detail,
\beq
\Gamma_{Z_i}^{f\bar f}=
\frac{{1}}{48\pi}M_{Z_i}\left[
9({g_{iV}^u}^2+{g_{iA}^u}^2)+9({g_{iV}^d}^2+{g_{iA}^d}^2)+3({g_{iV}^e}^2+
{g_{iA}^e}^2)+3({g_{iV}^{\nu}}^2+{g_{iA}^{\nu}}^2)
\right]
\eeq
\beq
\Gamma_{Z_1}^{WW}={1\over{3\pi}}\left ({1\over{16}}\right )^2{M_{Z1}^3\over
{M_W^2}}(1-z^4)(1+z^2)
\label{z1width}
\eeq
\beq
\Gamma_{Z_2}^{W_1W}={1\over{3\pi}}\left ({1\over{16}}\right )^2{M_{Z2}^3\over
{M_W^2}}z^4(1-z^2)^3[1+10z^2+z^4]
\label{z2width}
\eeq
with i=1,2, where in this case we have
included the $g'$ coupling in the definition of $g_{1,2V}^f$ and $g_{1,2A}^f$. 
In the above formulas $M_{Z1}$ and $M_{Z2}$ are the masses of the
two extra gauge bosons, $Z_{1,2}$, while $z$ is their ratio, i.e. 
$z=M_{Z1}/M_{Z2}$.
The direct consequence of this peculiarity is that the $Z_{1,2}$ leptonic 
branching ratio acquires a not trivial dependence on the $Z_{1,2}$ boson 
mass which reflects in an intrinsic mass dependence of the $c_u$ and $c_d$ 
coefficients. In addition, there is an external source of variation with mass.
As all vector and axial couplings in the four-site model can be expressed in 
terms of the three independent free parameters ($g_{2V}^e$, $M_{Z2}$, $z$), 
$c_u$ and $c_d$ are completely specified by these quantities as well: 
$c_{u,d}=c_{u,d}(g_{2V}^e, M_{Z2}, z)$. This means that at fixed masses, 
$M_{Z2}$ and $z=M_{Z1}/M_{Z2}$, these coefficients get constrained by the 
EWPT bounds acting on $g_{2V}^e$. As these limits vary with mass 
(see Fig.~\ref{fig:parameter_space}), $c_u$ and $c_d$ acquire this extra $M_{Z2}$ 
dependence. The net result opens up a parameter space in the 
$c_d-c_u$ plane which will be dispalyed at due time. 
   
As emphasized in \cite{Carena:2004xs}, the $c_d-c_u$ plane parametrization is 
a model-independent way to create a direct correspondance between the 
experimental bounds on $pp(\bar p) \to Z' \to \ell^+\ell^-$ cross sections 
and the parameters of the Lagrangian. An experimental limit on 
$\sigma(pp(\bar p) \to Z' \to \ell^+\ell^-)$ for a given $Z'$ mass gives in 
fact a linear relation between $c_u$ and $c_d$,
\beq
c_u = a - b c_d
\eeq
where $a,b$ can be regarded as known numbers given by:
\be
a=\frac{48s}{\pi} \frac{\sigma_{\ell^+\ell^-}^{exp}}{w_u}, \ \ \ \ 
b=\frac{w_d}{w_u}.
\eeq
where $\sigma_{\ell^+\ell^-}^{exp}$ represents the 95$\%$ C.L. upper bound on 
the experimental Drell-Yan cross section which can be derived from observed 
data.

In practice, it is more convenient to use a log-log scale resulting in the
limits appearing as contours for a fixed $Z'$ mass in the $c_d-c_u$ plane.
We use this representation in the next subsections.


\subsection{Higher-order corrections}

At higher-orders, the expression for $Z'$ production given by  Eq.~(\ref{eq:ll}) strictly
speaking is no longer valid. However, as it was  shown in Ref.~\cite{Carena:2004xs}, the
additional terms which are not  proportional to $c_u$ and $c_d$ in Eq.~(\ref{eq:ll}) can
be neglected  at NNLO. Therefore, Eq.~(\ref{eq:ll}) gives a quite accurate description
of  the approach we are discussing here even at NNLO.

In the following, we take into account QCD NNLO effects as implemented in 
the WZPROD program~\cite{Hamberg:1990np,vanNeerven:1991gh,ZWPROD}
{as a correction to the total $Z'$ production cross section
in the NWA.\footnote{{We would like to note, that study of 
NLO effects for kinematical distributions involving 
leptons from $Z'$ decay~\cite{Coriano:2008uz,Fuks:2008yp,Fuks:2007gk}
is beyond the scope of this paper.}}
}
 We have 
adopted this package for simulating the $Z^\prime$ production, and have 
linked it to an updated set of Parton Density Functions (PDF). This set 
includes in particular the most recent versions of 
CTEQ6.6~\cite{Kretzer:2003it,Nadolsky:2008zw} and 
MSTW08~\cite{Martin:2009iq} PDF's, which we use in our analysis. We can provide the 
complete code upon request. 

\begin{figure}[]
\includegraphics[width=0.5\textwidth]{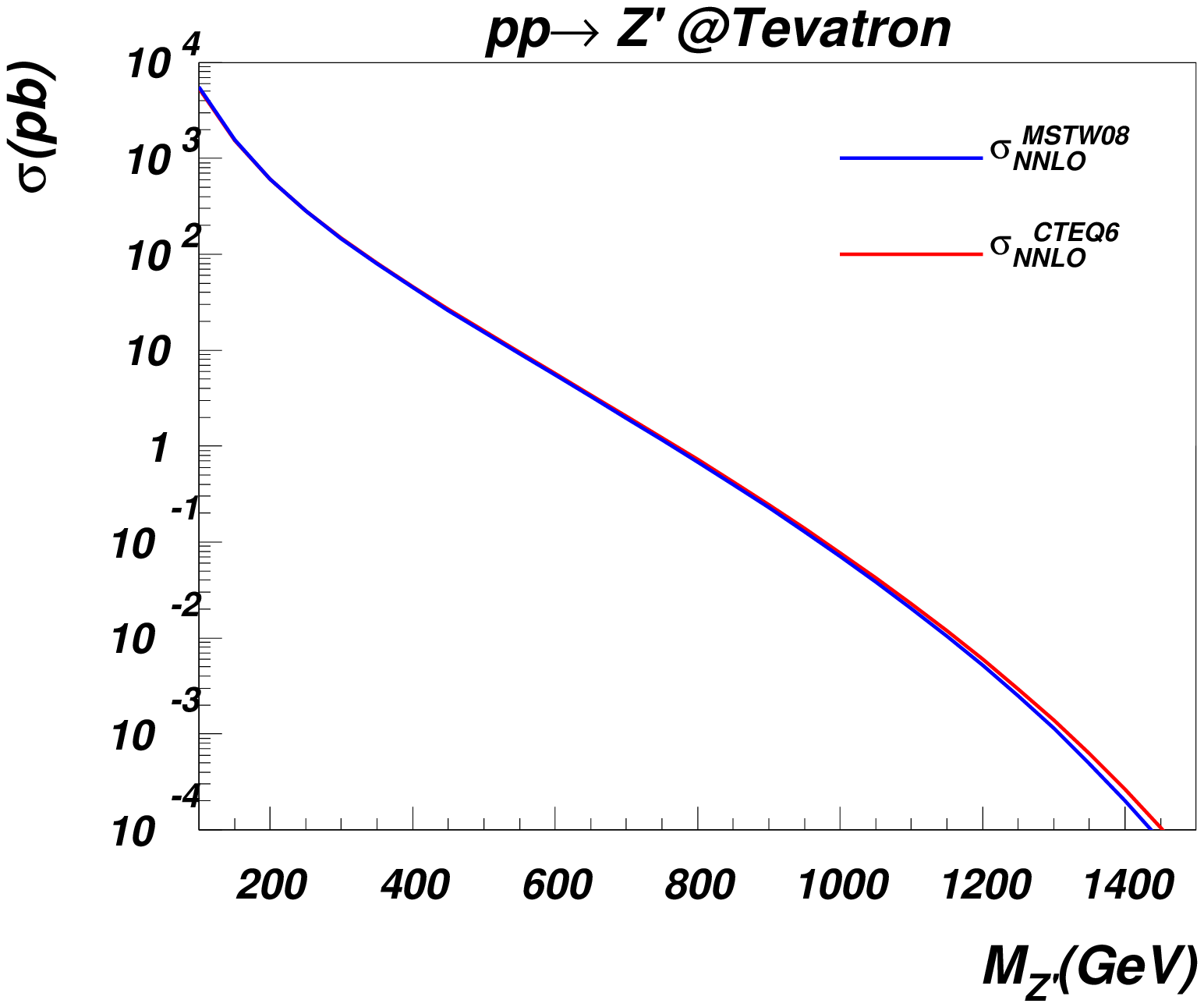}%
\includegraphics[width=0.5\textwidth]{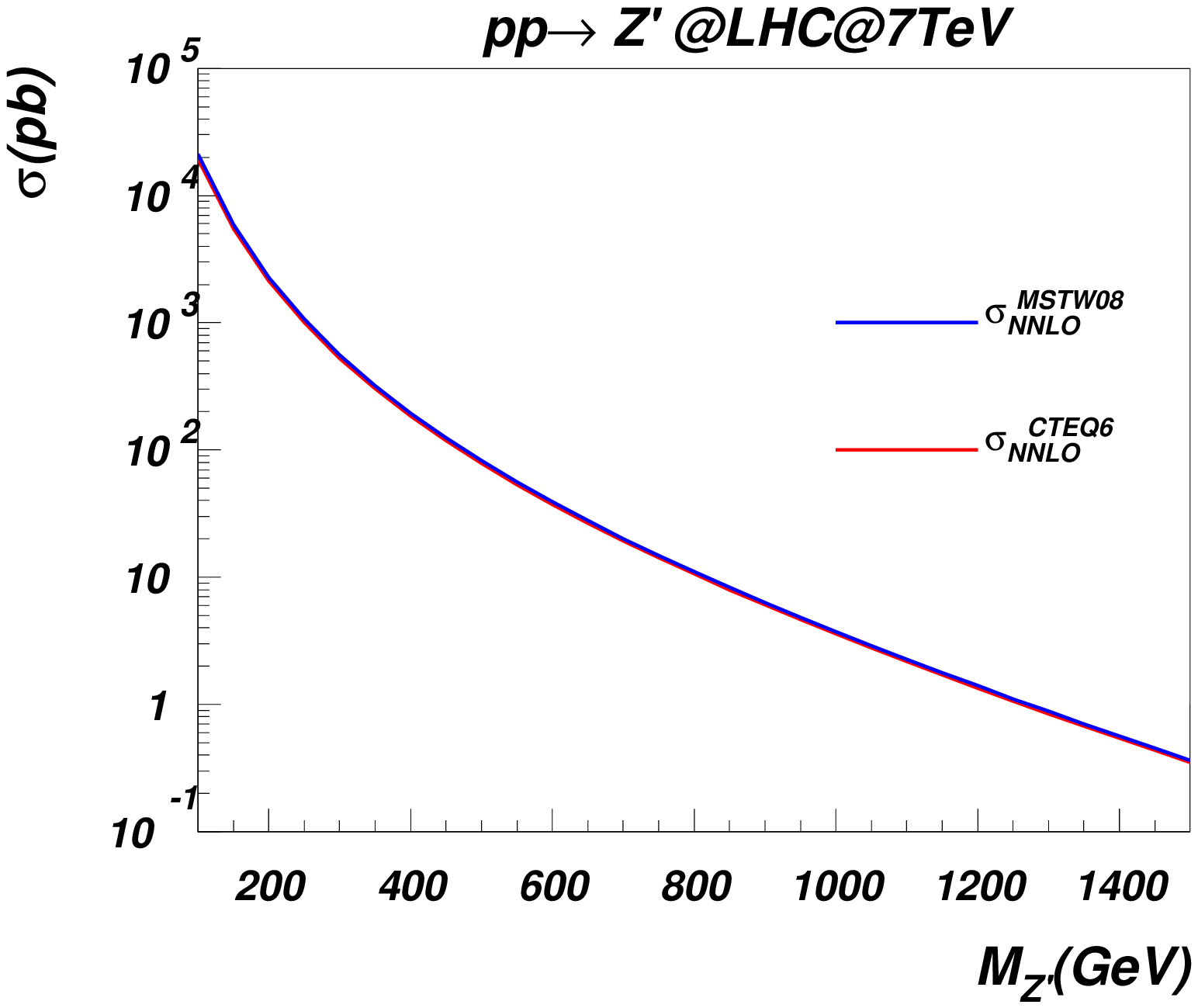}
\caption{\label{fig:snlo}
$\sigma(p\bar p(p)\to Z')_{NNLO}$ for Standard Model-like $Z'$ production at 
the Tevatron (left panel) and the LHC@7TeV (right panel) for CTEQ6.6 and 
MSTW08 PDF's}
\end{figure}

The QCD NNLO $Z^\prime$ production cross sections obtained using CTEQ6.6 and 
MSTW08 PDFs are in a good agreement. Their difference is in fact at the 2-3\% 
level over a wide $Z^\prime$ mass spectrum as shown in Fig.~\ref{fig:snlo}, 
where we plot the total $p\bar{p}(p) \to Z'$ cross section at the Tevatron 
and LHC@7TeV versus the $Z'$ mass. Here, we have taken as factorization 
scale the value $Q=M_{Z^\prime}$. The further detailed analysis of the cross 
section variation with the scale is outside of the scope of the current paper.

\begin{figure}[]
\includegraphics[width=0.5\textwidth]{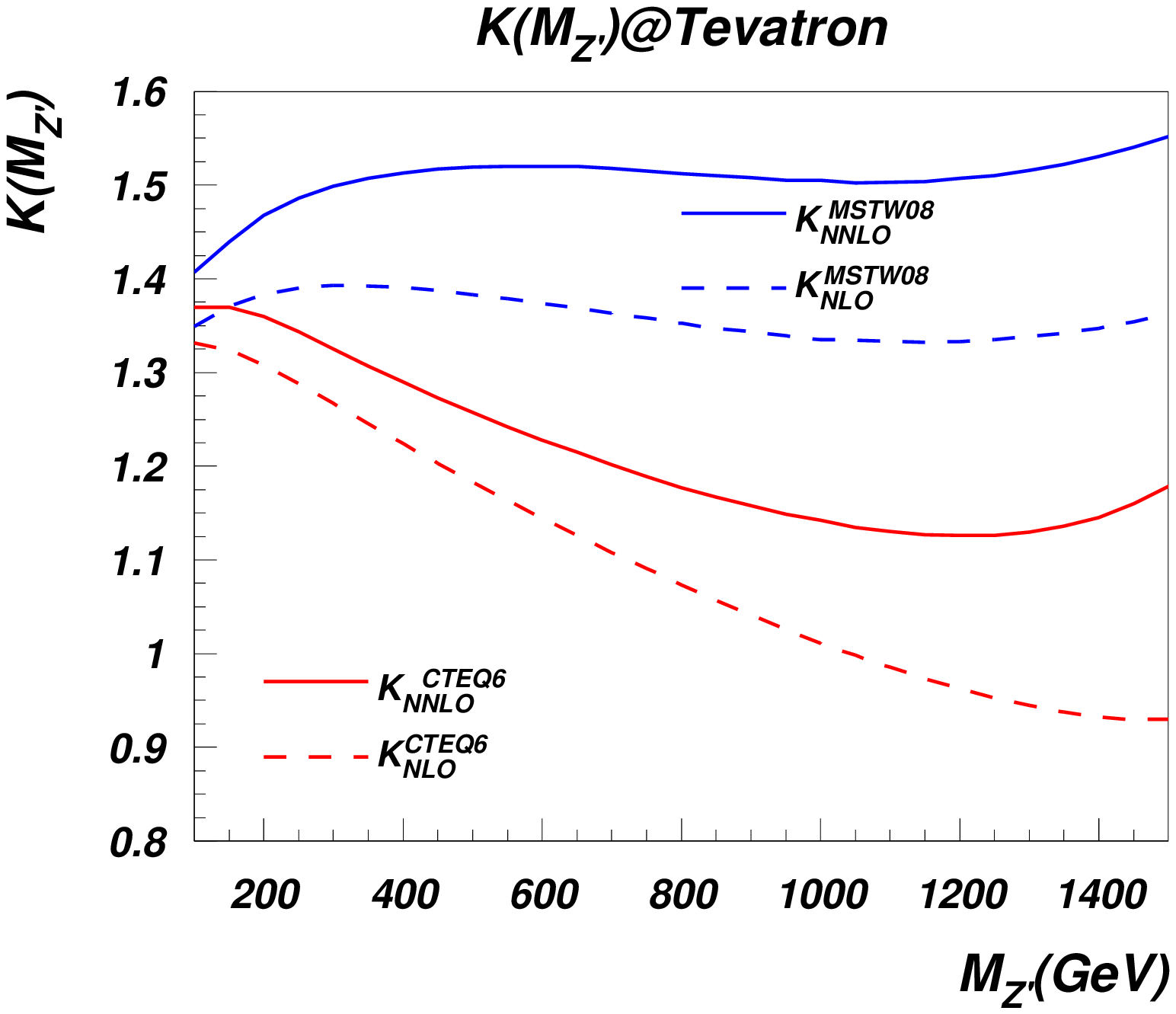}%
\includegraphics[width=0.5\textwidth]{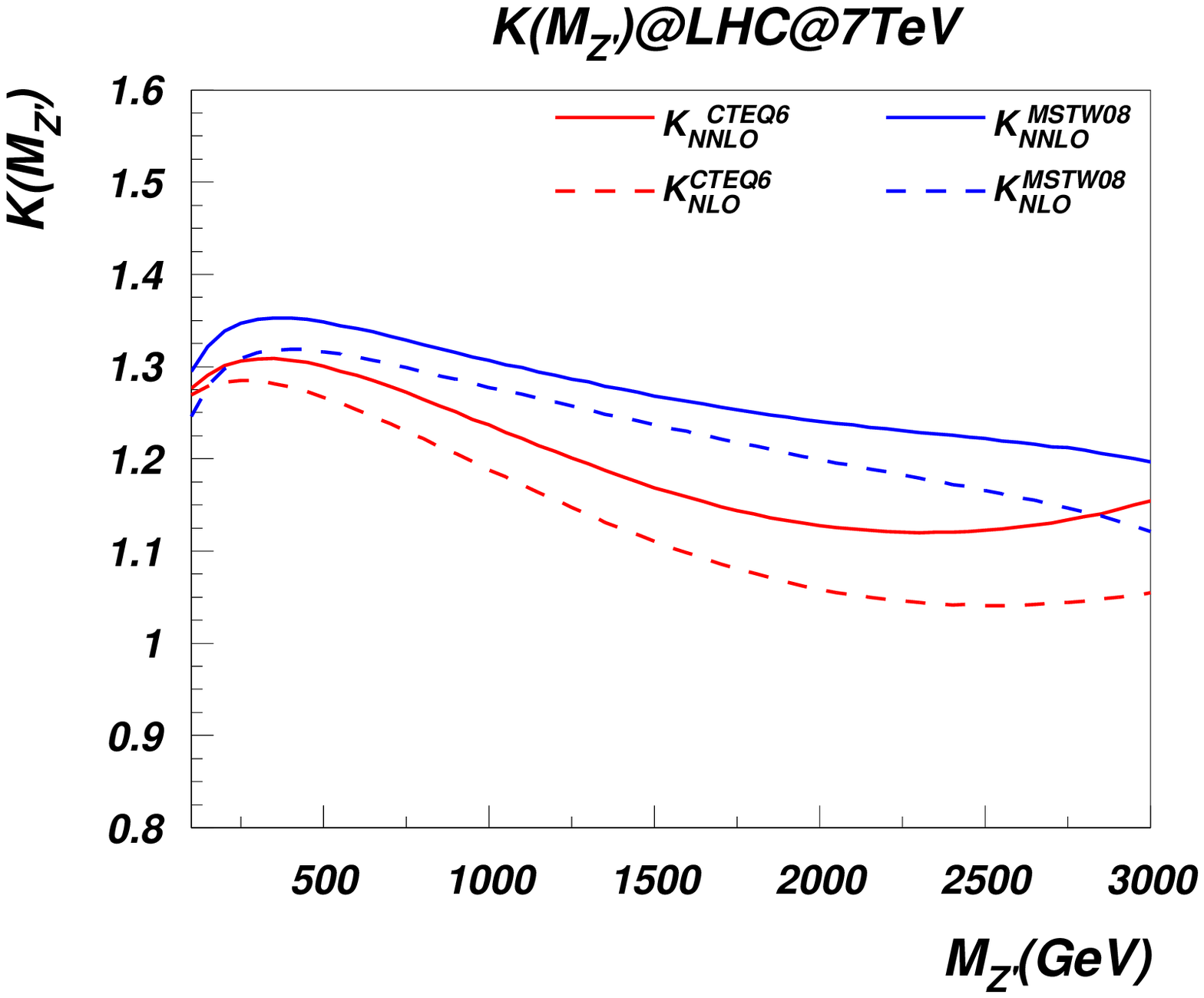}
\caption{\label{fig:kf}
NNLO and NLO $K$-factors defined by Eq.(\ref{eq:kf}) for Standard Model-like
$Z'$ production at the Tevatron (left panel) and the LHC@7TeV (right panel).}
\end{figure}

It is also convenient to define customary NLO and NNLO $K$-factors which can 
be useful for experimentalists in establishing $Z^\prime$ exclusion limits:
\begin{eqnarray}
K_i&=&{\sigma(pp(\bar{p})\to Z')_i}\over{\sigma(pp(\bar{p})\to Z')_0},
\label{eq:kf}
\end{eqnarray}
where the index $i=1,2$ corresponds to NLO and NNLO $K$-factors, respectively.
As an example, in Fig.~\ref{fig:kf} we present the values of these 
$K_i$-factors for Standard Model-like $Z'$ production at the Tevatron (left 
panel) and the LHC@7TeV (right panel) for CTEQ6.6 and MSTW08 PDF's.

Oppositely to what happens for the aforementioned exact NNLO $Z^\prime$ 
production cross section, where the agreement between CTEQ6.6 and MSTW08 PDF
predictions is optimal, in this case there is a noticable difference in the 
behavior of $K_{NLO}$ and $K_{NNLO}$ factors as a function of the $Z^\prime$ 
mass when convoluting the $Z^\prime$ production cross section with CTEQ6.6 or 
MSTW08 PDF's. This difference is related to the way of fitting the LO PDF's 
of CTEQ and MSTW collaborations (see e.g.~\cite{Lai:2009ne,Martin:2009iq}). 
Furthemore, both $K_{NLO,NNLO}^{MSTW08}$ and $K_{NLO,NNLO}^{CTEQ6}$ factors 
display a strong dependence on the $Z^\prime$ mass. As an example, 
$K_{NNLO}^{CTEQ6}$ varies between 10-40\% at the Tevatron and 10-30\% at the 
LHC@7TeV for potentially accessible $Z'$ masses.

Applying a universal $K$-factor can be highly misleading. As shown above, the 
$K_{NLO,NNLO}$-factor has indeed a two-fold source of dependence: PDF 
set and energy scale (i.e. $M_{Z^\prime}$). A uniform setup must be fixed when
comparing experimental limits on different models.

Since Eq.~(\ref{eq:ll}) gives an accurate description even at NNLO 
\cite{Carena:2004xs}, and noting that QCD NNLO corrections are universal for 
up- and down-quarks, one can effectively apply the same 
$K_{NNLO}$-factor derived for SM-like $Z'$ to generic $Z'$ models 
without loosing of generality.
{Owing to the remarkable $Z'$ mass dependence of 
the $K_{NNLO}$-factor, we first convolute the LO $Z^\prime$ production cross 
section with the respective LO PDF's and then we multiply it by 
$K_{NNLO}(M_{Z^\prime})$.}

For convenience and clarity, we provide in Tables~\ref{tab:snlo} and 
~\ref{tab:snlo2} shown in Appendix~\ref{append}
the values of $K_{NNLO}$-factors and cross sections for the SM-like $Z'$-boson 
production process at the Tevatron and the LHC@7TeV: $p\bar p(p)\to Z'+X$.
The first table contains the results obtained with MSTW08 PDF, the latter 
with CTEQ6.6 PDF. The quoted numbers correspond to the curves visualised in 
Figs.\ref{fig:snlo},\ref{fig:kf}.

In narrow width approximation, the two-fermion cross section is the 
product of the production cross section and the respective branching ratio.
When considering the complete $Z'$-boson production and decay in the Drell-Yan 
channel, one has to keep in mind that QCD NNLO corrections also affect the 
$Z^\prime$ branching ratio even for purely leptonic decays, 
$Br(Z'\to \ell^+\ell^-)$, since the $Z^\prime$ total decay width will be 
corrected at NNLO. This reflects into an higher order correction to the $c_u$ 
and $c_d$ coefficients, through $Br(Z'\to \ell^+\ell^-)$ which explicitly 
enters the expression for $c_u$ and $c_d$ given in Eq.(\ref{eq:cucd}).
The NNLO Drell-Yan cross section can be thus written as:
\begin{eqnarray}
\sigma_{\ell^+\ell^-}^{NNLO} &=& 
\frac{\pi}{48 s}
\left[ c_u^{NNLO} w_u(s,M_{Z'}^2)^{NNLO} + c_d^{NNLO}w_d(s,M_{Z'}^2)^{NNLO}\right]\nonumber\\
&=&
K_{NNLO}^{PDF}K_{NNLO}^{BR}\frac{\pi}{48 s}
\left[ c_u w_u(s,M_{Z'}^2) + c_d w_d(s,M_{Z'}^2)\right]
= K_{NNLO}^{PDF}K_{NNLO}^{BR} \sigma_{\ell^+\ell^-}^{LO}
\label{eq:cucd-nnlo}
\end{eqnarray}

The leading NLO QCD correction to the total $Z'$ width is known to be 
$\alpha_s/\pi$~\cite{Gorishnii:1988bc,Kataev:1992dg}. This gives an 
enhancement of the order of 2-3\% to the $Z'$ width for $M_{Z'}$ in the range 
500-2000 GeV. The $Br(Z'\to \ell^+\ell^-)$ will thus decrease accordingly by
$(2-3\%)\times Br(Z'\to hadrons)$. The net result corresponds to a $1-2\%$ 
deplection of the leptonic branching ratio within the SM-like $Z'$ model. An 
effect of the same order is expected for the other classes of $Z'$ models 
under consideration.
In the current study, we neglect this effect and use the following formula
for establishing limits on $Z'$ models:
\begin{eqnarray}
\sigma_{\ell^+\ell^-}^{NNLO} &\simeq& 
K_{NNLO}^{PDF} \sigma_{\ell^+\ell^-}^{LO}.
\label{eq:cucd-nnlo-2}
\end{eqnarray}

\subsection{Finite width effects}

So far we have have discussed the $Z'$ boson production using
narrow width approximation. However, the experimental search for an extra 
$Z'$ boson and the descrimination of the SM backgrounds could strongly depend 
on the realistic $Z^\prime$ width. Moreover the theoretical prediction of the $Z'$ 
production cross section also depends on its width as we discuss below.
\begin{figure}[htbp]
\centering
\includegraphics[width=0.50\textwidth]{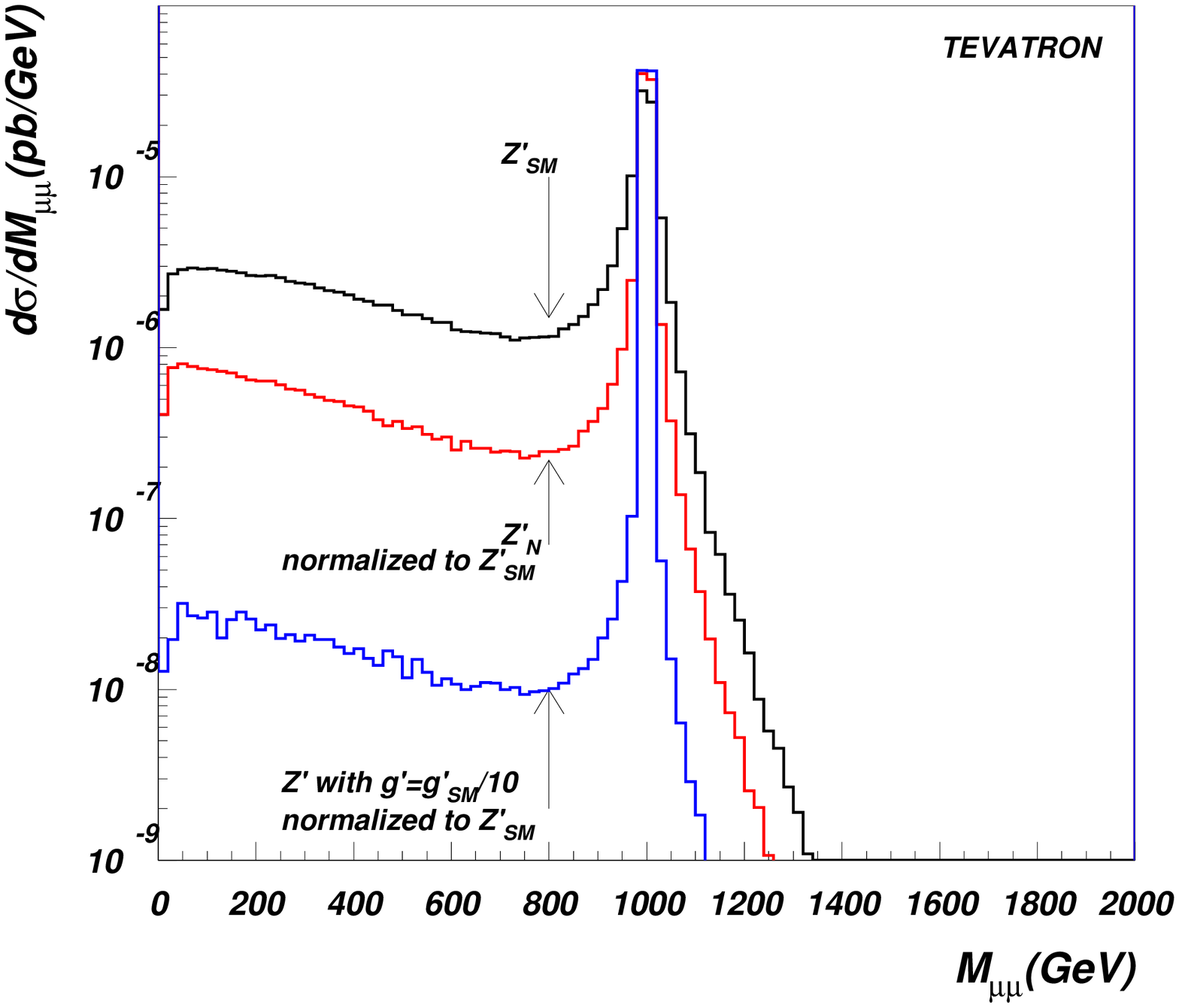}%
\includegraphics[width=0.50\textwidth]{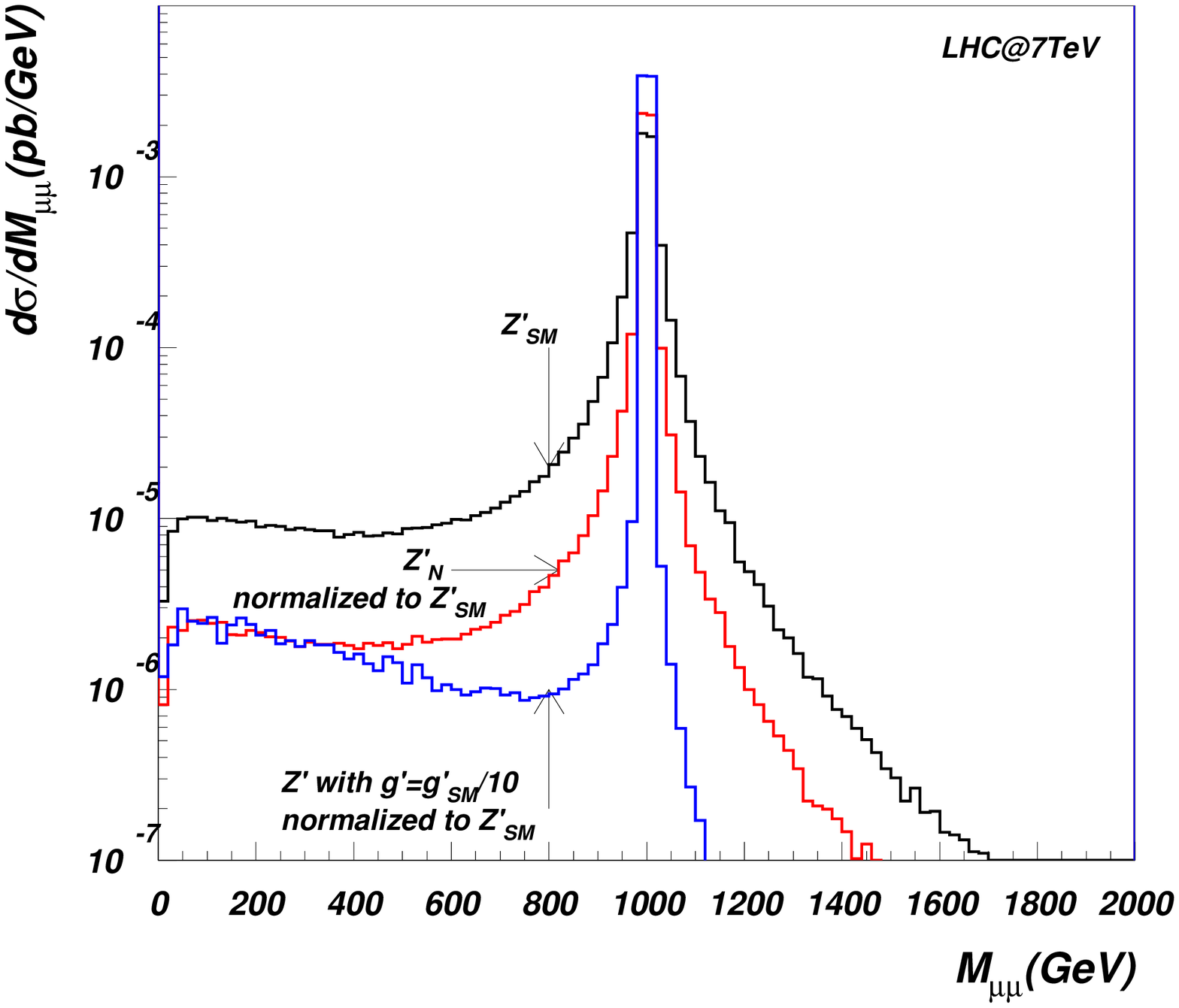}
\caption{\label{fig:m_tev_lhc} Di-lepton invariant mass distribution for the 
$Z'$ boson production in various models at the Tevatron (left panel) and 
LHC@7TeV (right panel).}
\end{figure}

We start this discussion with  Fig.~\ref{fig:m_tev_lhc} where we present the  
di-lepton invariant  mass distribution for the $Z'$ boson production within 
various models at the Tevatron (left panel) and the LHC@7TeV (right panel). 
We consider three representative models: the SM-like $Z'$ model (black line), 
the N-type $E_6$ model defined in Table~\ref{models1} (red line), and the 
weakly coupled SM-like $Z^\prime$ model where the $Z'$ boson gauge coupling to SM 
fermions is reduced by a factor 10 (blue line). From top to bottom, the last 
two distributions are normalized to the integral under the first one. 
We first consider the SM-like $Z'$ model distribution at the Tevatron. 
It is  important to
stress that the total cross section of  $p\bar{p}\to Z' \to \ell^+\ell^-$
process integrated over the entire $M_{\ell^+\ell^-}$ range is actually is
almost as twice as large as the SM-like $Z'$ in the narrow width approximation.
{The main reason for this effect is the specific shape of the $M_{\ell^+\ell^-}$ 
distribution in the region of small $M_{\ell^+\ell^-}$ far away
from $M_{Z'}$. This region is exhibited by a non-negligible tail
due to the steeply rising PDF in the region of low
$M_{\ell^+\ell^-}$ even though the  $Z'$ boson isextremely far off mass-shell
in this region. The integral over this region can even double
the cross section evaluated in the NWA in the case of $Z'$ production at the Tevatron.}

This effect, which is related to the off-shellness of the extra gauge boson, 
varies according to the total $Z^\prime$ width. In the $Z'_N$ model, it brings an 
additional 20\% contribution to the narrow width approximation cross section 
at the Tevatron. In the weakly coupled SM-like $Z'$ model, the far 
off-shellness effects are effectively negligible (below 1\%). 

We can see that in general experimental limits would and should strongly 
depend on the particular $Z'$ model predicting a specific $Z'$ width. 
On the other hand, if one requires a di-lepton mass window cut around the 
$Z'$ mass, one can establish a quasi model-independent experimental upper limit
on $\sigma (p\bar{p}\to Z' \times Br(Z' \to \ell^+\ell^-)$ versus $M_{Z'}$
and apply this limit to constraint different classes of models.
\begin{figure}[htbp]
\centering
\includegraphics[width=0.50\textwidth]{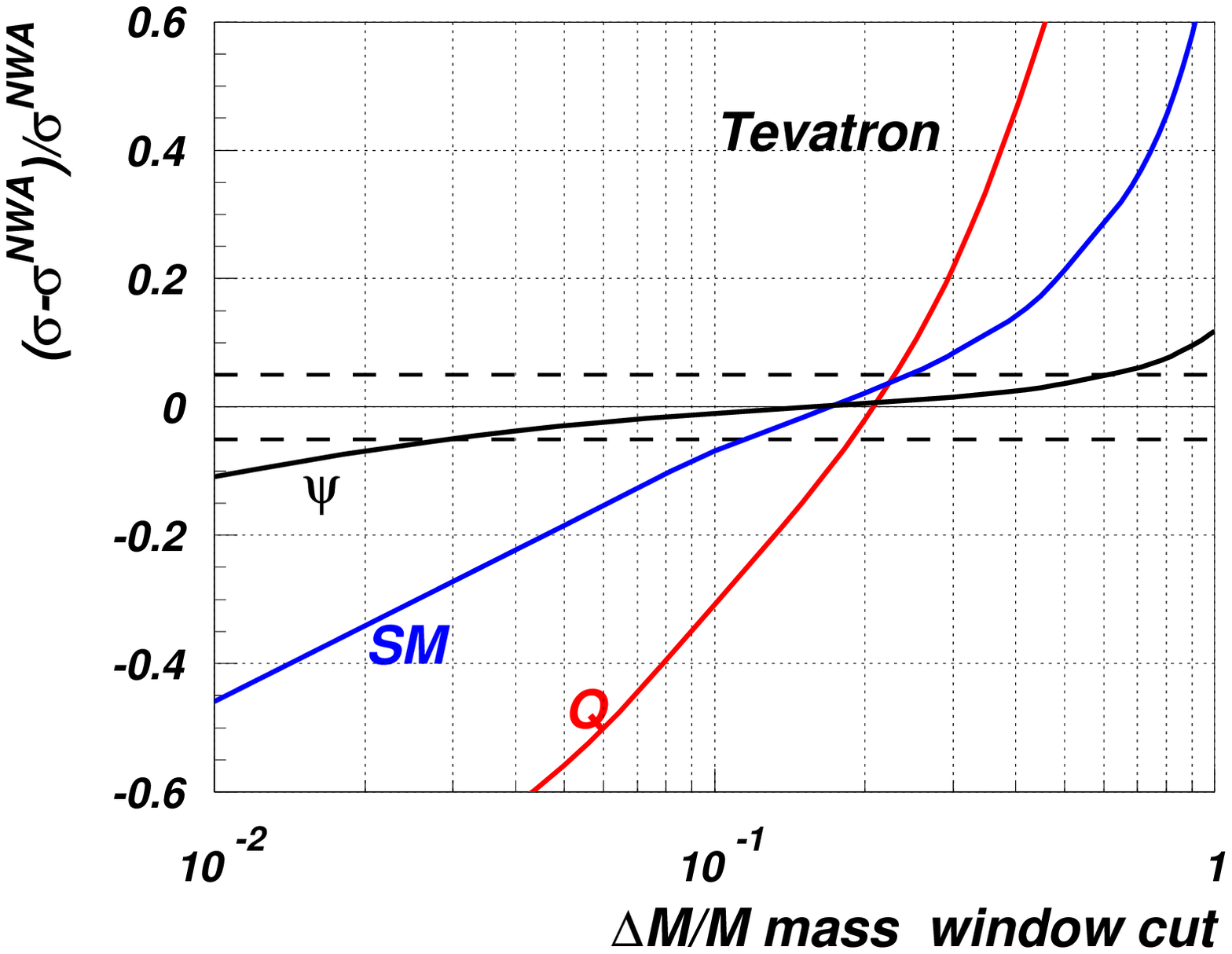}%
\includegraphics[width=0.50\textwidth]{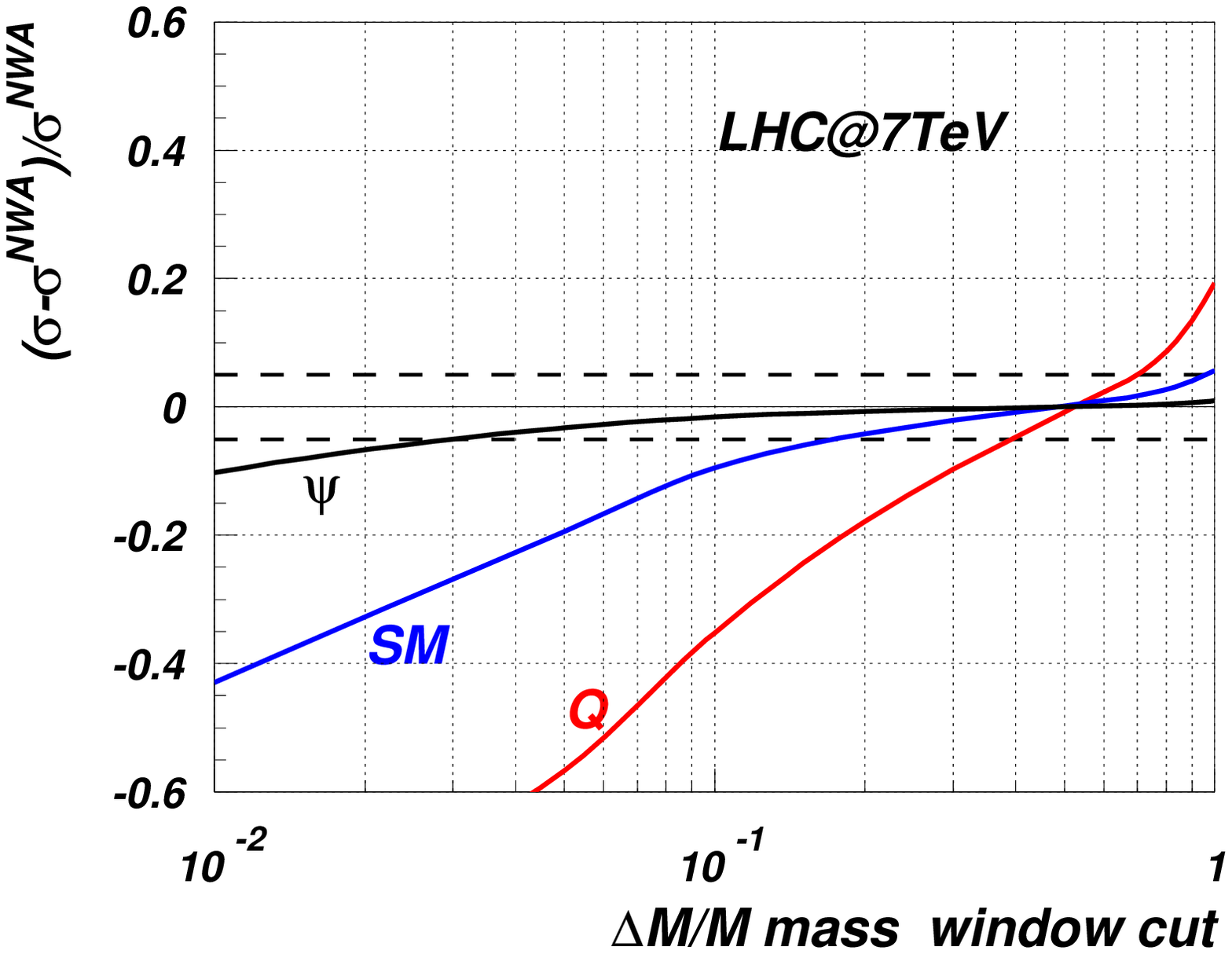}
\caption{\label{fig:nwa} Relative difference between the full cross section 
for $pp(\bar{p})\to Z' \to \ell^+\ell^-$ evaluated taking into account the 
finite $Z^\prime$ width ($\sigma$) and the cross section computed in narrow width 
approximation ($\sigma^{NWA}$). The relative difference is presented as a 
function of the $\Delta M/M_{Z'}$ symmetric mass window cut 
$|M_{\ell^+\ell^-}-M_{Z'}|<\Delta M$ applied to the full cross section 
($\sigma$). Three different representative models are considered for 
$M_{Z'}=1$~TeV (see Table~\ref{models1}).
}
\end{figure}

In Fig.~\ref{fig:nwa} we present the effect of a symmetric mass window cut 
around $M_{Z'}$ for the SM-like $Z'$ model and two other representative 
models (see Table~\ref{models1}) at the Tevatron and the LHC@7TeV. We fix the 
$Z^\prime$ mass to be $M_{Z'}=1$~TeV. We plot the relative difference between the 
full cross section for the process $pp(\bar{p})\to Z' \to \ell^+\ell^-$ 
evaluated taking into account the finite $Z^\prime$ width ($\sigma$) and the cross 
section computed in narrow width approximation ($\sigma^{NWA}$).  The
relative difference is presented as a function if the $\Delta M/M_{Z'}$
symmetric mass window cut ($|M_{\ell^+\ell^-}-M_{Z'}|<\Delta M$) applied to 
the full cross section $\sigma$.
One can see that for the SM-like $Z^\prime$ model at the Tevatron, a $\Delta M/M$ 
cut in the 9-25\% range brings the agreement between $\sigma$ and 
$\sigma^{NWA}$ down to 5\% level while $\Delta M/M \simeq 15\%$ exactly 
matches $\sigma$  and $\sigma^{NWA}$.
At the LHC, the corresponding range of the $\Delta M/M$ cut is 15-80\%, and  
$\sigma$ and $\sigma^{NWA}$ are matched for $\Delta M/M\simeq 45\%$. The 
$\psi$ model $Z'$ has a narrower width, making the choice of the 
cut more insensitive, while the $Q$ model $Z'$ width is broader leading to a 
more sensitive choice of the mass window cut to reproduce the narrow width 
approximation. Note that all lines cross the abscissa at about the same 
value of $\Delta M/M_{Z'}$, meaning that there will be an optimal mass window 
cut consistent with all models.
The choice of the mass window cut to gain agreement with the narrow width 
approximation also depends on $M_{Z'}$. This dependence is defined by proton
parton densities and is therefore model-independent. The net effect is 
again to make all the lines cross the abscissa at about the same value of 
$\Delta M/M_{Z'}$, where this point depends on $M_{Z'}$.
Therefore for every given mass one can work out a
quasi model-independent mass window cut where the full cross section matches 
the narrow width approximation. The additional advantage of this choice is that
 in the selected mass window around the $Z^\prime$ mass the 
model-dependent interference effect between $Z^\prime$ signal and SM background 
is highly suppressed.

The experimental
limits would  be quasi model-independent if one would  apply  
this cut on the
$M_{\ell^+\ell^-}$ around the $M_Z'$:
it brings in agreement the
cross section calculated in the narrow width approximation and in the finite
width approximation
as well as removes model-dependent shape of the
$M_{\ell^+\ell^-}$ distributions in the region 
of low $M_{\ell^+\ell^-}$ especially for the case of 
large $Z'$ width effects as, for example, take place
for SM-like $Z'$.
Moreover, the cut on $M_{\ell^+\ell^-}$ around the $M_Z'$
plays an important role 
in reducing an effect of $Z'$ interference with $Z/\gamma$
down to a few\% level, which again,
allows to establish and use experimental limits in model-independent way.

For example, in case of SM-like $Z'$ production at the Tevatron,
the relative interference, which is defined as
$R_i=[\sigma(p\bar{p}\to Z'/Z/\gamma \to \ell^+\ell^-)
    -\sigma(p\bar{p}\to           Z' \to \ell^+\ell^-)
    - \sigma(p\bar{p}\to Z/\gamma \to \ell^+\ell^-)]/
    \sigma(p\bar{p}\to           Z' \to \ell^+\ell^-)
$
is as large as about $-19$
(meaning $-1900\%$ of interference(!)) for  $M_{\ell^+\ell^-}>100$~GeV cut
but it drops down to $-6\%$ for $|M_{\ell^+\ell^-}-M_{Z'}|<0.15 M_{Z'}$
cut which matches NWA and finite width cross sections.
The effect of the mass window cuts is also quite large 
for the   case of SM-like $Z'$ production at the LHC,
where interference is about $-300\%$ for  $M_{\ell^+\ell^-}>100$~GeV cut
and only about $-2\%$ for $|M_{\ell^+\ell^-}-M_{Z'}|<0.15 M_{Z'}$ cut.

We can see, that there is a strong motivation to use an invariant mass window cut
for conducting a model-independent analysis.
The size of this cut, if one aims to match the NWA and finite width cross sections,
is collider dependent: it is about $15\%$ of  $M_{Z'}$
for the Tevatron and about $40\%$  of  $M_{Z'}$ for the LHC\@7TeV.

In this
paper we are using results of experimental analysis which are based  on
$M_{\ell^+\ell^-}$ mass window cut similar to what we are advocating.
This would allow us to
use precise NNLO model predictions and perform a respective 
model-independent interpretation of the experimental limits.

\section{Benchmark Models \label{III}}

In this section, we extend and classify the benchmark $Z^\prime$ models present in 
the literature. We divide such classes into two main types: perturbative  
and strongly coupled gauge theories.  

\subsection{Perturbative gauge theories}

\subsubsection{$E_6$ Models}
In these models one envisages that at the GUT scale the gauge group is $E_6$.
The gauge group $E_6$ is broken at the GUT scale to $SO(10)$ and a $U(1)_{\psi}$ gauge group,
\be
E_6 \rightarrow SO(10)\times U(1)_{\psi}.
\ee
The $SO(10)$ is further broken at the GUT scale to $SU(5)$ and a $U(1)_{\chi}$ gauge group,
\be
SO(10) \rightarrow SU(5) \times U(1)_{\chi}.
\ee
Finally the $SU(5)$ is broken at the GUT scale to the Standard Model (SM) gauge group,
\be
SU(5) \rightarrow SU(3)_C\times SU(2)_L\times U(1)_Y.
\ee
All these breakings may occur at roughly the GUT scale. The question which concerns us
here is what happens to the two Abelian gauge groups $U(1)_{\psi}$ and $U(1)_{\chi}$
with corresponding generators $T_{\psi}$ and $T_{\chi}$.
Do they both get broken also at the GUT scale, or may one or other of them survive down to the
TeV scale? In general it is possible for some linear combination of the two to survive down to the TeV scale,
\be
U(1)'=\cos \theta \ U(1)_{\chi} + \sin \theta \ U(1)_{\psi},
\ee
where $-\pi/2 < \theta \leq \pi/2$.
More correctly the surviving $E_6$ generator $Q_{E_6}$ should be written as,
\be
Q_{E_6}=\cos \theta \ T_{\chi} + \sin \theta \ T_{\psi}.
\ee
Some popular examples of such $U(1)'$ are shown in Table \ref{models1}.

The resulting heavy $Z'$ couples as $g'Q_{E_6}Z'$.
Note that in $E_6$ models it is reasonable to assume that the $Z'$ gauge coupling $g'$ is equal to
the GUT normalized
$U(1)_Y$ gauge coupling of the SM, $g_1(M_Z)=(e/c_W)\sqrt{5/3}\approx 0.462$ 
where $e=0.3122(2)$ and $c_W=\sqrt{1-s_W^2}$ where the $\overline{MS}$ value is $s_W^2=0.2312$.
Thus we take $g' \approx 0.46$.
GUT normalization also implies that the $T_{\psi}$ charges of the fermions in the $SO(10)$ $16$ representation
for the $\psi$ case are all equal to $1/\sqrt{24}$, while for the $\chi$ case the $T_{\chi}$ charges of the
$SU(5)$ representations $(10,\overline{5},1)$ are $(-1/\sqrt{40}, 3/\sqrt{40}, -5/\sqrt{40})$.
Recalling that $g_{V,A}^f=\epsilon_L^f\pm \epsilon_R^f$, and $10\rightarrow Q,u^c,e^c$ and
$\overline{5} \rightarrow L,d^c$, and that $u^c,d^c,e^c$ have the opposite charges to $u_R,d_R,e_R$,
this results in the values of the $g_{V,A}^f$ charges for the $U(1)_{\psi}$ and $U(1)_{\chi}$
cases as shown in Table \ref{models1}. The general charges as a function of $\theta$ are then simply
given as,
\be
g_{V,A}^f(\theta)=\cos \theta \ g_{V,A}^f(\chi ) + \sin \theta \ g_{V,A}^f(\psi ),
\ee
where the numerical charges for the popular models quoted in the literature 
are listed in Table \ref{models1}.

\subsubsection{Generalised Left-Right Symmetric Models (GLRs)}

These models are motivated by the left-right (LR) extensions of the SM gauge group with the
symmetry breaking,
\beq
SU(2)_L\times SU(2)_R \times U(1)_{B-L} \rightarrow SU(2)_L \times U(1)_Y
\eeq
which, from the point of view of $Z'$ models essentially involves the symmetry breaking,
\beq
U(1)_R \times U(1)_{B-L} \rightarrow U(1)_Y
\eeq
where $U(1)_R$ involves the generator $T_{3R}$ corresponding to the third component
of $SU(2)_R$, while $U(1)_{B-L}$ involves the generator $T_{B-L}=(B-L)/2$.
The hypercharge generator is then just given by $Y=T_{3R}+T_{B-L}$.
Assuming a left-right symmetry, the gauge couplings of $SU(2)_{L,R}$ are then 
equal, $g_L=g_R$ and the resulting heavy $Z'_{LR}$ then couples as 
$g_1Q_{LR}Z'$ where
\beq
Q_{LR}=\sqrt{\frac{3}{5}}\left(\alpha T_{3R} -\frac{1}{\alpha}T_{B-L} \right)
\eeq
with $\alpha =\sqrt{\cot^2\theta_W-1} \approx 1.53$ and $g_1\approx 0.462$ as 
before.

The left-right symmetric models therefore motivate a $U(1)_{LR}$ which is a 
particular linear combination of $U(1)_R$ and $U(1)_{B-L}$ with a specific 
gauge coupling.
From this perspective the special cases where the $Z'$ corresponds to a pure $U(1)_R$
or a pure $U(1)_{B-L}$ are not well motivated. Nevertheless these types of $Z'$ have been well
studied in the literature and so it is useful to propose a generalization of the LR models
which includes these special cases. To this end we propose a generalized left-right (GLR) symmetric
model in which the $Z'$ corresponds to a general linear combination of the generators of 
$U(1)_R$ and $U(1)_{B-L}$,
\be
Q_{GLR}=\cos \phi \  T_{3R} + \sin \phi \  T_{B-L},
\ee
where $-\pi/2 < \phi \leq \pi/2$. The gauge coupling $g'$ is fixed so that 
for a particular value of $\phi$ the $Z'$ of the GLR may be identified with 
the $Z'$ of the LR symmetric model above. To be precise, we identify, for a 
particular value of $\phi$:
\be
g_1Q_{LR}\equiv g' Q_{GLR}
\ee
which implies $\tan\phi = -1/\alpha^2$ which corresponds to $\phi = -0.128 \pi$
for $\alpha\approx 1.53$ and we find $g'=0.595$. Keeping $g'=0.595$ fixed, we 
are then free to vary $\phi$ over its range where $\phi = -0.128\pi$ gives 
the LR model, but other values of $\phi$ define new models.

Clearly $\phi =0$ gives a $U(1)_R$ model while $\phi =\pi/2$ gives a $U(1)_{B-L}$ model.
In the GLR model the
value of $\phi = \pi /4$ also defines a $Z'$ which couples to hypercharge $Y=T_{3R}+T_{B-L}$
(not to be confused with the sequential SM $Z'$ which couples like the $Z$).
The couplings of the $Z'$ for the special cases of the GLR models are give in Table \ref{models1}.
The general charges as a function of $\phi$ are then simply
given as,
\be
g_{V,A}^f(\phi)=\cos \phi \ g_{V,A}^f(R) + \sin \phi \ g_{V,A}^f(B-L),
\ee
where the numerical charges for particular models are shown in Table 
\ref{models1}.

\begin{table}[]
\centering
\begin{tabular}{lccccccccc}
\hline
$U(1)'$ & Parameter & $g_V^u$ & $g_A^u$ & $g_V^d$ & $g_A^d$ & $g_V^e$ & $g_A^e$ &
$g_V^{\nu}$ & $g_A^{\nu}$ \\
\hline
\hline
 $E_6$ $(g'=0.462)$  & $\theta$  &      &    &   &   &    &    &   &  \\
\hline
$U(1)_{\chi}$ & 0
& 0    &  -0.316  &  -0.632 & 0.316  & 0.632   & 0.316   &  0.474 &   0.474 \\
$U(1)_{\psi}$ & $0.5\pi$
& 0    &  0.408  &      0   & 0.408   &    0   & 0.408   &  0.204 &   0.204 \\
$U(1)_{\eta}$  & -$0.29\pi$
& 0    &  -0.516  &  -0.387 & -0.129  & 0.387   & -0.129 &  0.129 &   0.129 \\
$U(1)_S$  & $0.129\pi$
&    0    &  -0.129  &  -0.581 & 0.452   & 0.581   & 0.452   & 0.516 & 0.516 \\
$U(1)_I$  & $0.21\pi$
&    0    &   0      &   0.5   & -0.5    & -0.5    & -0.5    &  -0.5 & -0.5 \\
$U(1)_N$   & $0.42\pi$
&    0    &  0.316  &  -0.158 & 0.474   & 0.158   & 0.474   &  0.316 & 0.316 \\
\hline
\hline
 GLR  $(g'=0.595)$    & $\phi$ &      &    &   &   &    &    &   &  \\
\hline
$U(1)_{R}$  & 0
& 0.5    &  -0.5  &  -0.5 & 0.5  & -0.5   & 0.5   &  0       &   0 \\
$U(1)_{B-L}$  & $0.5\pi$
& 0.333    &  0  &  0.333 & 0  & -1   & 0   &  -0.5       &   -0.5 \\
$U(1)_{LR}$ & $-0.128\pi$
& 0.329    &  -0.46  &  -0.591 & 0.46  & 0.068   & 0.46   &  0.196 &   0.196 \\
$U(1)_{Y}$ & $0.25\pi$
& 0.833    &  -0.5  &  -0.167 & 0.5  & -1.5   & 0.5   &  -0.5       &   -0.5 \\
\hline
\hline
GSM   $(g'=0.760)$ & $\alpha$   &      &    &   &   &    &    &   &  \\
\hline
$U(1)_{SM}$ & $-0.072\pi$
& 0.193    &  0.5  &  -0.347 & -0.5  & -0.0387   & -0.5   &  0.5  &   0.5 \\
$U(1)_{T_{3L}}$ & $0$
& 0.5    &  0.5  &  -0.5 & -0.5  & -0.5   & -0.5   &  0.5       &   0.5 \\
$U(1)_Q$ & $0.5\pi$
& 1.333  &  0 & -0.666  & 0   & -2.0   &  0       &   0 & 0  \\
\hline
\hline
\end{tabular}
\caption{Benchmark model parameters and couplings. The angles $\theta$, $\phi$, 
$\alpha$ are defined in the text.
\label{models1}}
\end{table}

\subsubsection{Generalised Sequential Models (GSMs)}

No study is complete without including the sequential standard model (SSM)
$Z'_{SSM}$ which is defined to have identical couplings as for the usual $Z$,
namely given by $\frac{g_2}{c_W}Q_ZZ'_{SSM}$ and $Q_Z=T_{3L}-s_W^2Q$ where
$s_W^2=0.2312$  and $\alpha_2 (M_Z) = g_2^2/(4\pi)\approx 0.0338$ imply that
$\frac{g_2}{c_W} \approx 0.74$.  Similar to the GLR models, it is useful to define
a generalised version of the SSM called GSM where the heavy gauge boson $Z'_{GSM}$
then couples as $g'Q_{GSM}Z'_{GSM}$ where  $Q_{GSM}$ corresponds to a general
linear combination of the generators of $U(1)_{T_{3L}}$ and  $U(1)_Q$,
\be
Q_{GSM}=\cos \alpha \ T_{3L} + \sin \alpha \ Q,
\ee
and where $-\pi/2 < \alpha\leq\pi/2$. The gauge coupling $g'$ is fixed so 
that for a particular value of $\alpha$ the $Z'_{GSM}$ of the GSM may be 
identified with the $Z'_{SSM}$ of the SSM above.
To be precise, we identify, for a particular value of $\alpha$:
\be
\frac{g_2}{c_W}Q_Z \equiv g' Q_{GSM}.
\ee
This implies that the GSM reduces to the SSM case for $g'= \frac{g_2}{c_W} \sqrt{1+s_W^4} \approx  0.76$ 
and $\tan \alpha = -0.23$ which corresponds to $\alpha = -0.072 \pi$.
Keeping $g'=0.76$ fixed, we are then free
to vary $\alpha$ over its range where $\phi = -0.072 \pi$ gives the usual SSM, but other values 
of $\alpha$ define new models.
Clearly $\alpha =0$ gives a $U(1)_{T_{3L}}$ model while $\alpha =\pi/2$ gives a $U(1)_{Q}$ model.

The couplings of the $Z'$ for the special cases of the GLR models are give in Table \ref{models1}.
The general charges as a function of $\alpha$ are then simply
given as,
\be
g_{V,A}^f(\alpha)=\cos \alpha \ g_{V,A}^f(L) + \sin \alpha \ g_{V,A}^f(Q),
\ee
where the numerical charges for particular models are shown Table \ref{models1}.

\begin{figure}[]
\centering
\includegraphics[width=0.46\textwidth]{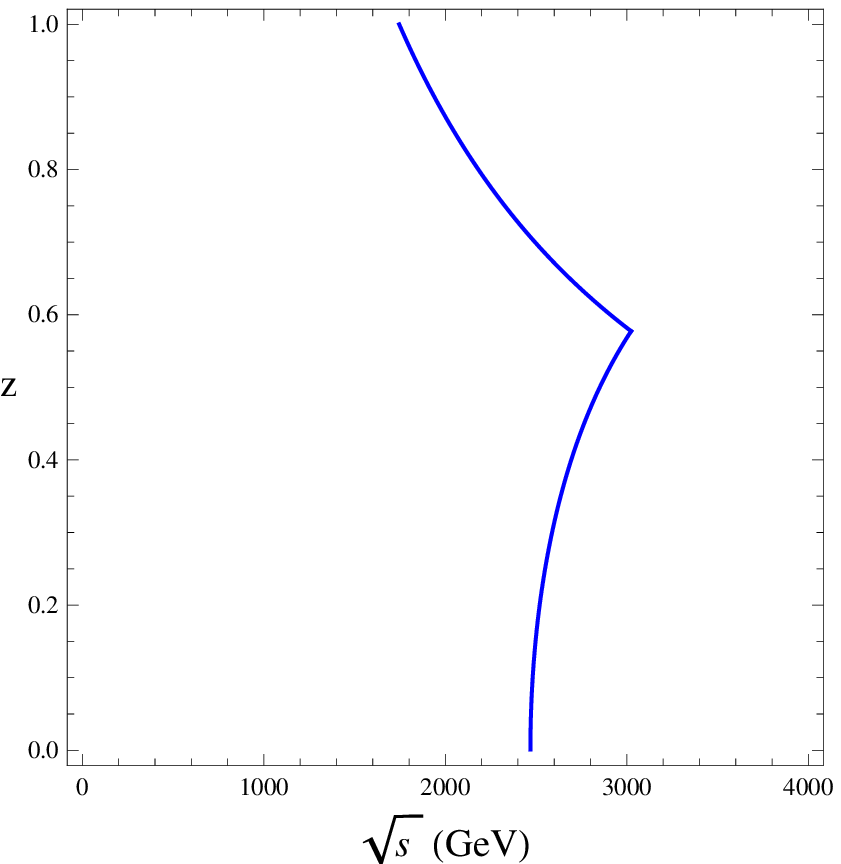}%
\includegraphics[width=0.51\textwidth]{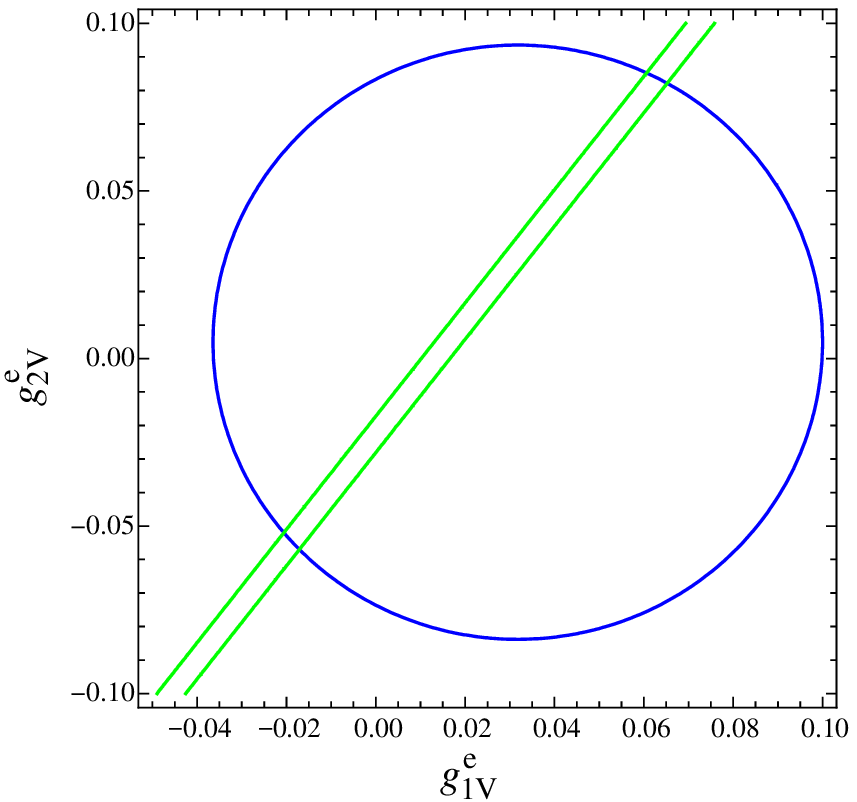}
\caption{Left: Unitarity bound as a function of the energy scale for
different $z=M_{Z1}/M_{Z2}$ values. The perturbative region is on the left of 
the contour plot. Right: 95$\%$ C.L. bounds on the vector couplings of the
$Z_{1,2}$-bosons to SM electrons from $\epsilon_1$ (blue) and
$\epsilon_3$ (green). We consider the representative case: $M_{Z1}=1$ TeV,
$M_{Z2}=1.3$ TeV.}
\label{fig:1}
\end{figure}

\subsection{Strongly coupled gauge theories}

Strongly interacting gauge theories provide an alternative mechanism for the
electroweak symmetry breaking (EWSB). The EWSB is not driven by a light Higgs 
boson anymore, but it happens in a dynamical way. Such theories date back to 
decades. However, even if they predict the existance of new gauge bosons in 
order to delay at high energy the perturbative unitarity violation in vector 
boson scattering amplitudes, they are not considered when performing
searches of $Z^\prime$ bosons in the dilepton Drell-Yan channel. The reason is that 
historically the predicted new resonances must be fermiophobic in order to 
evade EWPT constraints. However, in recent years, new models have been 
proposed that are able to satisfy the EWPT bounds without imposing such a 
strong condition. Both the Minimal Walking Technicolour \cite{Foadi:2007ue,
Belyaev:2008yj} and the 
four site Higgsless model 
\cite{Accomando:2008jh,Accomando:2010ir,Accomando:2008dm} 
predict extra $Z^\prime$ bosons with sizeable 
couplings to SM matter. Hence, they could be tested in the favoured Drell-Yan 
channel at the Tevatron and during the early stage of the LHC.   

\subsubsection{The Four Site Higgsless Model}
\label{4site-Model}

Higgsless models emerge naturally from local gauge theories in five 
dimensions. Their major outcome is delaying the unitarity violation of 
vector-boson scattering (VBS) amplitudes to higher energies, compared to the 
SM without a light Higgs, by the exchange of 
Kaluza-Klein excitations \cite{SekharChivukula:2001hz}. Their common drawback 
is to reconcile unitarity with the ElectroWeak Precision Test (EWPT) bounds.
Within this framework, and in the attempt to solve this dichotomy, many models 
have been proposed 
\cite{Csaki:2003dt,Agashe:2003zs,Csaki:2003zu,Barbieri:2003pr,Nomura:2003du,
Cacciapaglia:2004zv,Cacciapaglia:2004rb,Cacciapaglia:2004jz,Contino:2006nn}. 

In this paper, we consider the four site Higgsless model 
\cite{Casalbuoni:2005rs} as representative of strongly coupled theories. This 
model belongs to the class of the so called deconstructed theories 
\cite{ArkaniHamed:2001ca,Arkani-Hamed:2001nc,Hill:2000mu,Cheng:2001vd,
Abe:2002rj,Falkowski:2002cm,Randall:2002qr,Son:2003et,deBlas:2006fz} 
which come out from the discretization of the fifth dimension on a lattice, 
and are described by chiral lagrangians with a number of gauge-group replicas 
equal to the number of lattice sites. The simplest version of this class of 
models is related to the old BESS model 
\cite{Casalbuoni:1985kq,Casalbuoni:1986vq}, a lattice with only three sites 
and $SU(2)_L\times SU(2)\times U(1)_Y$ gauge symmetry (for it, sometimes 
called three-site Higgsless model).  In order to reconcile unitarity and 
EWPT-bounds, this minimal version predicts indeed the new triplet of vector 
bosons to be almost fermiophobic. Hence, only di-boson production, vector 
boson fusion and triple gauge boson production processes can be used to test 
these models. All these channels require high energy and luminosity and will 
be proper for a future upgrade of the LHC 
\cite{Birkedal:2004au,Belyaev:2007ss,He:2007ge}. 

In the strongly coupled scenario, the four site Higgsless model represents a 
novelty in this respect 
\cite{Accomando:2008jh,Accomando:2010ir,Accomando:2008dm}. Its 
phenomenological consequences are quite similar to those of the Minimal 
Walking Technicolour \cite{Foadi:2007ue,Belyaev:2008yj}. The four site model, 
based on the
\begin{equation}
SU(2)_L\times SU(2)_1\times SU(2)_2\times U(1)_Y
\end{equation}
gauge symmetry, predicts two neutral and four charged extra gauge bosons, 
$Z_{1,2}$ and $W^\pm_{1,2}$, and is capable to satisfy EWPT constraints 
without 
necessarily having fermiophobic resonances. Within this framework, the more 
promising Drell-Yan processes
become particularly relevant for the extra gauge boson search at the LHC.\\
The four site Higgsless model is described by four free parameters: 
$g_{1V}^e$, $g_{2V}^e$, $M_{Z1}$, $M_{Z2}$ that is the two vector 
couplings between $Z_{1,2}$-bosons and SM electrons and the two $Z_{1,2}$ 
masses (charged and neutral gauge bosons are degenerate). 

In terms of the mass eigenstates, the Lagrangian describing the neutral 
current interaction is given 
by\footnote{For details see \cite{Accomando:2008jh,Accomando:2010ir}.}
\begin{equation}\label{NC}
{\mathcal L}_{NC}=\frac{1}{2}\bar{f}\gamma^\mu\left[(g_{1V}^f-g_{1A}^f\gamma_5)Z_{1\mu}+ (g_{2V}^f-g_{2A}^f\gamma_5) Z_{2\mu} \right] f
\end{equation}
where $g_{1,2V}^f$, $g_{1,2A}^f$ are the vector and axial couplings of the 
extra $Z_{1,2}$ gauge bosons to ordinary matter. In the above formula, we have
included the $g'$ coupling in the definition of $g_{1,2V}^f$ and $g_{1,2A}^f$. 

The energy range, where the perturbative regime is still valid is plotted in 
the left panel of Fig.\ref{fig:1} for different values of the ratio 
$z=M_{Z1}/M_{Z2}$. Owing to the exchange of the extra gauge bosons, the 
perturbative unitarity violation can be delayed up to an energy scale of 
about $\sqrt{s}\simeq 3$ TeV. Hence, the mass spectrum of the new particles 
is constrained to be within a few TeV. 

In the past, the only way to combine the need of relatively low mass extra 
gauge bosons with EWPT was to impose the new particles to be fermiophobic. In 
the four-site Higgsless model, this strong assumption is not necessary 
anymore. In the right panel of Fig.\ref{fig:1}, we plot the bounds on the 
vector couplings of the $Z_{1,2}$-bosons to SM electrons coming from the EWPT 
expressed in terms of the $\epsilon_{1,3}$ parameters \cite{Barbieri:2004qk} 
($\epsilon_2$ is uneffective due to SU(2)-custodial symmetry). The outcome is 
that $\epsilon_3$ constraints the relation between the two couplings, while 
$\epsilon_1$ limits their magnitude. 

allows to reconcile unitarity and EWPT bounds, leaving a calculable and not  
fine-tuned parameter space, where the new gauge bosons are not fermiophobic.

\begin{figure}[]
\centering
\includegraphics[width=0.49\textwidth]{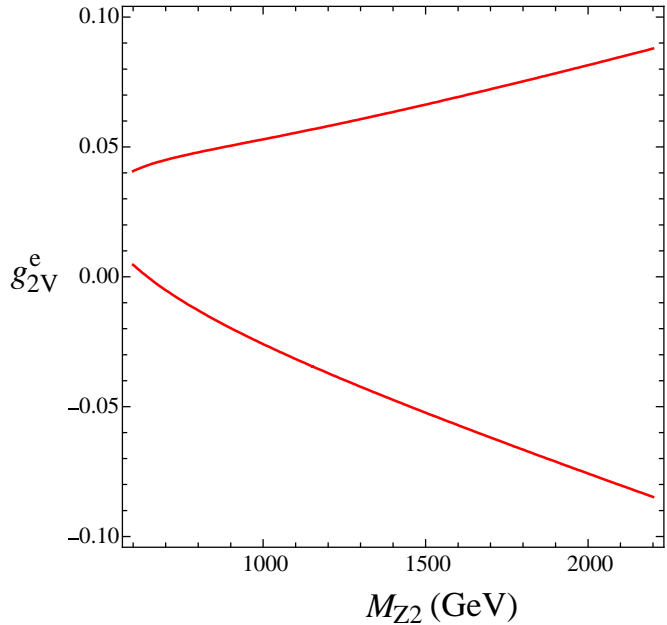}     
\caption{Parameter space in the plane $(g_{2V}^e, M_{Z2})$ where $g_{2V}^e$ 
is the $Z_2$-boson vector coupling to SM electrons, and $M_{Z2}$ is the 
$Z_2$-boson mass. The red solid lines restrict the area allowed by EWPT and 
unitarity. One sample case has been considered: $z$=0.8.}
\label{fig:parameter_space}
\end{figure}

Using the linear relation shown in the right panel of Fig.\ref{fig:1}, we can 
express the $Z_1$-boson vector coupling to SM electrons as a function of the 
$Z_2$-boson vector coupling to SM electrons (by computing back the 
bare-parameters of the Lagrangian, all other $Z_{1,2}$-boson-fermion 
couplings can be simultaneously derived). In this way, the number of 
independent free 
parameters describing the four-site Higgsless model gets reduced to three. We 
choose the following physical observables: $M_{Z2}$, $g_{2V}^e$, and $z$. In 
terms of these new variables, the parameter space allowed by EWPT and 
perturbative unitarity is shown in Fig.\ref{fig:parameter_space} for one 
representative $z$-value: $z$=0.8. The outcome is that one can reconcile 
unitarity and EWPT bounds, leaving a calculable and not   
fine-tuned parameter space, where the new gauge bosons are not fermiophobic.  
 
Compared to the popular extra $U^\prime (1)$ theories summarized in Table 
\ref{models1}, the four-site Higgsless model does not predict fixed values 
for the couplings of the extra gauge bosons to ordinary matter. One has 
indeed a parameter space bounded but still large enough to accomodate rather 
sizeable $Z_2$-boson couplings to SM fermions. They can range from zero to 
the order of SM couplings. We use this framework in the next sections, 
when discussing four-site model properties and limits.

\section{Application of Model Independent Approach 
to the benchmark models \label{IV}}

\subsection{Current limits from Tevatron}

\begin{table}
\centering
\begin{tabular}{lcccccccc}
\hline
$U(1)'$ & $Br({e^+e^-})$ & $c_u$ & $c_d$ & $c_u/c_d$ & $\Gamma_{Z'}/M_{Z'}$ & 
$M^{\mathrm{D}}_{Z'} (GeV)$ & $M^{\mathrm{I}}_{Z'} (GeV)$ & $|\theta_{ZZ'}|$\\
\hline
\hline
\multicolumn{9}{l}{$E_6$ $(g'=0.462)$}   \\
\hline
$U(1)_{\chi}$ &   0.0606 & $6.46\times 10^{-4}$ & $3.23\times 10^{-3}$    &  
0.2   &  0.0117  & 915 & 1141$^e$ & $1.6\times 10^{-3}$   \\
$U(1)_{\psi}$ &  0.0444 & $7.90\times 10^{-4}$ & $7.90\times 10^{-4}$     &  
1     &  0.0053  & 915 & 481$^c$  & $1.8\times 10^{-3}$   \\
$U(1)_{\eta}$ &  0.0371 & $1.05\times 10^{-3}$ & $6.59\times 10^{-4}$     &  
1.6   &  0.00636 & 940 & 434$^c$  & $4.7\times 10^{-3}$   \\
$U(1)_S$ &    0.0656 & $1.18\times 10^{-4}$ & $3.79\times 10^{-3}$        &  
0.31  &  0.0117  & 847 & 1257$^e$ & $1.3\times 10^{-3}$   \\
$U(1)_I$  &    0.0667 & $0$           & $3.55\times 10^{-3}$              &  
0     &  0.0106  & 795 & 1204$^e$ & $1.2\times 10^{-3}$   \\
$U(1)_N$  &     0.0555 & $5.94\times 10^{-4}$ & $1.48\times 10^{-3}$      &  
0.40  &  0.00635 & 892
 & 623$^e$  & $1.5\times 10^{-3}$   \\
\hline
\hline
\multicolumn{9}{l}{GLR $(g'=0.595)$} \\
\hline
$U(1)_{R}$ &  0.0476 & $4.21\times 10^{-3}$ & $4.21\times 10^{-3}$     &  1  
& 0.0247 & 1065 & 442$^e$ & -   \\
$U(1)_{B-L}$ &  0.154 & $3.02\times 10^{-3}$ & $3.02\times 10^{-3}$    &  1  
& 0.015 & 1035 & - & -   \\
$U(1)_{LR}$ &  0.0246 & $1.39\times 10^{-3}$ & $2.44\times 10^{-3}$    &  
0.57  &  0.0207 & 970 & 998$^e$ & $1.3\times 10^{-3}$   \\
$U(1)_{Y}$ &  0.125 & $1.04\times 10^{-2}$ & $3.07\times 10^{-3}$      &  
3.4  &  0.0235 & 1135 & - & -   \\
\hline
\hline
\multicolumn{9}{l}{SM $(g'=0.760)$}    \\
\hline
$U(1)_{SM}$      & 0.0308 & $2.43\times 10^{-3}$  & $3.13\times 10^{-3}$    
&  0.776   &  0.0297   &  1020 & 1787$^c$ & $9\times 10^{-4}$   \\
$U(1)_{T_{3L}}$  & 0.0417 & $6.02\times 10^{-3}$  & $6.02\times 10^{-3}$    
&  1.00    &  0.045  &  1095     &  -       &  -  \\
$U(1)_Q$         & 0.125  & $6.42\times 10^{-2}$  & $1.60\times 10^{-2}$    
&  4.01    &  0.1225   &  1275     &   -      &  -  \\
\hline
\hline
\end{tabular}
\caption{Model predictions and current constraints. The direct limits above
on the $Z'$ mass, $M^{\mathrm{D}}_{Z'}$, are the result of the analysis 
performed in this paper while the best indirect limits, 
$M^{\mathrm{I}}_{Z'}$, come from either electroweak (e) fits or contact (c) 
interactions at LEP2 \cite{Erler:2009jh}.
\label{models2}}
\end{table}


As discussed above, collider limits on the $Z^\prime$-boson mass can be 
presented via contours in the $c_u-c_d$ plane, with every contour 
corresponding to a well defined $M_{Z^\prime}$. Simultaneously, in the same 
plane, one can also show the values of $c_{d,u}$ couplings allowed by a 
specific $Z^\prime$ model. As a result, one can immediately visualize and  
derive a mass bound on the $Z'$-boson predicted by that particular model.

\begin{figure}[htb]
\includegraphics[width=0.515\textwidth]{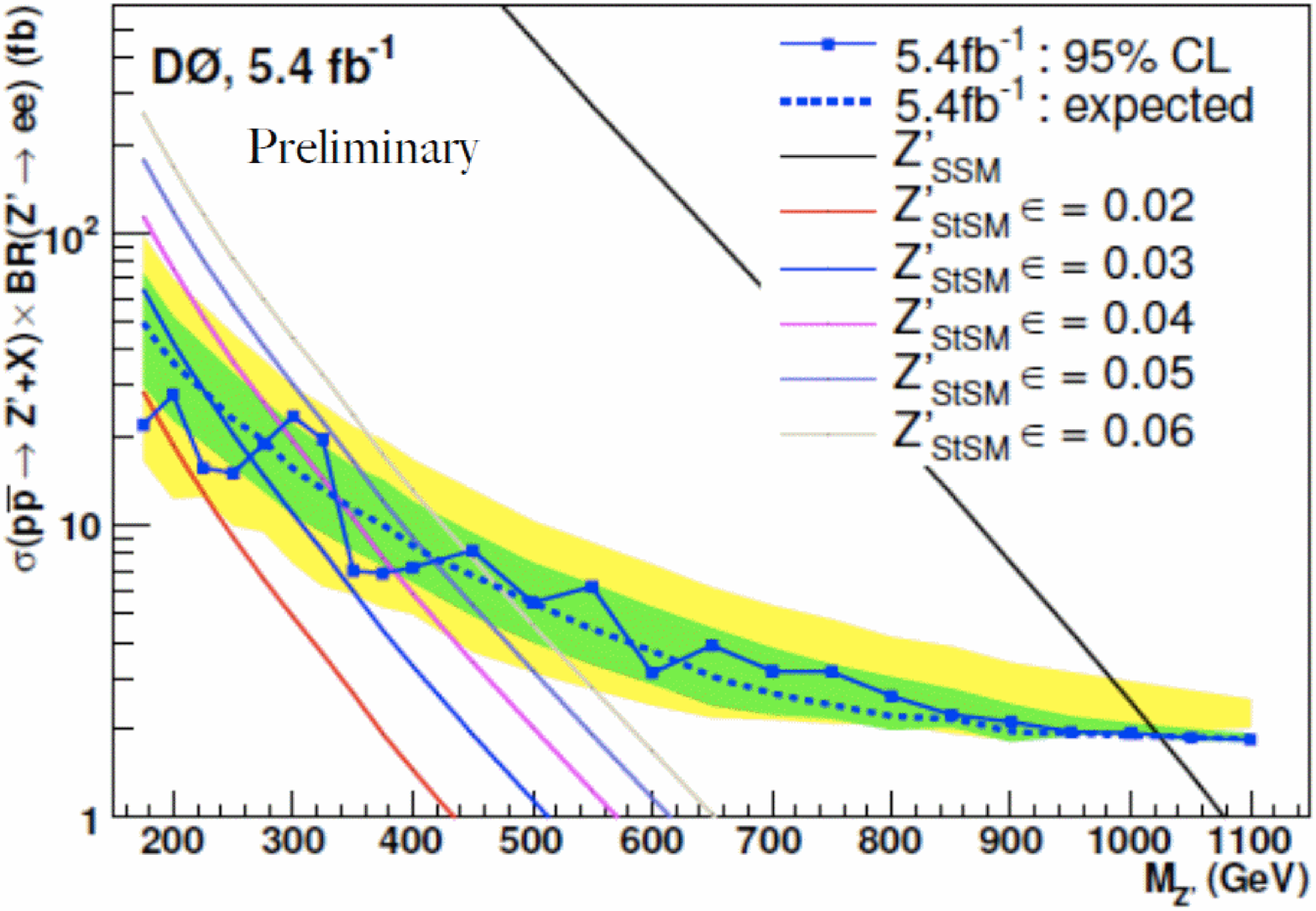}%
\includegraphics[width=0.515\textwidth]{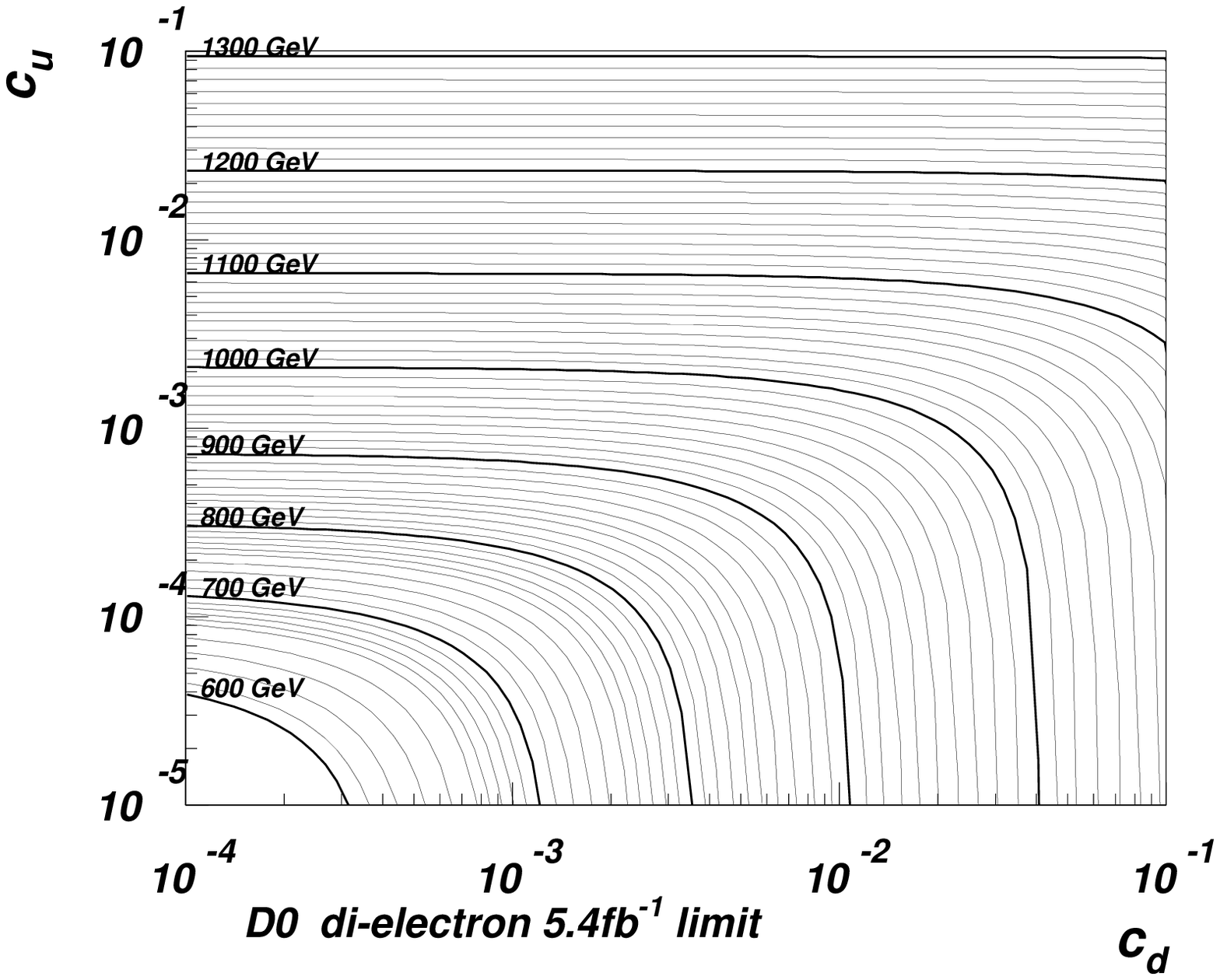}
\caption{\label{fig:d010}
Left: 95\% C.L. upper bound on $\sigma(p\bar p\to Z^\prime \to e^+e^-)$ 
obtained at the Tevatron with integrated luminosity L=5.4fb$^{-1}$ by the D0 
collaboration. Also shown, $Z^\prime$ mass bounds within the extra 
$U(1)`$ models displayed in the legend. 
Right: same 95\% C.L. upper bound on $\sigma(p\bar p\to Z^\prime\to e^+e^-)$ 
as above, but translated in the $c_u-c_d$ plane into contours corresponding 
to different $M_{Z^\prime}$ masses. The low mass models refer to the 
Stueckelberg Extensions of the SM (StSM) discussed in 
\cite{Feldman:2006ce,Feldman:2006wb}.}
\end{figure}

We start presenting our results at the Tevatron which is running at 
$\sqrt{s}=1.96$~TeV. We use the most recent 95\% C.L. upper bound on 
$\sigma(p\bar p\to Z^\prime\to\ell^+\ell^-)$ reported at the ICHEP 2010 
conference by the D0 collaboration for the di-electron channel 
\cite{ICHEP10,Abazov:2010ti} {where the $\Delta M/M_{Z'}\simeq 15\%$
cut was used in the analysis}.
This limit is shown in Fig.\ref{fig:d010} (left panel), together with its 
'translation' into the $c_u-c_d$ plane for different $M_{Z^\prime}$ masses
as shown in Fig.\ref{fig:d010} (right panel).

\begin{figure}[htb]
\includegraphics[width=0.5\textwidth]{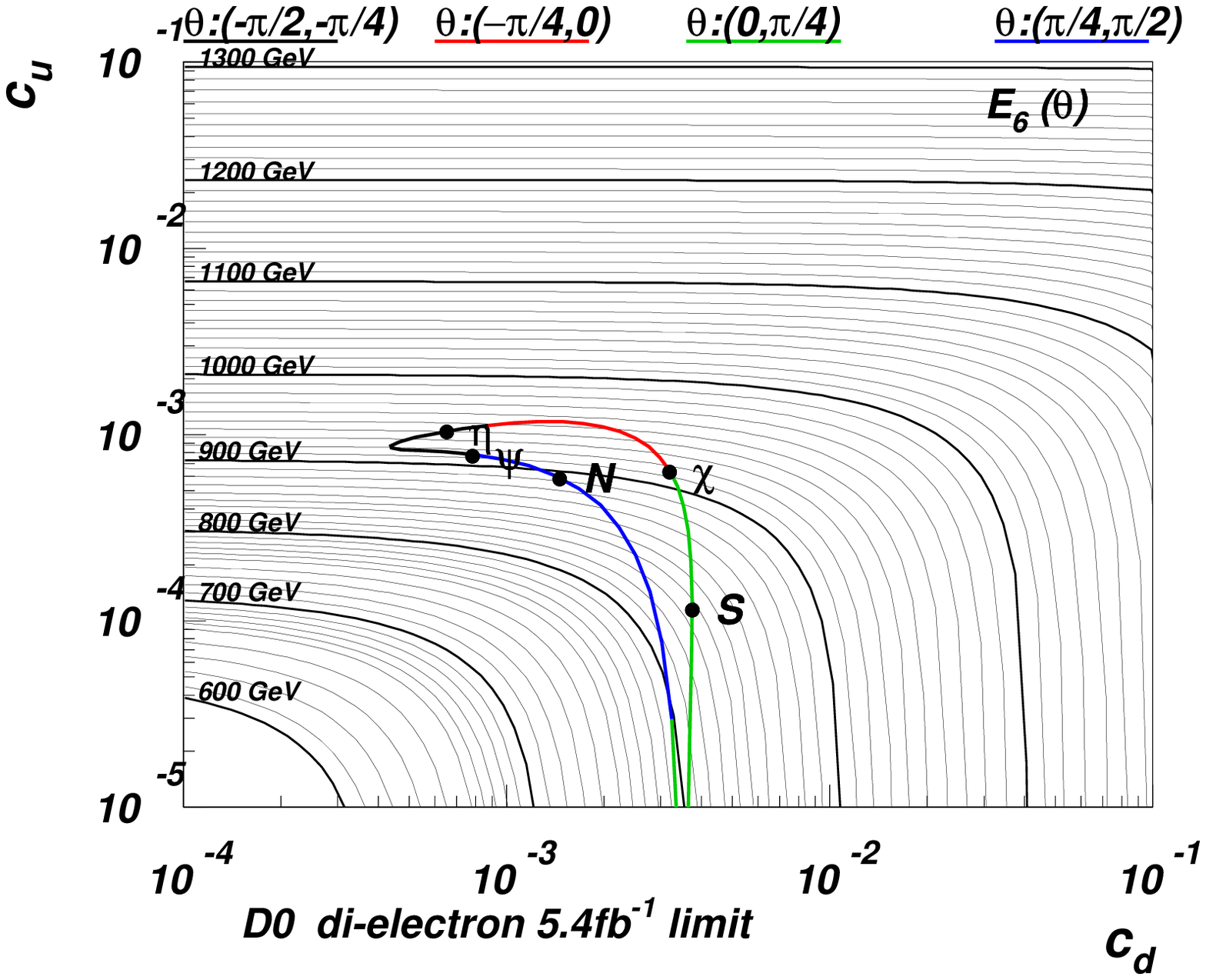}%
\includegraphics[width=0.5\textwidth]{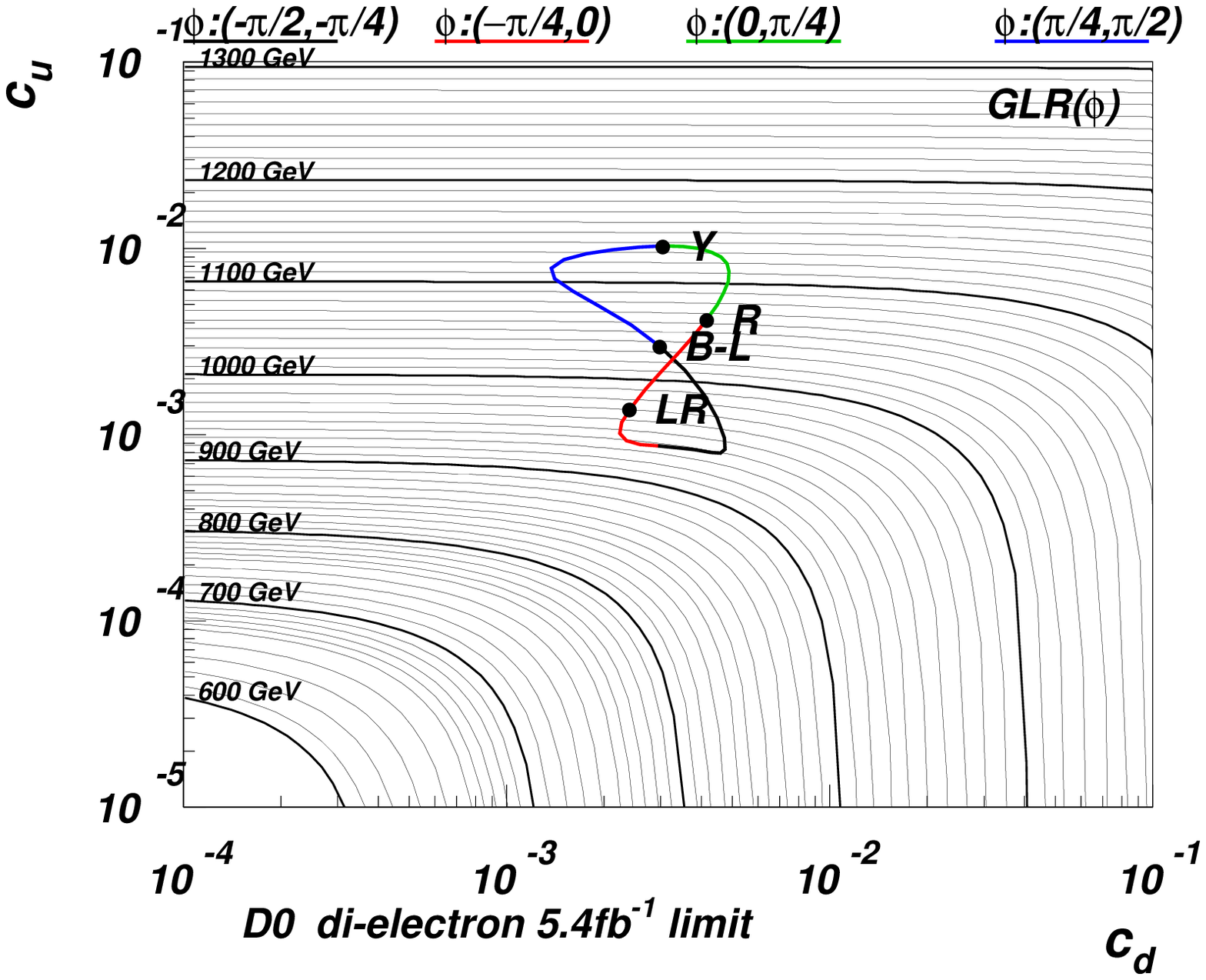}\\
\includegraphics[width=0.5\textwidth]{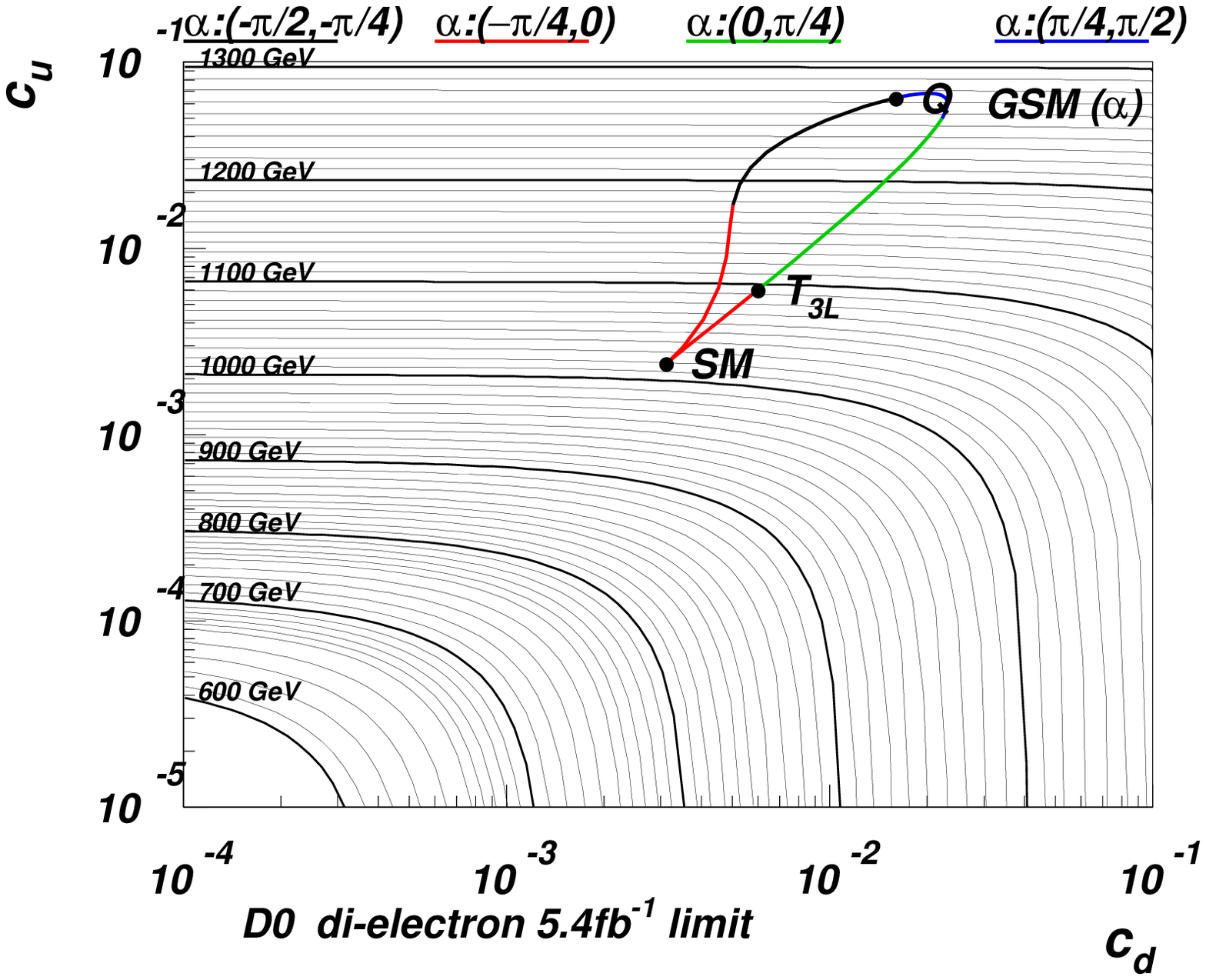}%
\includegraphics[width=0.5\textwidth]{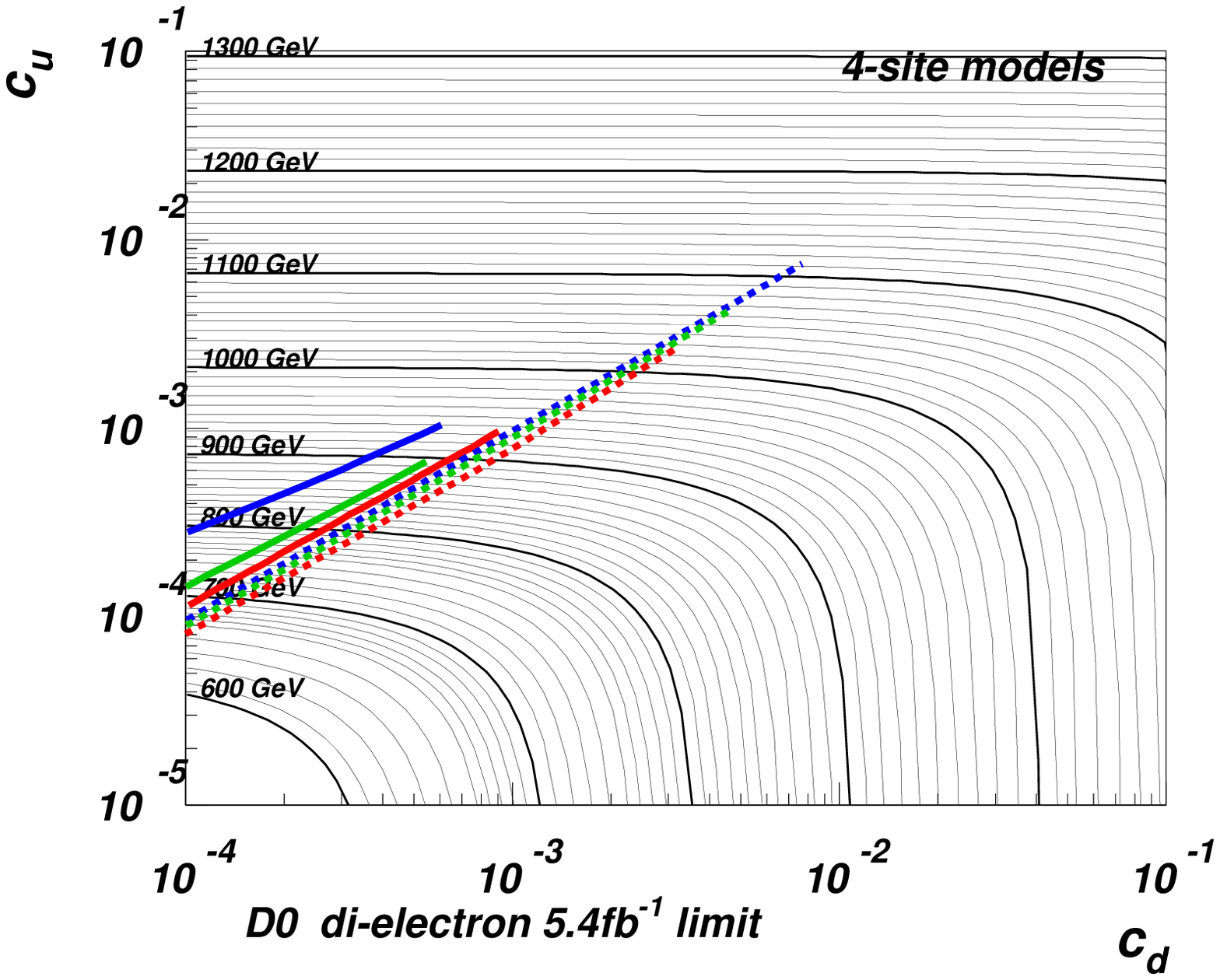}
\caption{\label{fig:d010-limits}
95\%CL limits on $M_{Z'}$ in the $c_u-c_d$ plane based on the 2010 analysis 
of the di-electron channel performed by the D0 collaboration at 
L=5.4fb$^{-1}$. Top-left, top-right, bottom-left, bottom-right panels 
present results for $E_6$ Models, Generalised Left-Right Models (GLR),
Generalised Sequential Models (GSM) and Four Site Higgsless Model (4S), 
respectively. The colour code corresponds to four equidistant intervals for 
the mixing angle within the $[-\pi/2,\pi/2]$ range for $E_6$, GLR, GSM models. 
The black dots on these contours correspond to the benchmark models listed in 
Table~\ref{models1}. For the 4S model, the colour denotes different values 
for $M_{Z_1}$ and $M_{Z_2}$. The line style distinguishes the $Z_1$ mass 
(solid  line) from the $Z_2$ mass (dashed line). 
For the $Z_1$ boson, the following mass values have been chosen: 
$M_{Z1}$= 480 (red), 800 (green) and 1600 (blue) GeV. For the chosen sample of 
free parameters, $z=0.8$, the  
corresponding values for the $Z_2$ bosons are: $M_{Z2}$= 600, 1000 and  
2000 GeV shown with the same colour coding.}
\end{figure}

In the following, we use Fig.\ref{fig:d010} (right panel) to interpret 
current limits from Tevatron, and to derive mass bounds on the $Z^\prime$ boson 
predicted in the classes of models described in the previous section.
The results are shown in Fig.~\ref{fig:d010-limits}. The top-left panel 
displays the contour representing $E_6$ Models in the $c_u-c_d$ plane, the 
top-right panel shows the Generalised Left-Right Models (GLR), the 
bottom-left panel contains the Generalised Sequential Models (GSM), and 
finally the bottom-right one gives the Four Site Higgsless Model (4S).
In the first three mentioned panels, the colour code corresponds to four 
equidistant intervals for the mixing angle in the $[-\pi/2,\pi/2]$ range
for $E_6$, GLR, GSM models, represented by continuous and closed contours.
The black dots on these contours denote the popular benchmark models quoted in 
Table~\ref{models1}. In the bottom-right panel, which shows the parameter space
of the Four Site Model, the colour indicates different mass values for 
$M_{Z_1}$ and $M_{Z_2}$. The line style distinguishes the $Z_1$ mass (solid 
line) from the $Z_2$ mass (dashed line).
For the $Z_1$ boson, the following mass values have been chosen:
$M_{Z1}$= 480 (red), 800 (green) and 1600 (blue) GeV. For the chosen sample of
free parameters, $z=0.8$, the 
corresponding values for the $Z_2$ bosons are: $M_{Z2}$= 600, 1000 and 
2000 GeV shown with the same colour coding.

Several comments are in order. The first remarkable fact is that there is 
almost no overlap between contours for $E_6$, GLR, GSM models. This means 
that, if a $Z'$ boson will be discovered and its cross section will be 
measured with a reasonable accuracy, these classes of $Z'$ models can be
well distinguished using just this basic information. The second remark 
concerns the experimental sensitivity to the $Z^\prime$ production within
different models. Comparing the four plots, it is clear that the highest 
sensitivity is to the GSM class of models. In particular, the $Q$-model 
can be excluded at the Tevatron up to masses $M_{Z'}\geq 1260$ GeV. Among
the GLR models, which are second in terms of the experimental sensitivity to
a $Z'$ boson, the Y-model can be already excluded up to $M_{Z'}\geq 1125$ GeV
with the current Tevatron data. Interestingly, the lowest experimental 
sensitivity is to $E_6$ models, that is one of the most popular class of $Z'$ 
models. Within this class, the strongest limit can be derived for 
$\theta\in[-\pi/4,0]$ providing the mass bound $M_{Z'}\geq 955$, as one
can read from the red-coloured part of the $E_6$ contour.

The 4S class of $Z'$ models must be considered separately. First of all, it 
predicts two $Z'$ bosons with two different masses. Secondly, the parameter 
space of the 4S model is described by more than just one parameter, so the 
model would be represented by an area in the $c_u-c_d$ plane rather
than by a contour. In order to interpret Fig.\ref{fig:d010-limits} 
(bottom-right panel) and following analogous figures correctly, a 
clarification is needed. In the 4S model, the two extra gauge bosons can decay
into both SM fermion pairs and boson pairs. While the contribution to the total width coming from the decay into fermion pairs is linear in the extra gauge   
boson mass, the contribution from the diboson decay grows with the third 
power of the extra gauge boson mass (see Eq. \ref{width_ff}).
As a consequence, and oppositely to the other perturbative gauge models, the 
$Z_{1,2}$ boson branching ratio into lepton pairs acquires a mass dependence 
(see Sec.~\ref{sec:nwa} for details). This reflects into a mass dependence of 
the $c_u$ and $c_d$ parameters which parametrize the 4S model. So,
Fig.\ref{fig:d010-limits} (bottom-right) should be interpreted as the full 
parameter space of the four site model projected into the $c_u-c_d$ plane. 
To simplify the visualisation of this area, we have varied the vector coupling 
between the $Z_2$ boson and SM electrons, $g_{2V}^e$, within the allowed 
region of Fig.~\ref{fig:parameter_space} for the sample scenarios: z=0.8 and 
$M_{Z2}$= 600, 1000 and 2000 GeV. This setup should give a full 
representation of the parameter space, $M_{Z2}$= 600 GeV being the minimum 
allowed mass and $M_{Z2}$= 2000 GeV being close to the maximum value of the 
mass permitted by unitarity. 
The parameter space for these values of $M_{Z2}$ and the respective 
$M_{Z1}=0.8\times M_{Z2}$ is presented in
Fig.~\ref{fig:d010-limits}(bottom-right) by coloured lines (see caption).
Whenever the coloured line describing a given $M_{Z1,Z2}$ value for the 
4S model crosses the black contour corresponding to the same mass value, that 
crossing point would give the experimental sensitivity to a $Z_{1,2}$ boson 
with that mass. The portion of the coloured line above this crossing point 
would represent the excluded region in the $c_u-c_d$ parameter space. 
To clarify the interpretation, let us consider the following examples. 
For $M_{Z_2}=1000$ GeV and $M_{Z_1}=800$ GeV (green lines), one can see that 
the parameter space for the $Z_1$ boson (solid green line) 
is greatly excluded in the region of the $c_u-c_d$ plane above the black 
contour line labelled by 800 GeV. Only a small portion of the $c_u-c_d$ 
parameter space is instead excluded for the $Z_2$ boson (dashed
green line). The $c_{u,d}$ couplings are indeed bounded to be 
$c_{u,d}\le 10^{-3}$, as one can see from the crossing point between the 
dashed green line and the black countour labelled by 1000 GeV. 
Since, the two extra gauge bosons would be simultaneously produced, from the 
discussed green lines one should deduce that the most restrictive bound on 
the $c_{u,d}$ couplings comes from the $Z_1$ boson. If not observed, the 
crossing point of the solid green line representing $M_{Z1}=800$ GeV with 
the black contour at 800 GeV would give the bound: $c_{u,d}\le 2 10^{-4}$.  
Consider now a second scenario: $M_{Z2}=2000$ GeV and $M_{Z1}=1600$ GeV.
This is represented by blue lines. In this case, the Tevatron has no 
sensitivity at all to the two extra gauge bosons. The solid blue line 
representing $M_{Z1}=1600$ GeV, in fact, never crosses the corresponding 
black contour line labelled by 1600 GeV. And the same is true for the dashed 
blue line representing $M_{Z2}=2000$ GeV.

\begin{figure}[htb]
\includegraphics[width=0.50\textwidth]{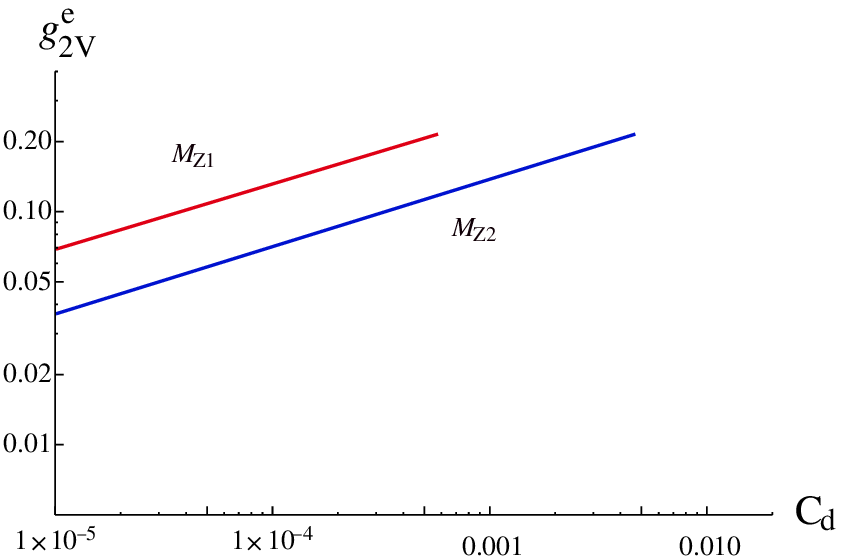}%
\includegraphics[width=0.50\textwidth]{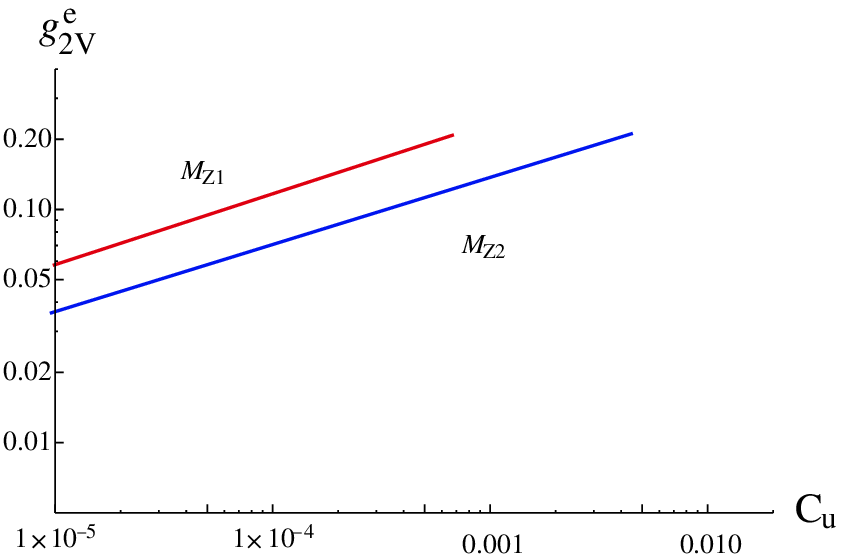}
\caption{\label{fig:4spars}
$g_{2v}$ fermion-boson coupling versus 
$c_u$(left) or  $c_d$(right) in four-site model}
\end{figure}

Even in more complicated multi-resonance scenarios, as in the 4S model, the 
$c_{u,d}$ representation allows one to visualize directly via the black 
contour lines, up to what mass value the experiment could be sensitive.
If no signal of new physics is observed, the bound on the mass can be 
translated into limits on the $c_{u,d}$ coefficients. From there, one can 
then trace back exactly the lagrangian parameters of the 4S model which are 
in turn excluded. In case of new physics discovery, the $c_{u,d}$ approach
allows one to uniquely determine the $c_{u,d}$ values corresponding to that
observed mass. Since the $c_{u,d}$ coefficients are strictly related to the 
new gauge boson couplings, this in turn enables one to extract informations 
on their size.
In the 4S model, the coupling between the $Z_2$ boson and SM electrons, 
$g_{2V}^e$, grows linearly with $c_u$ and $c_d$ as shown in 
Fig.~\ref{fig:4spars}. Any mass measurement therefore translates into a 
coupling determination. Moreover, $g_{2V}^e$ is one of the three free 
parameters of the model. The measurement of the mass of the new gauge 
bosons would therefore allow one to derive direct informations on the bare 
lagrangian parameters.

\begin{figure}[htb]
\includegraphics[width=0.46\textwidth]{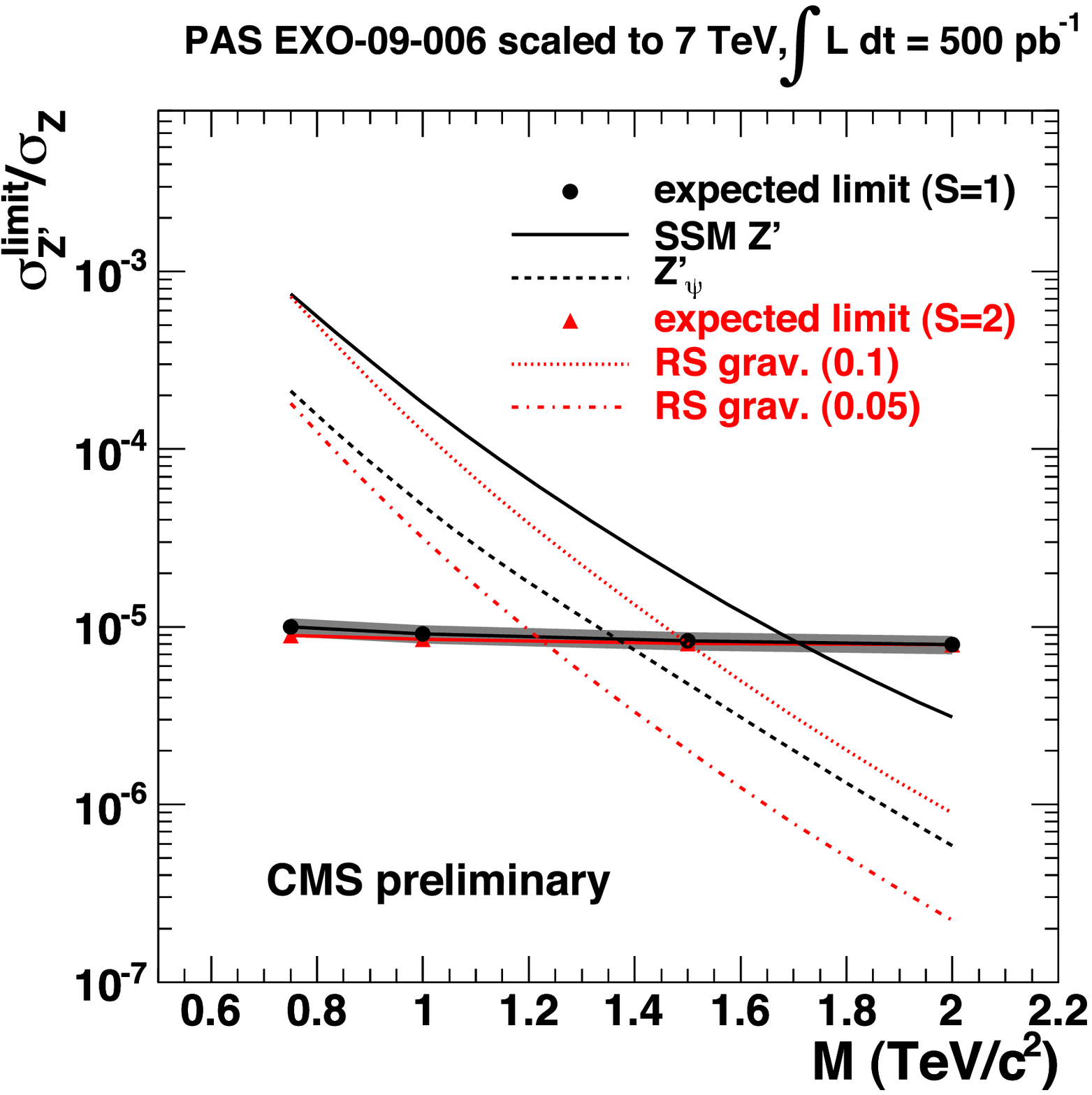}%
\includegraphics[width=0.54\textwidth]{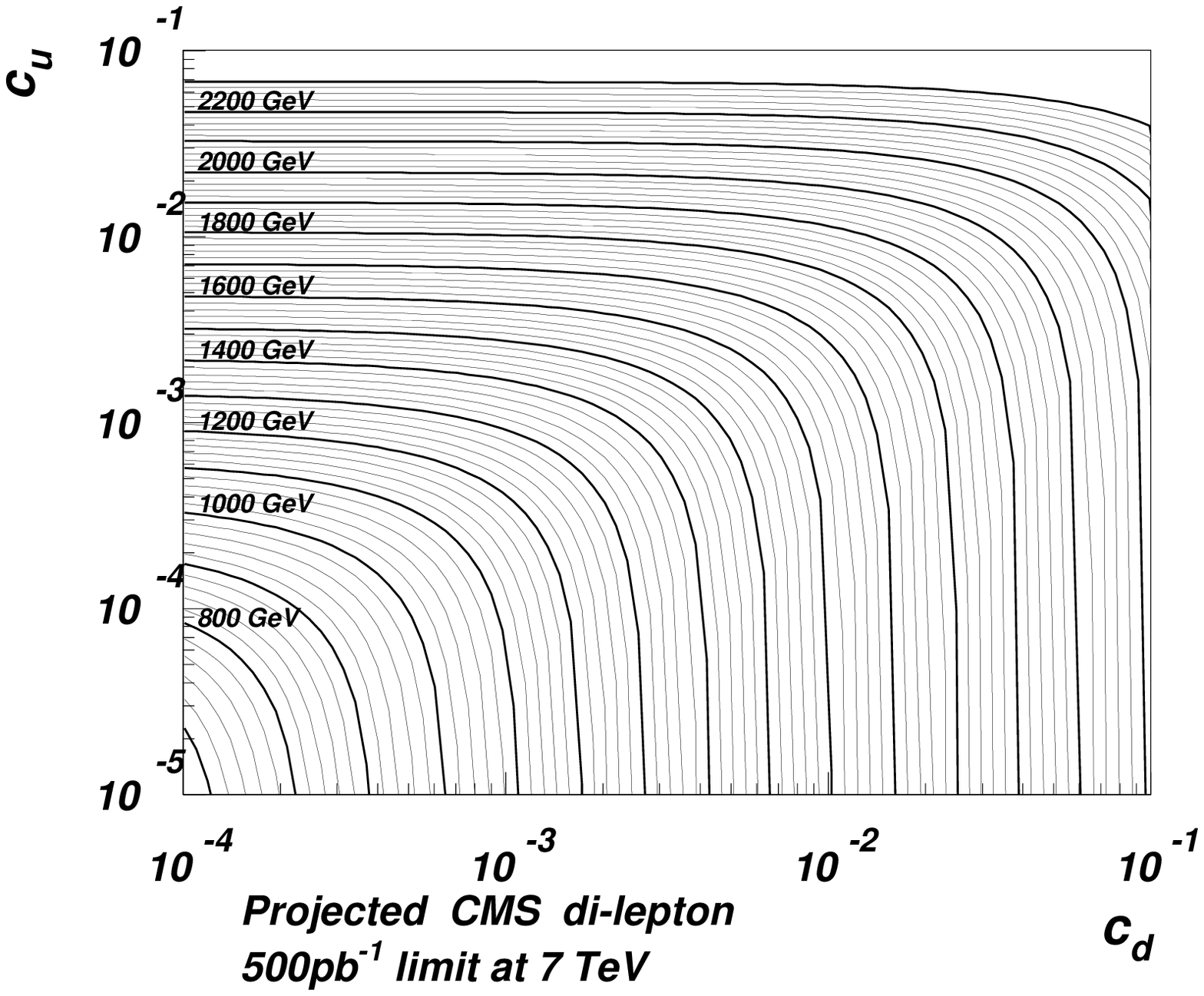}
\caption{\label{fig:cms07_500pb-exp} 
Left: CMS limit on the $Z^\prime$ boson production cross section in the 
di-electron channel, normalized to the SM $Z$ boson cross section, as a 
function of the $Z^\prime$ mass. The limit is projected at 500pb $^{-1}$ 
\cite{CMSEXO}.
Right: Limits in the $c_u-c_d$ plane, based on the projected LHC 
500pb $^{-1}$ limit shown in the right panel. In the $c_u-c_d$ representation,
the limits appear as contour lines corresponding to different $M_{Z'}$ values.}
\end{figure}

\subsection{Expected LHC potential at 7 TeV to probe $Z'$ models}

We now explore the LHC@7TeV potential to test the classes of $Z'$ models under 
consideration. We use the projected limits from LHC. In particular, we rely 
on the limits given by the CMS Exotica group for 500pb $^{-1}$ of integrated 
luminosity which hopefully will be available in about one year from now.
This limit is shown in Fig.\ref{fig:cms07_500pb-exp} (left panel), which
is taken from the public web page of the CMS Exotica group \cite{CMSEXO}.
The projected 500pb $^{-1}$ limit from CMS is given as a ratio
$\sigma_{Z'}/\sigma_{Z}$, where $\sigma_{Z}$ is the Z-boson production cross 
section in the $60<m_{ee}<120$ GeV window
{and we have converted this limit into the limit on the NNLO
production cross section for the $Z'$ boson
shown in terms of $c_{u,d}$ coefficients
in  $c_u-c_d$ plane for different $Z'$ 
masses given in the  right panel of Fig.\ref{fig:cms07_500pb-exp}}.
This representation is the analog of what done before at the Tevatron.
Comparing Fig.\ref{fig:d010} and Fig.\ref{fig:cms07_500pb-exp}, one can 
observe the strong gain in sensitivity one gets at the LHC@7TeV with 
500pb $^{-1}$, compared to the Tevatron with 5.4fb $^{-1}$, at fixed 
$c_{u,d}$ value.

\begin{figure}[htb]
\includegraphics[width=0.50\textwidth]{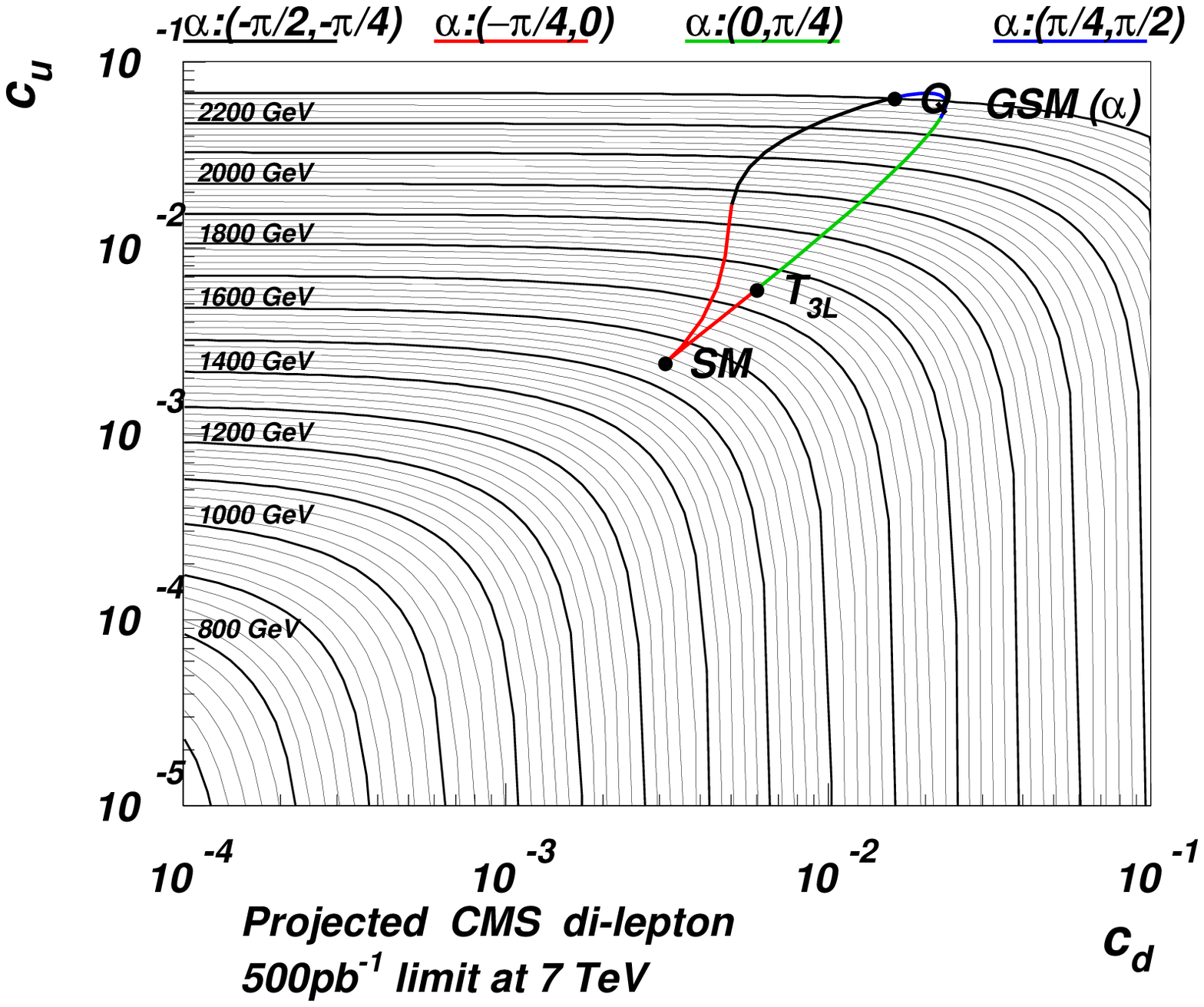}%
\includegraphics[width=0.50\textwidth]{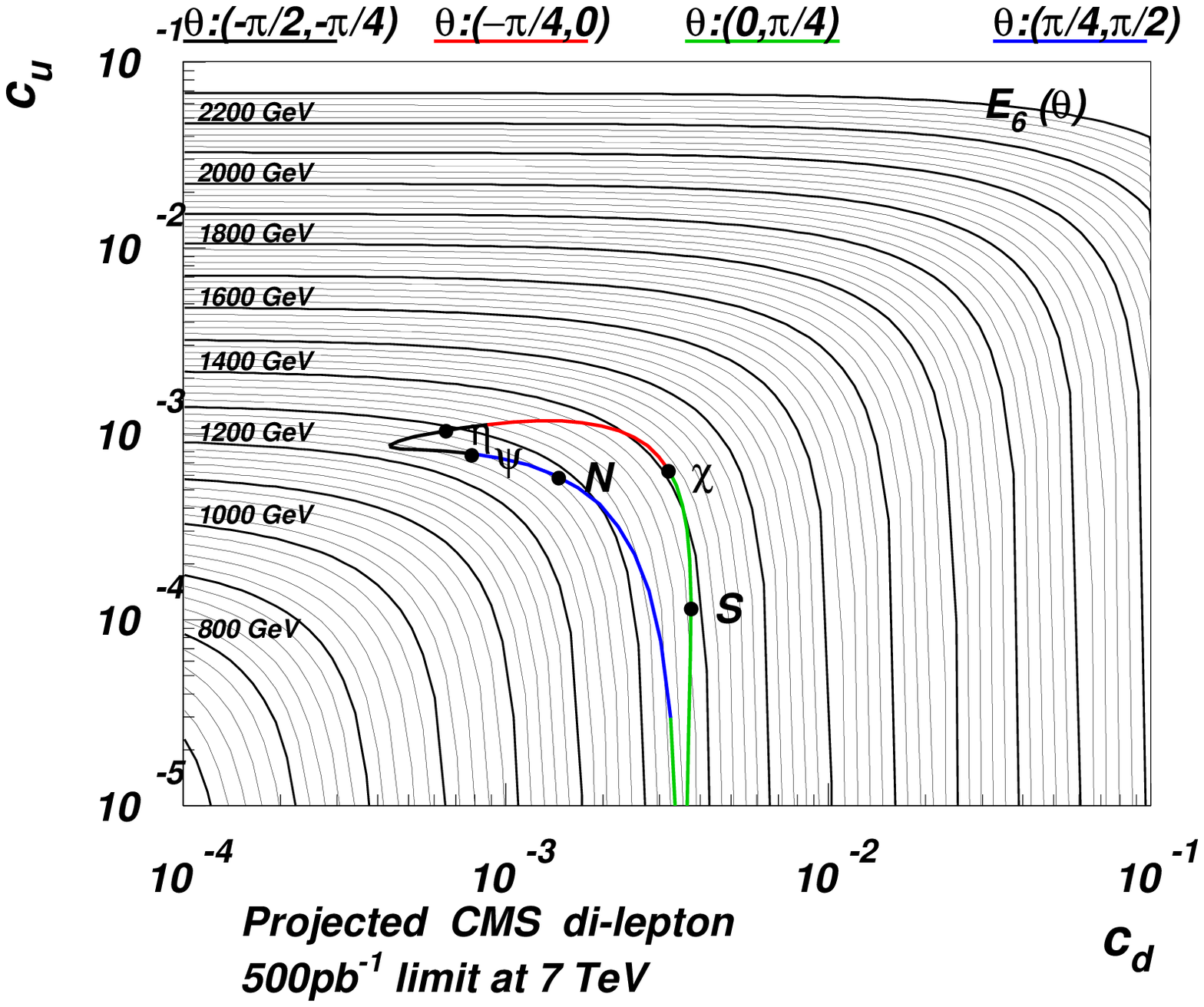}\\
\includegraphics[width=0.50\textwidth]{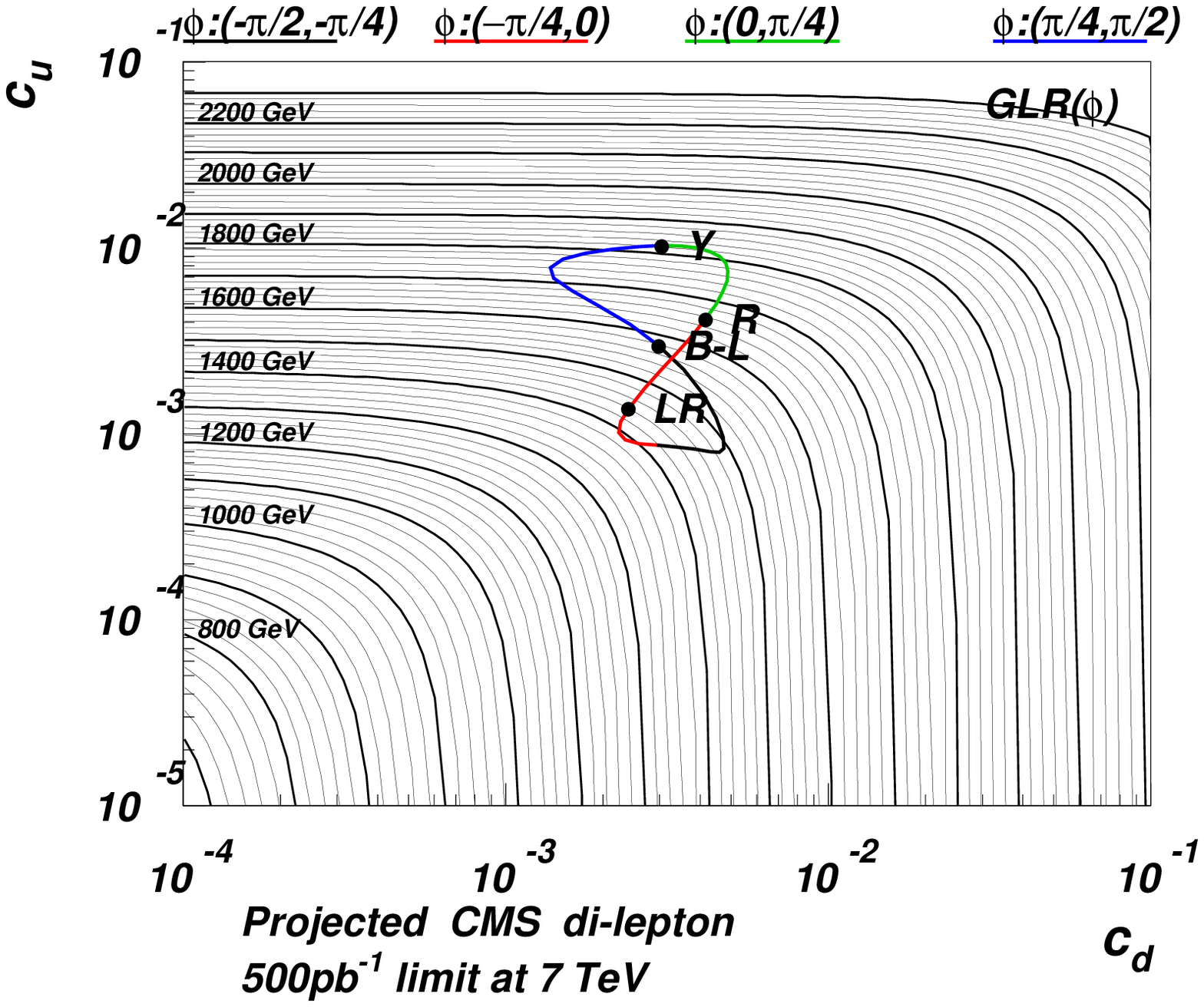}%
\includegraphics[width=0.50\textwidth]{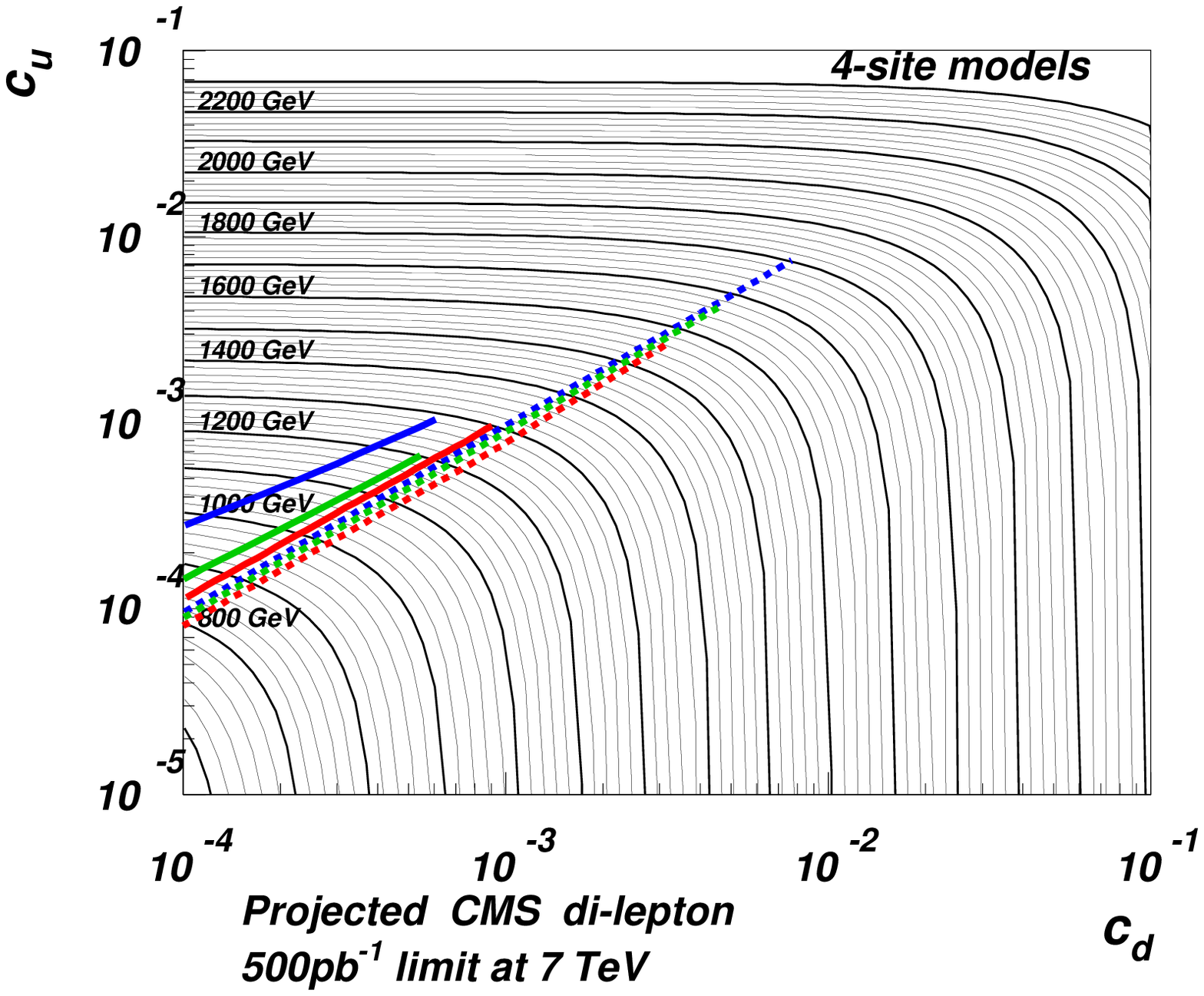}
\caption{\label{fig:cms07_500pb} 
Limits in $c_u-c_d$ plane are based on  projected CMS 500pb $^{-1}$ limit.
95\%CL limits on $M_{Z'}$ in $c_u-c_d$ plane based on 
on  projected CMS 500pb $^{-1}$
analysis of di-electron channel.
The legend scheme is  the same as in Fig.~\ref{fig:d010-limits}.}
\end{figure}

Now we can estimate the LHC@7TeV potential for deriving bounds on $Z'$ models 
at 500pb $^{-1}$. The results are shown in Fig.~\ref{fig:cms07_500pb}, where
the legend scheme is the same as in Fig.~\ref{fig:d010-limits}.
From Fig.~\ref{fig:cms07_500pb} and Fig.~\ref{fig:d010-limits}, one 
can see that for the models with small-intermediate values of $Z'$ boson 
couplings to SM fermions (that is $E_6$ models, GLR-models partly, and 
some GSM models), the LHC@7TeV can extend the limits on $M_{Z'}$ by about 
500 GeV when compared to the Tevatron. For example, the limit on the SM-like 
$Z'$ boson could be extended from 1020 GeV to about 1520 GeV.
On the other hand, the limits for larger $c_{u,d}$ coefficients and 
respectively larger masses could be extended in the near future up to
a 2 TeV scale, which is unreachable at the Tevatron.
For the Q-model, belonging to the GSM class, the mass bound could be improved 
from 1210 GeV (current Tevatron) to 2250 GeV. 
Regarding the 4S model, one can see that the scenario characterized by 
$M_{Z1}=800$ GeV and $M_{Z2}=1000$ GeV could be totally excluded in the 
$c_u-c_d$ plane shown in Fig.~\ref{fig:cms07_500pb}. The solid green line, 
representing the $Z_1$ boson parameter space, lies in fact beyond the 800 GeV
black contour line in the dispayed plane. No improvement, compared to the 
current Tevatron, would instead be possible for the scenario: 
$M_{Z_1}=1600$ GeV and $M_{Z_2}=2000$. The LHC@7TeV and 500 pb$^{-1}$ will 
not have sensitivity enough, as one can deduce from observing that the blue
lines never cross the black contour lines labelled by their respective mass 
values. Exploring this region of the 4S parameter space would require higher 
integrated luminosity and preferably higher collider energy.

\begin{figure}[htb]
\includegraphics[width=0.50\textwidth]{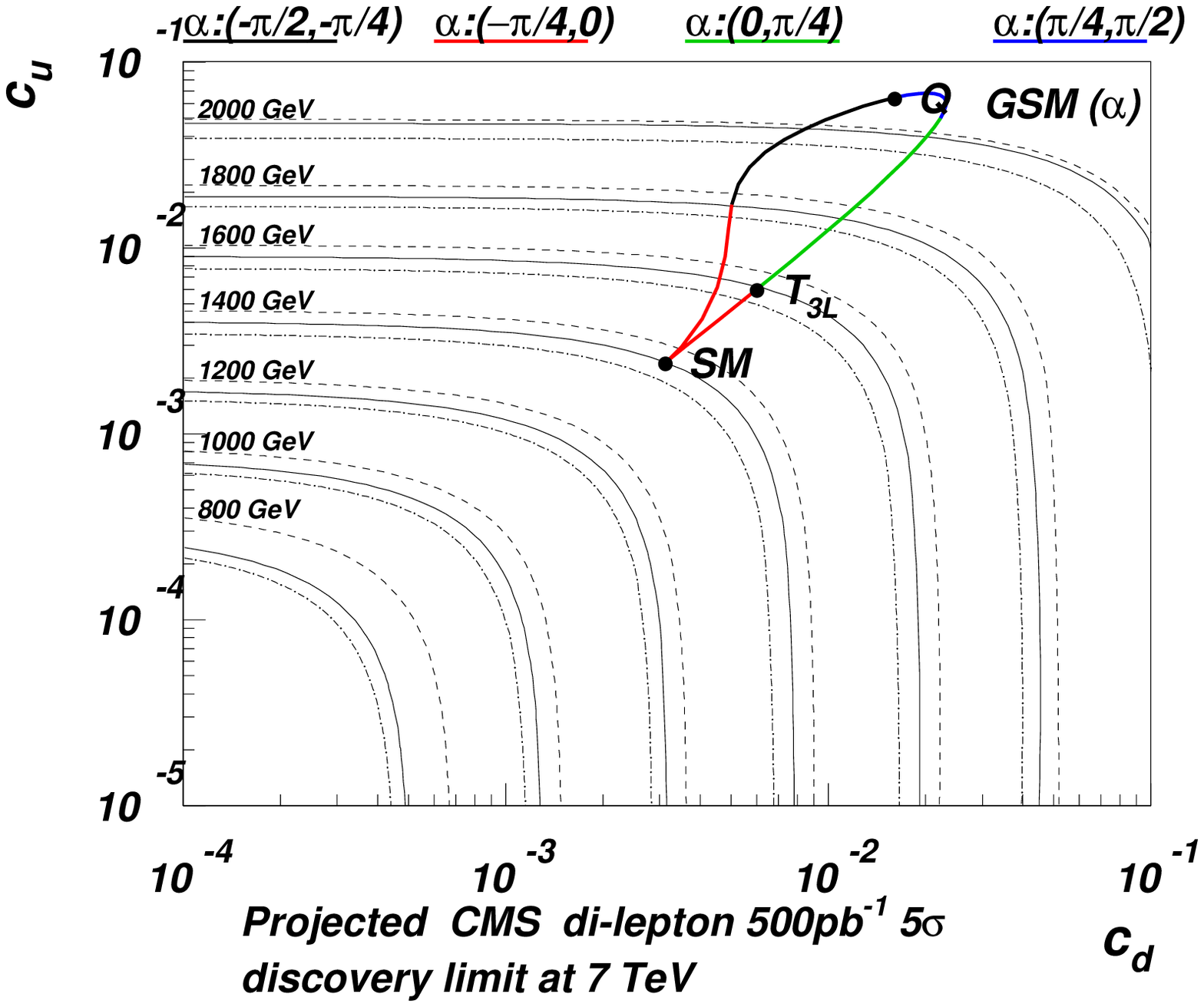}%
\includegraphics[width=0.50\textwidth]{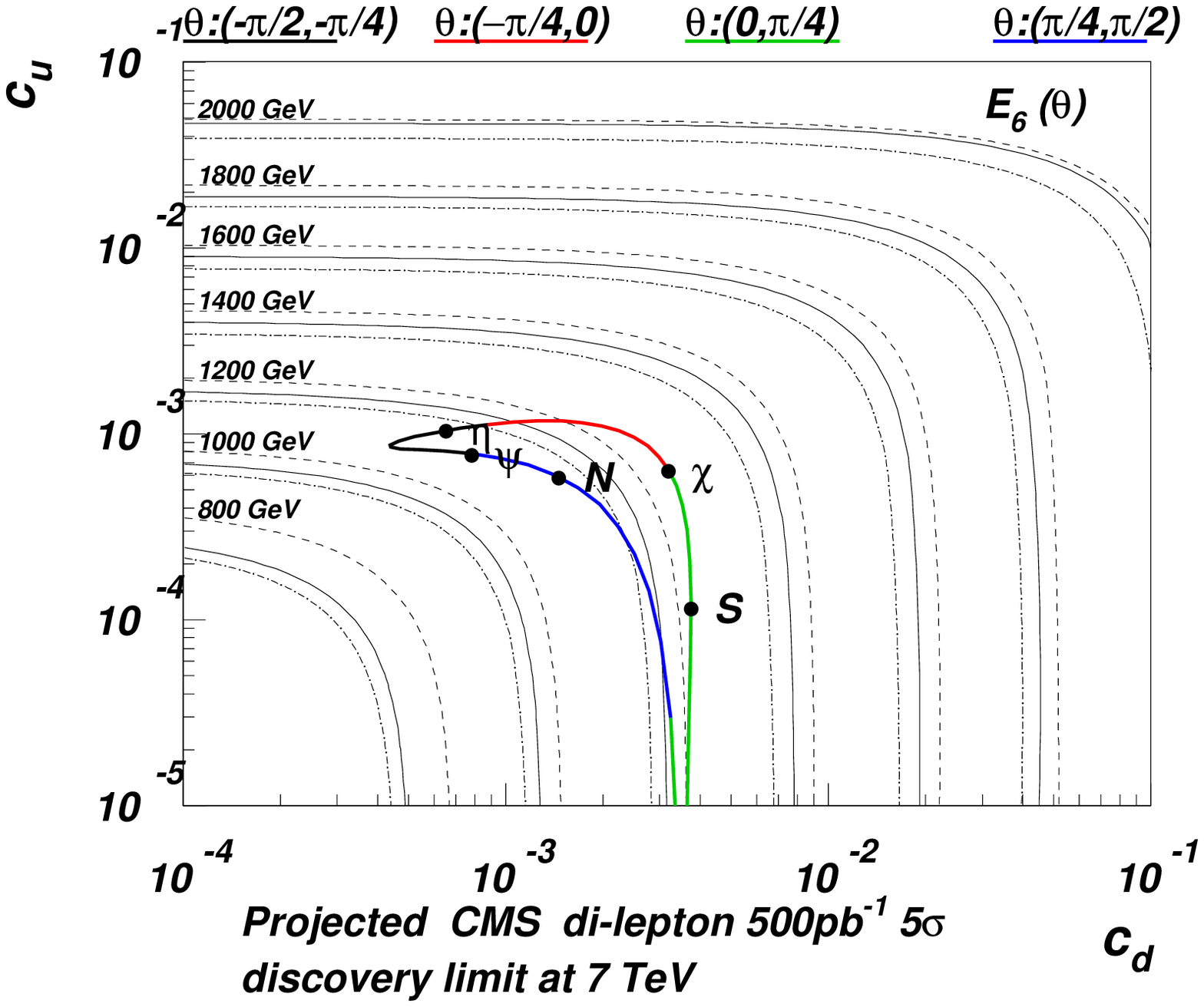}\\
\includegraphics[width=0.50\textwidth]{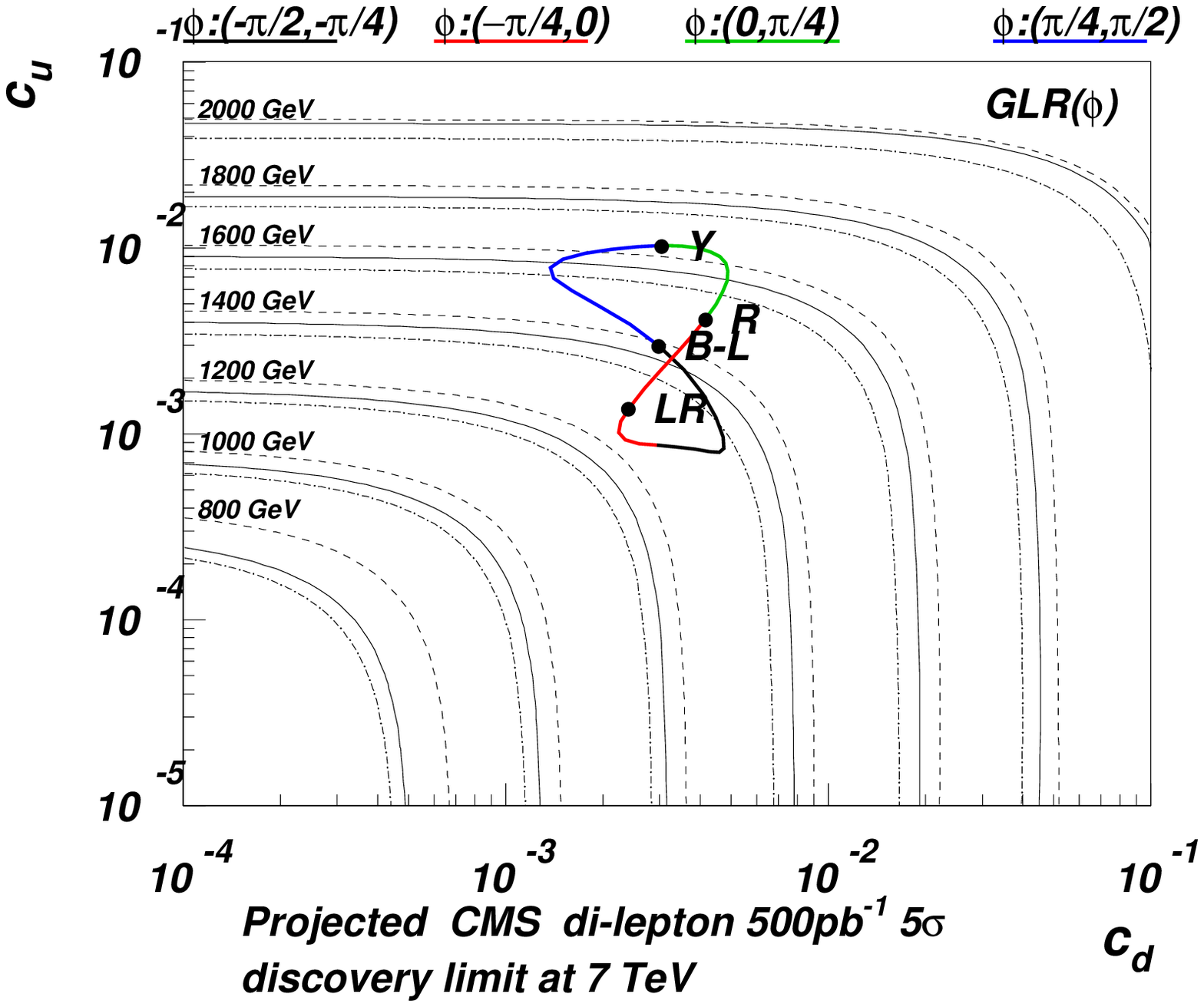}%
\includegraphics[width=0.50\textwidth]{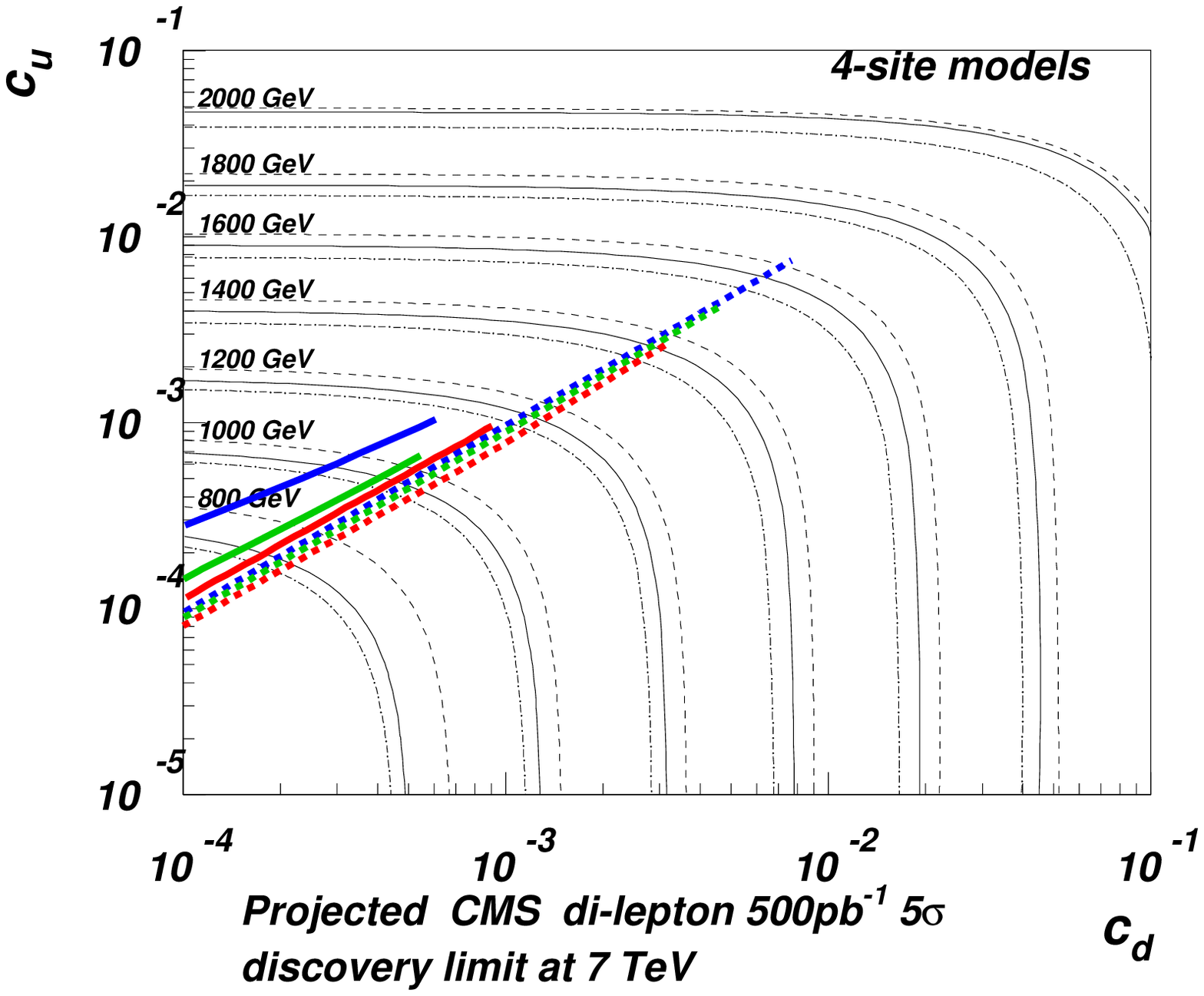}
\caption{\label{fig:cms07_500pb-disc}
$5\sigma$ Discovery limits in the $c_u-c_d$ plane, based on the 500pb$^{-1}$ 
projected analysis of the $Z^\prime$ production in the di-electron 
channel performed by the CMS collaboration.
The legend scheme is  the same as in Fig.~\ref{fig:d010-limits}.
The upper(dashed) and lower(dot-dashed) contours correspond
to the uncertainty in Fig.~\ref{fig:cms07_500pb-exp}(left)
reflected in the width of the band.}
\end{figure}

We also use Fig.~\ref{fig:cms07_500pb-exp} to estimate the LHC@7TeV discovery 
limits for 500pb $^{-1}$. In this analysis, we assume that the significance 
grows as $\sqrt{L}$, where $L$ is the total integrated luminosity and that 
the signal over background ratio is constant for the same selection cuts. The 
last assumption is motivated by the fact that for a chosen  invariant 
di-lepton mass window the $q\bar{q}$ parton densities are very similar for 
signal and background processes and defined mainly by the $\sqrt{\hat{s}}$ 
value.

The LHC discovery potential for various $Z'$
models  is shown in Fig.~\ref{fig:cms07_500pb-disc}
in the $c_u-c_d$ plane. The legend scheme is  the same as in 
Fig.~\ref{fig:d010-limits}.
The upper(dashed) and lower(dot-dashed) contours correspond
to the uncertainty in Fig.~\ref{fig:cms07_500pb-exp}(left)
reflected in the width of the band.
One  can see that discovery limits are typically 150-200 GeV lower than
exclusion ones.

\section{Impact of $Z^\prime$ width on search strategies \label{V}}

Invariant mass distributions may be examined in a number of ways for
evidence of resonant structures. The sensitivity of any particular
approach has a dependence on the intrinsic width of any possible
resonance. The simplest approach is to bin the invariant mass
distribution and determine the compatability of the number of events
in any bin with the Standard Model prediction. A p-value may be used
to quantify this compatability. In this approach the width of the bins
for optimal sensitivity depends on the intrinsic width of possible
resonances and the detector resolution. In the case where the width of
the resonance is much smaller than the detector resolution this
parameter defines the optimal bin size. For intrinsic widths comprable
to the detector resolution then the optimal width depends on both of these 
parameters. 

An alternative is to use a parameterization of the expected
distribution in the two alternative hypotheses of, a distribution
resulting only from Standard Model Physics and one resulting from the
addition of a new physics process. A comparision of some measure of
the quality of the fit in both cases allows a determinatin of the
probability of the presence of New Physics.  In this case the
functional form of the resonance structure is typically taken to be
some variant on a convolution of a Gaussian and a Breit-Wigner.  For a
resonance where the width of the Breit-Wigner is small in comparision
to the width of the Gaussian the sensitivity of the search is
insensitive to the width of the Breit-Wigner. Such circumstances
result in the greatest possible experimental sensitivity. For
resonances with large widths compared to the experimental resolution
then the Breit-Wigner width must be included as a further parameter in
any fits. In cases where the interference effect is large this must be
included in the fuctional form used to fit to the invariant mass
distribution. We show here that in the mass regions which will provide
the greatest sensitivity for low integrated luminosities at the LHC
the interference effect is negligible and may be ignored without
compromising the search sensitivity. 


The above descriptions of possible methods of searching for a
resonance in an invariant mass distribution illustrate that a
knowledge of the width of the resonances being searched for has an
impact on the search procedure used. In order to obtain the best sensitivity, 
it is thus important to have a knowledge of the magnitude of the widths from 
New Physics models. The results of such searches depend on the assumptions 
made in the analysis and can't be easily interpreted in other circumstances.

Besides that, in the case of a discovery the $Z^\prime$ boson mass and decay 
width will be determined from fits to the reconstructed invariant mass 
distribution of di-lepton candidate events. In this section, we thus focus on 
the prediction and the possible measurement of the total decay width, 
comparing the various classes of models under consideration.

\begin{figure}[]
\centering
\hspace*{-0.5cm}\includegraphics[width=0.4\textwidth]{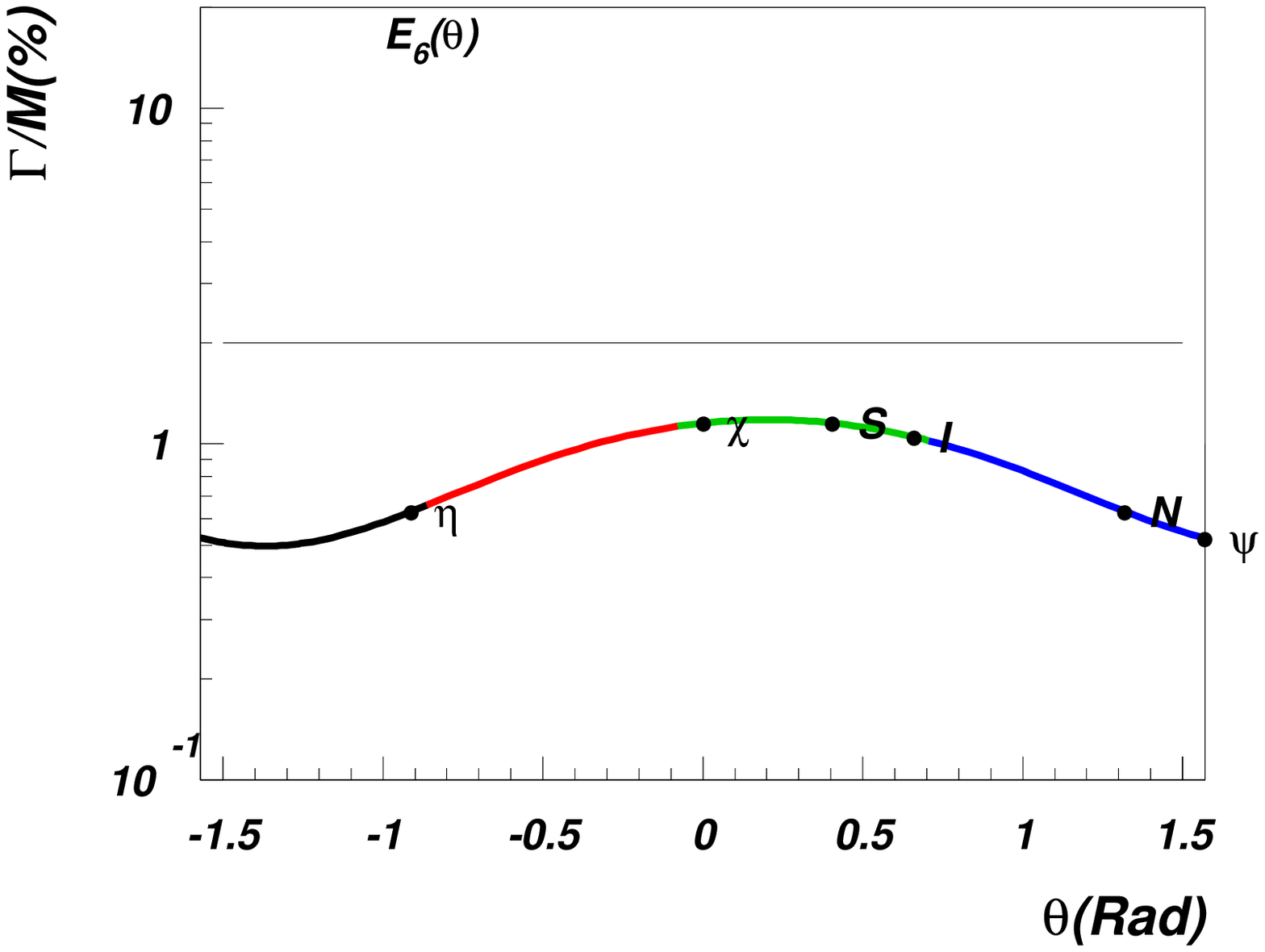}%
\hspace*{-0.9cm}\includegraphics[width=0.4\textwidth]{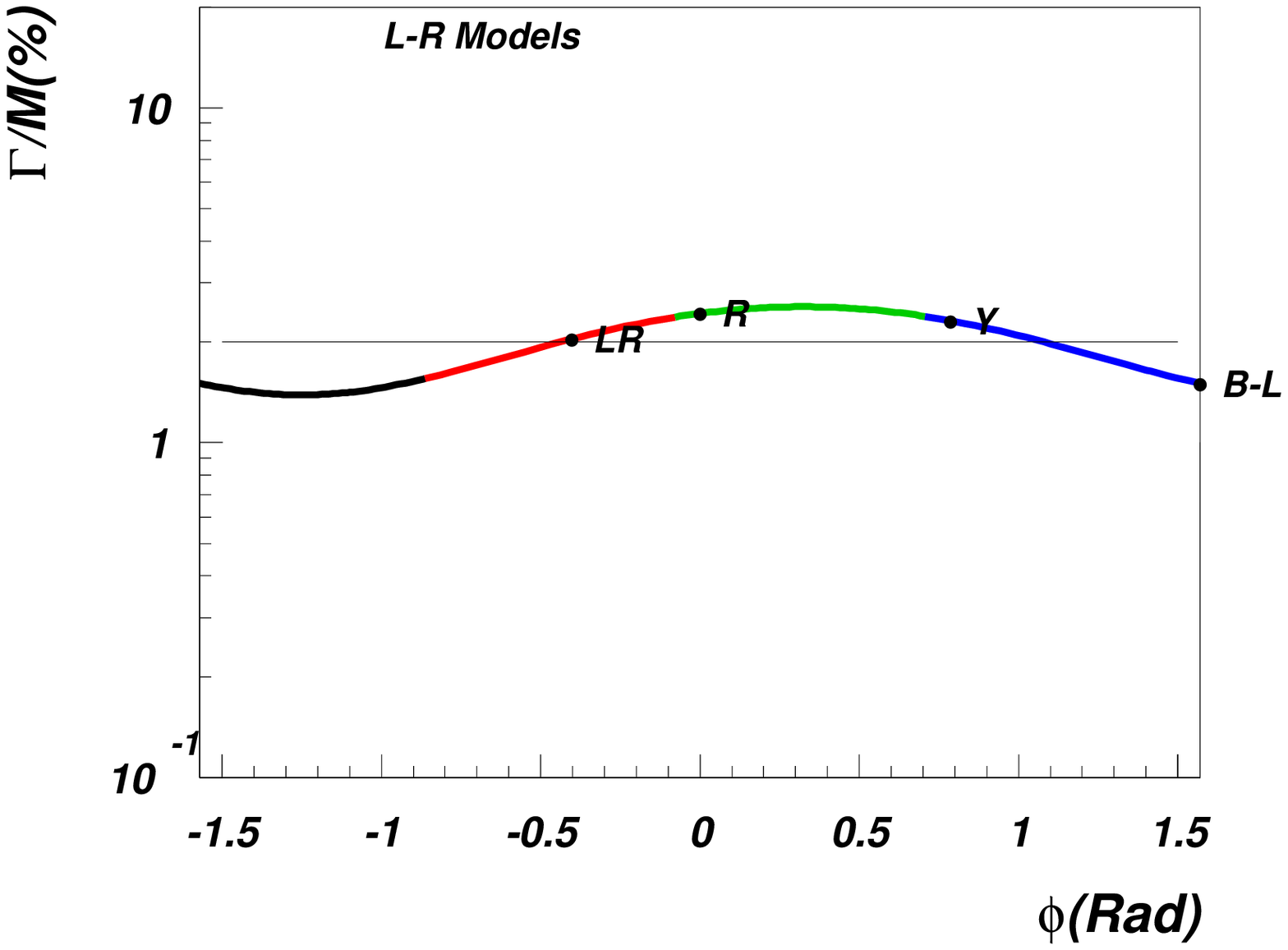}%
\hspace*{-0.9cm}\includegraphics[width=0.4\textwidth]{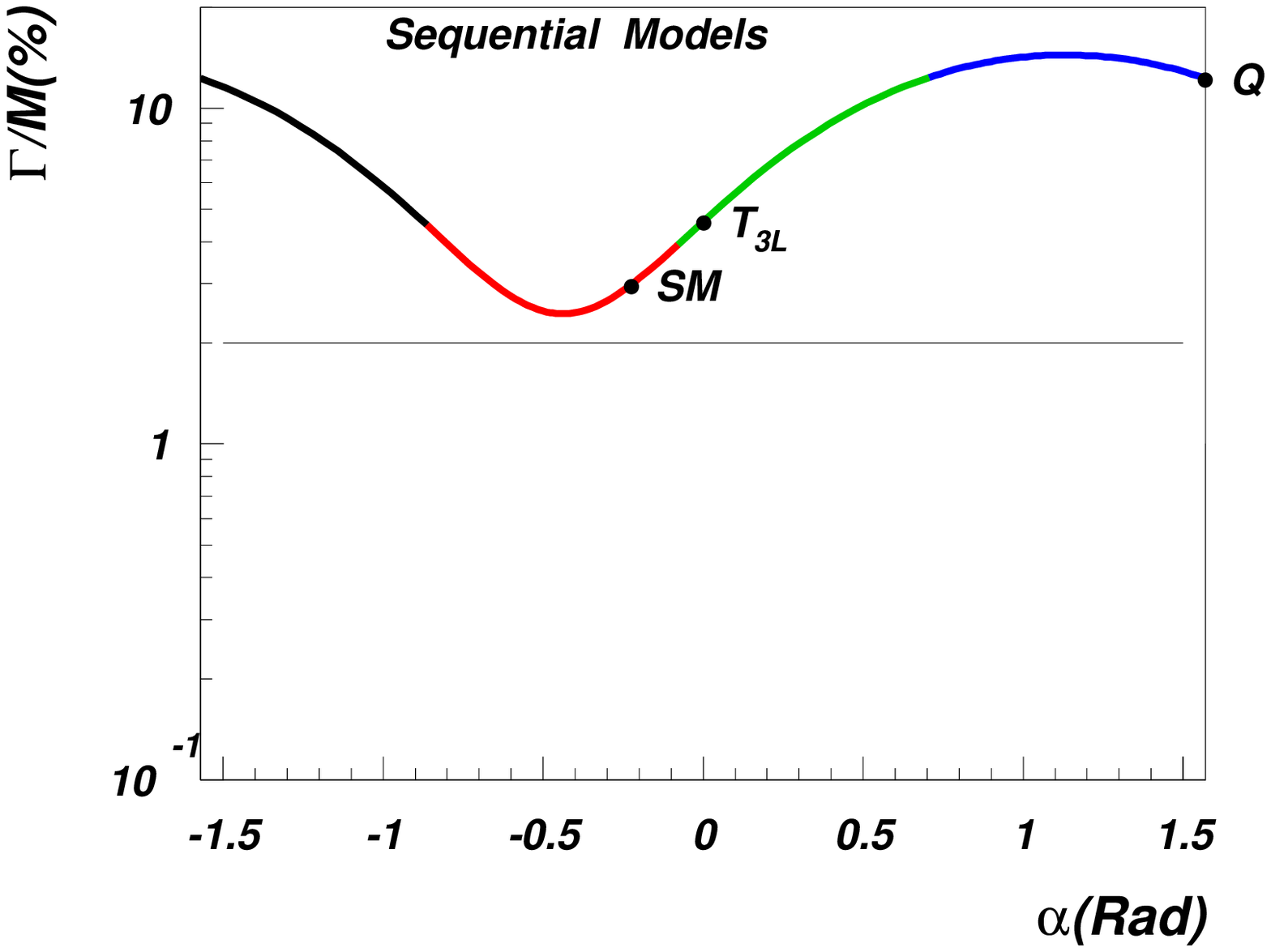}
\caption{$Z^\prime$ boson width over its mass as a function of the mixing 
angle parametrizing the $E_6$, GLR and GSM class of models given in 
Table~\ref{models1}. The colour code corresponds to four equidistant 
intervals in the $[-\pi/2,\pi/2]$ region. The black dots on the contours 
correspond to the benchmark models listed in Table~\ref{models1}.}
\label{fig:width2}
\end{figure}

\begin{figure}[]
\centering
\includegraphics[width=0.49\textwidth]{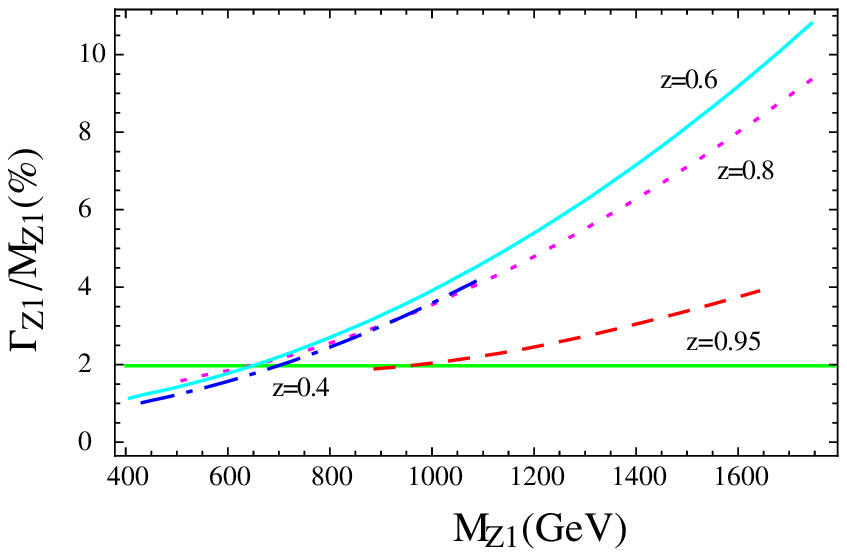}
\includegraphics[width=0.49\textwidth]{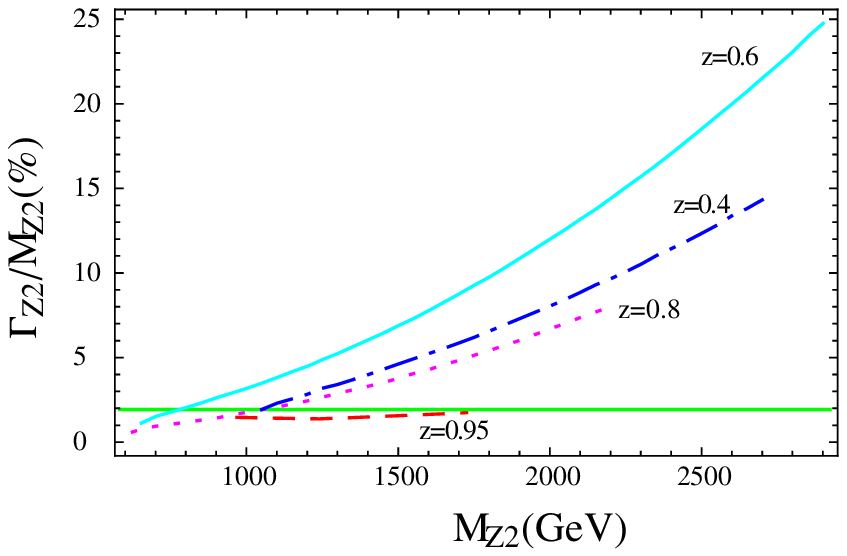}
\caption{Left: $Z_1$-boson width over its mass as a function of $M_{Z1}$ for
different values of the free $z$-parameter. Right: same for the $Z_2$-boson.
We assume the maximal $Z_2$-boson coupling to SM fermions.}
\label{fig:width-4s}
\end{figure}

All extra $U^\prime (1)$ theories summarized in Table \ref{models1} make the 
assumption, for sake of simplicity, that the $Z^\prime$-boson decays purely 
into SM fermions. Under this approximation, the total decay width is given by
Eq.\ref{total_width} and its value never exceeds a few percent of the 
corresponding mass ($\Gamma_{Z^\prime}/M_{Z^\prime}\le 3\%$), as shown in 
Table \ref{models2}. 
This property has a direct implication on the possibility to measure the 
$Z^\prime$ decay width at the LHC, being correlated to the experimental 
di-lepton mass resolution. 
If indeed $\Gamma_{Z^\prime}\ge R$, being $R$ the mass resolution, one can 
have direct access to the decay width of the observed spin-1 particle. During 
the early stage of the 7 TeV LHC, the expected di-electron mass resolution is 
about $R_{LHC}=2\%M$ \cite{Bayatian:2006zz,:1999fq}. As a consequence, within the majority
of models summarized in Table \ref{models2} the total $Z^\prime$-boson width 
is hardly measurable. This is visualized in Fig.~\ref{fig:width2}, where the 
ratio between $Z^\prime$ width and mass is plotted as a function of the 
mixing angle parametrizing the three classes of models listed in 
Table~\ref{models1}: $E_6$, GLR and GSM. All $E_6$ inspired models predict
a quite narrow $Z^\prime$ boson. For some benchmark model within the
generalized Left-Right class (GLR), the ratio becomes slightly bigger than 
the early LHC resolution. The scenario changes when we consider generalized 
sequential models (GSM). Here, the width over mass ratio is well above the 
early 2$\%$ resolution of the LHC. 
Another example of measurable decay width is given by the four-site Higgsless 
model. Here, in fact, in most part of the parameter space  
$\Gamma_{Z_{1,2}}/M_{Z_{1,2}}\ge 2\%$. This property is shown in 
Fig.\ref{fig:width-4s} for different values of the free $z$ parameter. In 
this case, the distinctive behaviour is due to the fact that the 
$Z_{1,2}$-bosons predicted by the four-site model decay preferebly into the 
diboson channel: $Z_1\rightarrow WW$ and $Z_2\rightarrow W_1W$. Thus, their 
width grows with the third power of the corresponding mass, as shown in 
Eqs.\ref{z1width},\ref{z2width}, making it wider compared to the other models.
This feature is common to all Higgsless and Technicolor models.

This discussion is appropriate when considering the $Z^\prime$ boson production
in the di-electron DY channel. For di-muon final states, the analysis would
change drastically. The estimated di-muon mass resolution during the early 
stage of the 7 TeV LHC is in fact around $R\simeq 8\%$ 
\cite{Bayatian:2006zz,:1999fq}.
Hence, only in a very restricted range of the GSM mixing angle the 
$Z^\prime$ width could be measurable. An exception is given by the four-site 
Higgsless model, where the $Z_2$ boson width could be determined in a large 
portion of the parameter space.  

The discussed differences between classes of models should be taken into 
account for improving search strategies and possibly measurements.


\section{summary and conclusions \label{VI}}

In this paper we have discussed the prospects for setting limits on or discovering spin-1 $Z'$
bosons using early LHC data at 7 TeV.  In order to facilitate the connection between experimental
data and theoretical models, we have advocated the narrow width approximation, in which the
leptonic Drell-Yan $Z'$ boson production cross-section only depends on the $Z'$ boson mass
together with two parameters $c_u$ and $c_d$. These variables provide a convenient way of
expressing the experimental limits or discovery information about the mass and cross-section
which enables a direct comparison to be made with the predictions arising from theoretical models. The
experimental limits on the $Z'$ boson cross-section may be expressed as contours in the $c_u-c_d$
plane, with a unique contour for each value of $Z'$ boson mass. If a discovery is made then the
measurement of the mass and cross-section corresponds to  some unique contour, or in practice a
unique band including the error. Such contours may be compared to the theoretical prediction of
$c_u$ and $c_d$ arising from particular models,  enabling limits to be set on models or a
discovery to discriminate between particular models. However the application of this strategy
requires the experimental cross-sections to be properly defined and the theoretical
cross-sections to be accurately calculated as we now discuss.

On the experimental side, we have seen that the use of the narrow width approximation requires an
appropriate di-lepton invariant mass window cut around the mass of the $Z'$ boson. The effect of
the cuts is rather subtle since it depends on both the width of the $Z'$ boson and the energy of
the collider, with higher widths and lower collider energies leading to more prominent signal
tails at low invariant masses. Fortunately we have seen that at LHC energies the suitable
experimental cut is rather insensitive, especially for models with lower $Z'$ boson widths, and
furthermore may be optimised at a unique value suitable for all models, although there is some
unavoidable dependence on the $Z'$ mass. One important conclusion is that whatever cut is chosen
should involve invariant masses well above 100 GeV otherwise interference effects will be
dominant. In summary, we have demonstrated that this cut plays a crucial role: it diminishes a possibly
huge   model-dependent interference effect, removes model-dependent shape of the
$M_{\ell^+\ell^-}$ distributions in the region  of low $M_{\ell^+\ell^-}$ especially for the case
of  large $Z'$ width and brings into agreement NWA and FWA cross sections.
On the theoretical side, we have evaluated cross-sections at  NNLO 
using updated  ZWPROD package. One should stress that
$K_{NNLO}$ factors are depend on the $Z'$ mass
and we have tabulated them for convenience for both the Tevatron
and LHC. Moreover the $K_{NNLO}$ are very PDF dependent and one should specify which PDF is being
used to apply a respective $K_{NNLO}$.

We have applied the approach above to two quite different types of 
$Z'$ models: perturbative gauge models and strongly coupled
models. Among the perturbative gauge models we have studied three classes:  $E_6$ models,
left-right symmetric models and sequential standard models. Each class of model is defined in
terms of a continuous {mixing angle} variable.  This enabled infinite classes of benchmark models to be
defined, rather than just a finite number of models, where for each class of model, the
respective angles serve to parametrize different orbits in the $c_u$-$c_d$ plane.
These orbits turn to be non-overlapping for these three classes of models. A limitation of this approach is that it
ignores the effect of the SUSY and exotics (and right-handed neutrinos) on the width
$\Gamma_{Z'}$. 
Assuming only SM particles in the final state we have calculated the widths of the
benchmark classes of models and seen that the perturbative models generally involve relatively
narrow widths (which however can be increased if SUSY and exotics are
included in the decays), while the strongly coupled models involve multiple $Z'$ bosons with
rather broad widths.
We have also commented on the significance of the width on search strategies
which if measured would a complementary for a discrimination of the underlying $Z'$ model
as well as it would allow to test non-SM $Z'$ decays menationed above.
Another limitation of our approach is that it ignores the effects of $Z-Z'$ mixing  which is
quite model dependent. However such effects must be small due to the constraints from electroweak
precision measurements, so such effects will not have a major effect on direct collider searches
considered here, although of course they will affect the precise vector and axial couplings (see
for example  \cite{King:2005jy,King:2005my,Howl:2007zi} where the $U(1)_N$ vector and axial
vector couplings are  calculated including the mixing effects).  As regards the strongly coupled
models, we only considered the four-site Higgless model which contains just two excitations of
the $Z$ (and $W$) bosons. However it is representative
of models  of  walking technicolour models and 
the KK excitations of the $Z$(and $W$) which is considered  for the first time in the
$c_u$-$c_d$ plane. It is clearly seen that the associated  $Z'$ bosons may easily be
distinguished from those of the perturbative gauge models.

In conclusion, our results support the use of the narrow width approximation in
which the leptonic Drell-Yan $Z'$ boson production cross-section only depends on
the $Z'$ boson mass together with two parameters $c_u$ and $c_d$. However, as
discussed in this paper, care must be taken concerning the experimental cuts and
the theoretical $K_{NNLO}$ factors tabulated here must be included correctly.
Providing the experimental cross-section is appropriately defined, according to
the recipe we provide in Fig.\ref{fig:nwa}, and the theoretical cross-sections
are properly calculated at NNLO, we have shown that such a strategy is safe,
convenient and provides the most unbiased way of comparing experiment to
theoretical models which avoids any built-in model dependent assumptions.  
The
experimental limits or  discovery bands may then be reliably confronted with the
theoretical predictions on the $c_u$-$c_d$ plane as shown in
Fig.\ref{fig:d010-limits},\ref{fig:cms07_500pb},\ref{fig:cms07_500pb-disc}, which
represent the main results of our study, leading to the new limits which we
derive here for the Tevatron and to the projected limits for LHC. 
The results
show that the LHC at 7 TeV with as little data as 500 pb$^{-1}$ can either
greatly improve on current Tevatron mass limits, or discover a $Z'$, with a
measurement of the mass and cross-section providing a powerful discriminator
between the benchmark models using this approach.

\acknowledgments{
We would like to thank Pavel Nadolsky, Alexander Pukhov, Douglas Ross
and Ian Tomalin
for stimulating discussions.
A.B. and S.K. would like to thank Chriss Hays and Sam Harper for help
with understanding details of the experimental analysis at CDF and CMS.
A.B. would like to acknowlegde a support of
International Joint Project  Royal Society Grant  \#JP090598.
L.F. would like to thank the School of Physics $\&$ Astronomy of the  
University of Southampton for hospitality.
}

\bibliographystyle{apsrev}
\bibliography{bib}

\newpage
\appendix
\section{$K_{NNLO}$-factors and cross sections for the $p\bar p(p)\to Z'+X$ 
production process at the Tevatron and the LHC@7TeV
\label{append}}

\begin{table}[htb]
\begin{tabular}{| l | l | l | l | l | l | l |}
\hline\hline
&\multicolumn{3}{c|}{}&\multicolumn{3}{c|}{}\\
$M_{Z'}$~(GeV) & \multicolumn{3}{c|}{Tevatron} 	    & \multicolumn{3}{c|}{LHC@7TeV}\\
\cline{2-7}
              & $\sigma_{LO}$~(pb) & $\sigma_{NNLO}$~(pb) & $K_{NNLO}$ 
              & $\sigma_{LO}$~(pb) & $\sigma_{NNLO}$~(pb) & $K_{NNLO}$    \\
\hline
$ 100 $ & $3.96\times 10^3	$ & $5.57\times 10^3	$ & $  1.41 	$ & $1.63\times 10^4	$ & $2.12\times 10^4	$ & $ 1.29  $ \\
$ 150 $ & $1.08\times 10^3	$ & $1.56\times 10^3	$ & $  1.44 	$ & $4.45\times 10^3	$ & $5.87\times 10^3	$ & $ 1.32  $ \\
$ 200 $ & $4.15\times 10^2	$ & $6.09\times 10^2	$ & $  1.47 	$ & $1.70\times 10^3	$ & $2.28\times 10^3	$ & $ 1.34  $ \\
$ 250 $ & $1.90\times 10^2	$ & $2.83\times 10^2	$ & $  1.49 	$ & $7.89\times 10^2	$ & $1.06\times 10^3	$ & $ 1.35  $ \\
$ 300 $ & $9.67\times 10	$ & $1.45\times 10^2	$ & $  1.50 	$ & $4.14\times 10^2	$ & $5.60\times 10^2	$ & $ 1.35  $ \\
$ 350 $ & $5.25\times 10	$ & $7.91\times 10	$ & $  1.51 	$ & $2.37\times 10^2	$ & $3.20\times 10^2	$ & $ 1.35  $ \\
$ 400 $ & $2.96\times 10	$ & $4.49\times 10	$ & $  1.51 	$ & $1.44\times 10^2	$ & $1.95\times 10^2	$ & $ 1.35  $ \\
$ 450 $ & $1.72\times 10	$ & $2.61\times 10	$ & $  1.52 	$ & $9.20\times 10	$ & $1.24\times 10^2	$ & $ 1.35  $ \\
$ 500 $ & $1.02\times 10	$ & $1.54\times 10	$ & $  1.52	$ & $6.10\times 10	$ & $8.22\times 10	$ & $ 1.35  $ \\
$ 550 $ & $6.05 		$ & $9.20		$ & $  1.52	$ & $4.16\times 10	$ & $5.60\times 10	$ & $ 1.34  $ \\
$ 600 $ & $3.62 		$ & $5.51		$ & $  1.52	$ & $2.92\times 10	$ & $3.91\times 10	$ & $ 1.34  $ \\
$ 650 $ & $2.17 		$ & $3.30		$ & $  1.52	$ & $2.08\times 10	$ & $2.79\times 10	$ & $ 1.34  $ \\
$ 700 $ & $1.29 		$ & $1.97		$ & $  1.52	$ & $1.52\times 10	$ & $2.02\times 10	$ & $ 1.33  $ \\
$ 750 $ & $7.68\times 10^{-1}	$ & $1.16 		$ & $  1.52	$ & $1.12\times 10	$ & $1.49\times 10	$ & $ 1.33  $ \\
$ 800 $ & $4.52\times 10^{-1}	$ & $6.83\times 10^{-1} $ & $  1.51	$ & $8.35		$ & $1.11\times 10	$ & $ 1.32  $ \\
$ 850 $ & $2.63\times 10^{-1}	$ & $3.97\times 10^{-1} $ & $  1.51	$ & $6.30		$ & $8.32		$ & $ 1.32  $ \\
$ 900 $ & $1.51\times 10^{-1}	$ & $2.28\times 10^{-1} $ & $  1.51	$ & $4.80		$ & $6.32		$ & $ 1.32  $ \\
$ 950 $ & $8.52\times 10^{-2}	$ & $1.28\times 10^{-1} $ & $  1.50	$ & $3.69		$ & $4.84		$ & $ 1.31  $ \\
$1000 $ & $4.72\times 10^{-2}	$ & $7.11\times 10^{-2} $ & $  1.51	$ & $2.86		$ & $3.73		$ & $ 1.31  $ \\
$1050 $ & $2.56\times 10^{-2}	$ & $3.85\times 10^{-2} $ & $  1.50	$ & $2.23		$ & $2.90		$ & $ 1.30  $ \\
$1100 $ & $1.35\times 10^{-2}	$ & $2.03\times 10^{-2} $ & $  1.50	$ & $1.74		$ & $2.26		$ & $ 1.30  $ \\
$1150 $ & $6.95\times 10^{-3}	$ & $1.04\times 10^{-2} $ & $  1.50	$ & $1.37		$ & $1.78		$ & $ 1.29  $ \\
$1200 $ & $3.45\times 10^{-3}	$ & $5.20\times 10^{-3} $ & $  1.51	$ & $1.09		$ & $1.40		$ & $ 1.29  $ \\
$1250 $ & $1.65\times 10^{-3}	$ & $2.49\times 10^{-3} $ & $  1.51	$ & $8.63\times 10^{-1} $ & $1.11		$ & $ 1.29  $ \\
$1300 $ & $7.52\times 10^{-4}	$ & $1.14\times 10^{-3} $ & $  1.52	$ & $6.88\times 10^{-1} $ & $8.82\times 10^{-1} $ & $ 1.28  $ \\
$1350 $ & $3.25\times 10^{-4}	$ & $4.95\times 10^{-4} $ & $  1.52	$ & $5.50\times 10^{-1} $ & $7.03\times 10^{-1} $ & $ 1.28  $ \\
$1400 $ & $1.32\times 10^{-4}	$ & $2.02\times 10^{-4} $ & $  1.53	$ & $4.41\times 10^{-1} $ & $5.63\times 10^{-1} $ & $ 1.28  $ \\
$1450 $ & $4.97\times 10^{-5}	$ & $7.66\times 10^{-5} $ & $  1.54	$ & $3.55\times 10^{-1} $ & $4.51\times 10^{-1} $ & $ 1.27  $ \\
$1500 $ & $1.71\times 10^{-5}	$ & $2.65\times 10^{-5} $ & $  1.55	$ & $2.86\times 10^{-1} $ & $3.63\times 10^{-1} $ & $ 1.27  $ \\
$1550 $ & $			$ & $	   		$ & $	    	$ & $2.31\times 10^{-1} $ & $2.92\times 10^{-1} $ & $ 1.27  $ \\
$1600 $ & $			$ & $	   		$ & $	   	$ & $1.87\times 10^{-1} $ & $2.36\times 10^{-1} $ & $ 1.26  $ \\
$1650 $ & $			$ & $	   		$ & $	   	$ & $1.52\times 10^{-1} $ & $1.91\times 10^{-1} $ & $ 1.26  $ \\
$1700 $ & $			$ & $	   		$ & $	   	$ & $1.23\times 10^{-1} $ & $1.55\times 10^{-1} $ & $ 1.26  $ \\
$1750 $ & $			$ & $	   		$ & $	   	$ & $1.00\times 10^{-1} $ & $1.26\times 10^{-1} $ & $ 1.25  $ \\
$1800 $ & $			$ & $	   		$ & $	   	$ & $8.17\times 10^{-2} $ & $1.02\times 10^{-1} $ & $ 1.25  $ \\
$1850 $ & $			$ & $	   		$ & $	   	$ & $6.67\times 10^{-2} $ & $8.31\times 10^{-2} $ & $ 1.25  $ \\
$1900 $ & $			$ & $	   		$ & $	   	$ & $5.44\times 10^{-2} $ & $6.78\times 10^{-2} $ & $ 1.25  $ \\
$1950 $ & $			$ & $	   		$ & $	   	$ & $4.45\times 10^{-2} $ & $5.53\times 10^{-2} $ & $ 1.24  $ \\
$2000 $ & $			$ & $	   		$ & $	   	$ & $3.64\times 10^{-2} $ & $4.51\times 10^{-2} $ & $ 1.24  $ \\
$2050 $ & $			$ & $	   		$ & $	   	$ & $2.98\times 10^{-2} $ & $3.69\times 10^{-2} $ & $ 1.24  $ \\
$2100 $ & $			$ & $	   		$ & $	   	$ & $2.44\times 10^{-2} $ & $3.02\times 10^{-2} $ & $ 1.24  $ \\
$2150 $ & $			$ & $	   		$ & $	   	$ & $2.00\times 10^{-2} $ & $2.47\times 10^{-2} $ & $ 1.23  $ \\
$2200 $ & $			$ & $	   		$ & $	   	$ & $1.64\times 10^{-2} $ & $2.02\times 10^{-2} $ & $ 1.23  $ \\
$2250 $ & $			$ & $	   		$ & $	   	$ & $1.34\times 10^{-2} $ & $1.65\times 10^{-2} $ & $ 1.23  $ \\
$2300 $ & $			$ & $	   		$ & $	   	$ & $1.10\times 10^{-2} $ & $1.35\times 10^{-2} $ & $ 1.23  $ \\
$2350 $ & $			$ & $	   		$ & $	   	$ & $9.04\times 10^{-3} $ & $1.11\times 10^{-2} $ & $ 1.23  $ \\
$2400 $ & $			$ & $	   		$ & $	   	$ & $7.41\times 10^{-3} $ & $9.08\times 10^{-3} $ & $ 1.23  $ \\
$2450 $ & $			$ & $	   		$ & $	   	$ & $6.08\times 10^{-3} $ & $7.44\times 10^{-3} $ & $ 1.22  $ \\
$2500 $ & $			$ & $	   		$ & $	   	$ & $4.98\times 10^{-3} $ & $6.09\times 10^{-3} $ & $ 1.22  $ \\
\hline\hline
\end{tabular}
\caption{\label{tab:snlo}
Values of $K_{NNLO}$-factors and cross sections for the $p\bar p(p)\to Z'+X$ 
production process at the Tevatron and the LHC@7TeV, obtained with ZWPROD 
package for  MSTW08 PDF.
}
\end{table}

\begin{table}[htb]
\begin{tabular}{| l | l | l | l | l | l | l |}
\hline\hline
&\multicolumn{3}{c|}{}&\multicolumn{3}{c|}{}\\
$M_{Z'}$~(GeV) & \multicolumn{3}{c|}{Tevatron} 	    & \multicolumn{3}{c|}{LHC@7TeV}\\
\cline{2-7}
              & $\sigma_{LO}$~(pb) & $\sigma_{NNLO}$~(pb) & $K_{NNLO}$ 
              & $\sigma_{LO}$~(pb) & $\sigma_{NNLO}$~(pb) & $K_{NNLO}$    \\
\hline
$ 100 $ & $3.94\times 10^3	$ & $5.40\times 10^3	$ & $  1.37	$ & $1.52\times 10^4	$ & $1.95\times 10^4	$ & $ 1.28  $ \\
$ 150 $ & $1.12\times 10^3	$ & $1.54\times 10^3	$ & $  1.37	$ & $4.23\times 10^3	$ & $5.46\times 10^3	$ & $ 1.29  $ \\
$ 200 $ & $4.47\times 10^2	$ & $6.08\times 10^2	$ & $  1.36	$ & $1.64\times 10^3	$ & $2.13\times 10^3	$ & $ 1.30  $ \\
$ 250 $ & $2.12\times 10^2	$ & $2.85\times 10^2	$ & $  1.34	$ & $7.66\times 10^2	$ & $1.00\times 10^3	$ & $ 1.31  $ \\
$ 300 $ & $1.11\times 10	$ & $1.47\times 10^2	$ & $  1.32	$ & $4.04\times 10^2	$ & $5.29\times 10^2	$ & $ 1.31  $ \\
$ 350 $ & $6.14\times 10	$ & $8.03\times 10	$ & $  1.31	$ & $2.32\times 10^2	$ & $3.04\times 10^2	$ & $ 1.31  $ \\
$ 400 $ & $3.54\times 10	$ & $4.57\times 10	$ & $  1.29	$ & $1.42\times 10^2	$ & $1.85\times 10^2	$ & $ 1.31  $ \\
$ 450 $ & $2.10\times 10	$ & $2.67\times 10	$ & $  1.27	$ & $9.08\times 10	$ & $1.18\times 10^2	$ & $ 1.30  $ \\
$ 500 $ & $1.26\times 10	$ & $1.58\times 10	$ & $  1.26	$ & $6.04\times 10	$ & $7.85\times 10	$ & $ 1.30  $ \\
$ 550 $ & $7.61 		$ & $9.46		$ & $  1.24	$ & $4.13\times 10	$ & $5.35\times 10	$ & $ 1.29  $ \\
$ 600 $ & $4.63 		$ & $5.68		$ & $  1.23	$ & $2.90\times 10	$ & $3.74\times 10	$ & $ 1.29  $ \\
$ 650 $ & $2.81 		$ & $3.42		$ & $  1.21	$ & $2.08\times 10	$ & $2.67\times 10	$ & $ 1.28  $ \\
$ 700 $ & $1.70 		$ & $2.05		$ & $  1.20	$ & $1.51\times 10	$ & $1.94\times 10	$ & $ 1.28  $ \\
$ 750 $ & $1.03\times 10^{-1}	$ & $1.22		$ & $  1.19	$ & $1.12\times 10	$ & $1.42\times 10	$ & $ 1.27  $ \\
$ 800 $ & $6.12\times 10^{-1}	$ & $7.21\times 10^{-1} $ & $  1.18	$ & $8.38		$ & $1.06\times 10	$ & $ 1.26  $ \\
$ 850 $ & $3.61\times 10^{-1}	$ & $4.22\times 10^{-1} $ & $  1.17	$ & $6.35		$ & $7.98		$ & $ 1.26  $ \\
$ 900 $ & $2.11\times 10^{-1}	$ & $2.44\times 10^{-1} $ & $  1.16	$ & $4.85		$ & $6.06		$ & $ 1.25  $ \\
$ 950 $ & $1.21\times 10^{-2}	$ & $1.39\times 10^{-1} $ & $  1.15	$ & $3.74		$ & $4.64		$ & $ 1.24  $ \\
$1000 $ & $6.80\times 10^{-2}	$ & $7.77\times 10^{-2} $ & $  1.14	$ & $2.90		$ & $3.59		$ & $ 1.24  $ \\
$1050 $ & $3.75\times 10^{-2}	$ & $4.26\times 10^{-2} $ & $  1.13	$ & $2.27		$ & $2.78		$ & $ 1.23  $ \\
$1100 $ & $2.02\times 10^{-2}	$ & $2.28\times 10^{-2} $ & $  1.13	$ & $1.78		$ & $2.18		$ & $ 1.22  $ \\
$1150 $ & $1.06\times 10^{-3}	$ & $1.19\times 10^{-2} $ & $  1.13	$ & $1.41		$ & $1.71		$ & $ 1.21  $ \\
$1200 $ & $5.36\times 10^{-3}	$ & $6.04\times 10^{-3} $ & $  1.13	$ & $1.12		$ & $1.35		$ & $ 1.21  $ \\
$1250 $ & $2.62\times 10^{-3}	$ & $2.96\times 10^{-3} $ & $  1.13	$ & $8.88\times 10^{-1} $ & $1.07		$ & $ 1.20  $ \\
$1300 $ & $1.23\times 10^{-3}	$ & $1.39\times 10^{-3} $ & $  1.13	$ & $7.10\times 10^{-1} $ & $8.48\times 10^{-1} $ & $ 1.19  $ \\
$1350 $ & $5.49\times 10^{-4}	$ & $6.24\times 10^{-4} $ & $  1.14	$ & $5.70\times 10^{-1} $ & $6.77\times 10^{-1} $ & $ 1.19  $ \\
$1400 $ & $2.31\times 10^{-4}	$ & $2.65\times 10^{-4} $ & $  1.15	$ & $4.58\times 10^{-1} $ & $5.42\times 10^{-1} $ & $ 1.18  $ \\
$1450 $ & $9.08\times 10^{-5}	$ & $1.05\times 10^{-5} $ & $  1.16	$ & $3.70\times 10^{-1} $ & $4.35\times 10^{-1} $ & $ 1.18  $ \\
$1500 $ & $3.28\times 10^{-5}	$ & $3.86\times 10^{-5} $ & $  1.18	$ & $2.99\times 10^{-1} $ & $3.50\times 10^{-1} $ & $ 1.17  $ \\
$1550 $ & $			$ & $	   		$ & $	    	$ & $2.42\times 10^{-1} $ & $2.82\times 10^{-1} $ & $ 1.16  $ \\
$1600 $ & $			$ & $	   		$ & $	   	$ & $1.97\times 10^{-1} $ & $2.28\times 10^{-1} $ & $ 1.16  $ \\
$1650 $ & $			$ & $	   		$ & $	   	$ & $1.60\times 10^{-1} $ & $1.85\times 10^{-1} $ & $ 1.15  $ \\
$1700 $ & $			$ & $	   		$ & $	   	$ & $1.31\times 10^{-1} $ & $1.50\times 10^{-1} $ & $ 1.15  $ \\
$1750 $ & $			$ & $	   		$ & $	   	$ & $1.06\times 10^{-1} $ & $1.22\times 10^{-1} $ & $ 1.14  $ \\
$1800 $ & $			$ & $	   		$ & $	   	$ & $8.70\times 10^{-2} $ & $9.92\times 10^{-1} $ & $ 1.14  $ \\
$1850 $ & $			$ & $	   		$ & $	   	$ & $7.12\times 10^{-2} $ & $8.09\times 10^{-2} $ & $ 1.14  $ \\
$1900 $ & $			$ & $	   		$ & $	   	$ & $5.83\times 10^{-2} $ & $6.60\times 10^{-2} $ & $ 1.13  $ \\
$1950 $ & $			$ & $	   		$ & $	   	$ & $4.77\times 10^{-2} $ & $5.40\times 10^{-2} $ & $ 1.13  $ \\
$2000 $ & $			$ & $	   		$ & $	   	$ & $3.92\times 10^{-2} $ & $4.41\times 10^{-2} $ & $ 1.13  $ \\
$2050 $ & $			$ & $	   		$ & $	   	$ & $3.21\times 10^{-2} $ & $3.61\times 10^{-2} $ & $ 1.13  $ \\
$2100 $ & $			$ & $	   		$ & $	   	$ & $2.64\times 10^{-2} $ & $2.96\times 10^{-2} $ & $ 1.12  $ \\
$2150 $ & $			$ & $	   		$ & $	   	$ & $2.17\times 10^{-2} $ & $2.43\times 10^{-2} $ & $ 1.12  $ \\
$2200 $ & $			$ & $	   		$ & $	   	$ & $1.78\times 10^{-2} $ & $2.00\times 10^{-2} $ & $ 1.12  $ \\
$2250 $ & $			$ & $	   		$ & $	   	$ & $1.46\times 10^{-2} $ & $1.64\times 10^{-2} $ & $ 1.12  $ \\
$2300 $ & $			$ & $	   		$ & $	   	$ & $1.20\times 10^{-2} $ & $1.35\times 10^{-2} $ & $ 1.12  $ \\
$2350 $ & $			$ & $	   		$ & $	   	$ & $9.88\times 10^{-3} $ & $1.11\times 10^{-2} $ & $ 1.12  $ \\
$2400 $ & $			$ & $	   		$ & $	   	$ & $8.12\times 10^{-3} $ & $9.10\times 10^{-3} $ & $ 1.12  $ \\
$2450 $ & $			$ & $	   		$ & $	   	$ & $6.67\times 10^{-3} $ & $7.48\times 10^{-3} $ & $ 1.12  $ \\
$2500 $ & $			$ & $	   		$ & $	   	$ & $5.48\times 10^{-3} $ & $6.15\times 10^{-3} $ & $ 1.12  $ \\
\hline\hline

\end{tabular}
\caption{\label{tab:snlo2}
Values of $K_{NNLO}$-factors and cross sections for the $p\bar p(p)\to Z'+X$ 
production process at the Tevatron and the LHC@7TeV, obtained with ZWPROD 
package for CTEQ6.6 PDF.
}
\end{table}

\end{document}